\documentclass[twocolumn]{aastex701} 
\usepackage{booktabs}
\usepackage{amsmath}
\usepackage{subcaption} 
\usepackage[percent]{overpic} 
\usepackage{cleveref}
\usepackage{graphicx}
\usepackage{comment}
\usepackage{multirow}
\usepackage{hyperref}
\hypersetup{
    colorlinks=true,       
    linkcolor=green,        
    citecolor=blue,       
    urlcolor=cyan,          
    pagebackref=true     
}

\newcommand{\kms}{{\,{\rm km}\,{\rm s}^{-1}}}
\newcommand{\obsunits}{\,{\rm K\,km\,s^{-1}\,pc^2}}
\newcommand{\skirt}{\texttt{SKIRT}}
\newcommand{\cloudy}{\texttt{CLOUDY}}
\newcommand{\fire}{\texttt{FIRE}}
\newcommand{\fireTwo}{\texttt{FIRE-2}}
\newcommand{\core}{\texttt{Core}}
\newcommand{\tripleLatte}{\texttt{Triple Latte}}
\newcommand{\massivefire}{\texttt{MassiveFIRE}}
\newcommand{\firebox}{\texttt{FIREbox}}

\newcommand{\websky}{\texttt{WebSky}}

\newcommand{\LcoOneZeroSolarLum}{\ensuremath{L_{{\rm CO}(1-0)}}}
\newcommand{\LcoOneZero}{\ensuremath{L'_{{\rm CO}(1-0)}}}
\newcommand{\LcoTwoOne}{\ensuremath{L'_{{\rm CO}(2-1)}}}

\newcommand{\Mstar}{\ensuremath{M_{\rm star}}}

\newcommand{\ZgasHl}{\ensuremath{Z_{{\rm gas}(R50)}}}

\newcommand{\Mhalo}{\ensuremath{M_{\rm halo}}}
\newcommand{\sfrHundredMyr}{\ensuremath{{\rm SFR}_{100\,{\rm Myr}}}}
\newcommand{\Mgas}{\ensuremath{M_{\rm gas}}}
\newcommand{\LFIR}{\ensuremath{L_{\rm FIR}}}
\newcommand{\RvisibleHl}{\ensuremath{R50}}

\newcommand{\cm}{{\rm cm}}
\newcommand{\s}{{\rm s}}
\newcommand{\g}{{\rm g}}
\newcommand{\pc}{{\rm pc}}
\newcommand{\kpc}{{\rm kpc}}
\newcommand{\ergs}{{\rm erg\,s}^{-1}}
\newcommand{\Myr}{\,{\rm Myr}}
\newcommand{\yr}{\,{\rm yr}}

\newcommand{\Rs}{{R_{\rm stop}}}

\newcommand{\lambdaCDM}{\ensuremath{\Lambda{\rm CDM}}}

\newcommand{\RBFInterpolator}{\texttt{RBFInterpolator}}
\newcommand{\LinearNDInterpolator}{\texttt{LinearNDInterpolator}}

\newcommand{\be}{\begin{equation}}
\newcommand{\ee}{\end{equation}}

\defcitealias{Paper2}{Paper~II}
\defcitealias{Paper3}{Paper~III}

\begin{document}

\title{Modeling CO Line Luminosity Using FIRE Simulations: I. Methodology and Validation of Molecular Gas and CO Abundance}

\author[0000-0002-3155-946X]{Doğa Tolgay}
\affiliation{Canadian Institute for Theoretical Astrophysics, University of Toronto, 60 St. George Street, Toronto, ON M5S 3H8, Canada}
\affiliation{University of Toronto Physics Department \\
McLennan Physical Laboratories, 60 St George St \\
     Toronto, ON M5S 1A7, CANADA}
\email{doga.tolgay@mail.utoronto.ca}

\author[0000-0002-8659-3729]{Norman W. Murray} 
\affiliation{Canadian Institute for Theoretical Astrophysics, University of Toronto, 60 St. George Street, Toronto, ON M5S 3H8, Canada}
\email{murray@cita.utoronto.ca}

\author[0000-0003-2618-6504]{Dongwoo Chung}
\affiliation{Department of Astronomy, Cornell University, Ithaca, NY 14853, USA}
\email{dongwooc@cornell.edu}

\author[0000-0003-2358-9949]{J. Richard Bond}
\affiliation{Canadian Institute for Theoretical Astrophysics, University of Toronto, 60 St. George Street, Toronto, ON M5S 3H8, Canada}
\email{bond@cita.utoronto.ca}

\author[0000-0002-2731-7708]{Nathan J. Carlson}
\affiliation{Canadian Institute for Theoretical Astrophysics, University of Toronto, 60 St. George Street, Toronto, ON M5S 3H8, Canada}
\affiliation{University of Toronto Physics Department \\
McLennan Physical Laboratories, 60 St George St \\
     Toronto, ON M5S 1A7, CANADA}
\email{njcarlson@physics.utoronto.ca}

\author[0000-0002-1109-1919]{Robert Feldmann} 
\affiliation{Department of Astrophysics, University of Zurich, Zurich CH-8057, Switzerland}
\email{robert.feldmann@uzh.ch}

\author[0000-0003-0603-8942]{Andrew Wetzel}
\affiliation{Department of Physics \& Astronomy, University of California, Davis, CA 95616, USA}
\email{awetzel@ucdavis.edu}

\author[0000-0003-4070-497X]{Guochao Sun} 
\affiliation{CIERA and Department of Physics and Astronomy, Northwestern University, 1800 Sherman Ave, Evanston, IL 60201, USA}
\email{guochao.sun@northwestern.edu}

\begin{abstract}
We present a post-processing pipeline that predicts CO line luminosities, molecular hydrogen masses, and CO abundances for galaxies in the Feedback in Realistic Environments (\texttt{FIRE-2}) simulations, as a first step toward CO response functions for line intensity mapping mocks. \texttt{SKIRT} gives the radiation field reaching each gas cell. \texttt{CLOUDY}, tabulated on a five-parameter grid (metallicity, density, turbulent velocity, radiation field, cloud scale) and interpolated, gives $L'_{\rm CO(1-0)}$ and the $\rm H_2$ and CO masses. In this first of three papers we validate the pipeline against observations. For Milky Way (MW) molecular clouds, the predicted $L'_{\rm CO(1-0)}$ rises sharply at $A_V\sim1$ and matches the observed values above that column. For a uniform $3\,Z_\odot$ the summed luminosity is within 35\% of the observed total. A factor of three increase in metallicity changes the total $L'_{\rm CO(1-0)}$ by an order of magnitude. Along MW sight lines, the predicted $N_{\rm CO}/N_{\rm H_2}$ reproduces the observed upturn within 0.3 dex in $N_{\rm H_2}$ but falls below the data at higher columns. At $z=0$ our molecular gas masses agree with xCOLD GASS for ${\rm SFR}\gtrsim0.3\,M_\odot\,\yr^{-1}$ and are converged across our simulation resolutions. At $z\gtrsim1$ observed gas masses exceed ours at fixed stellar mass, though selection effects and conversion-factor assumptions limit this comparison. CO luminosities are compared with extragalactic observations in \citetalias{Paper2}. Capturing the metallicity-dependent dust shielding of CO from photodissociation is essential for modeling CO emission in low-mass and high-redshift galaxies.
\end{abstract}



\section{Introduction} \label{sec:Introduction}

\subsection{Line intensity mapping as a probe of large scale structure}

Cosmic Microwave Background (CMB) experiments currently provide $1\%$ level constraints on the cosmological parameters of the standard cosmological model, Lambda Cold Dark Matter (\lambdaCDM), e.g., \citep{planck_2018}. The standard model explains many properties of our universe, but a number of recent observations, particularly those at redshifts much lower than those at which the CMB formed ($z\approx1,100$), hint at shortcomings of the theory. Perhaps the most notable is the `Hubble tension' \citep{Riess_2016_H0_tension, Poulin_2023_H0_tension}; some estimates of the Hubble constant $H_0$ in the local universe appear to disagree with estimates from measurements of the CMB. To resolve this and other discrepancies, and to constrain the possible evolution of the dark energy equation of state $w(z)$ \citep{Zhao_2017_H0_tension_wz, Valentino_2018_H0_tension_wz, Dinda_2018_LIM_and_cosmology_wz}, it is essential to map the three-dimensional Large Scale Structure (LSS) through cosmic time.

While traditional galaxy surveys offer high-resolution observations of the local Universe, they are restricted to small volumes and bright sources at high redshifts. Line Intensity Mapping (LIM) is emerging as a transformative alternative probe of the growth of large scale structure. By design, LIM experiments have low angular resolution, which prevents them from resolving individual sources. However, LIM surveys can map spatial fluctuations in the integrated line emission from many individually unresolved sources. These sources also include low luminosity and low-surface-brightness objects. The measured line intensity is a biased tracer of the baryonic matter distribution, but the LIM bias differs from that of wide-field optical surveys, so that LIM experiments compliment galaxy surveys. Because LIM experiments benefit from shorter observation times to map large patches of the Universe, they offer a highly cost-effective alternative to traditional galaxy surveys at similar redshifts. The broad frequency coverage of LIM experiments allows them to measure the redshifted line emission along line of sight as a function of cosmic time over a similarly broad range of redshifts, providing three-dimensional tomographic map of the large scale structure of the Universe \citep{Visbal2010_LIM, Kovetz_2017, Kovetz_2019, Bernal_2019, Chang_Lidz_2026_LIM}.   

This tomographic view allows LIM experiments to test theoretical frameworks within and beyond \lambdaCDM. In particular, LIM experiments have the potential to probe primordial non-Gaussianity, measuring the non-linearity parameter $f_{NL}$ with a resolution of $\sigma(f_{NL})\sim1$, e.g., \citet{2019ApJ...872..126M, planck_2018, Krolewski2024_desi_fnl, Chiarenza2026_planckAndDESIcross_fnl}. At the same time, emission lines such as CO \citep{Cleary_2022_comap_earlyScience1}, [\ion{C}{2}] \citep{Yue_cii_LIM, ccat_2023}, H$\alpha$ \citep{SPHEREx_2014, Silva_2018_ha_lim, spherex_2026}, and Ly$\alpha$ \citep{SPHEREx_2014, spherex_2026} can serve as tracers of the cosmic star formation rate (SFR), enabling LIM experiments to provide a more complete census of the star-forming galaxy population across cosmic time.

\subsection{CO LIM Mocks and the Response Functions}


While LIM experiments are poised to make contributions to cosmology, galaxy formation, and gas physics, confirmed carbon monoxide (CO) LIM detections remain elusive. In this context, mock maps, or simulated sky maps, play a crucial role in enabling statistical studies of cosmology and astrophysics. These mock maps include both astrophysical signals and observational effect. They can be used to forecast the scientific potential of upcoming LIM experiments and, once the detections are made, to constrain the physics of the ISM and inform estimates of observational uncertainties. 

In this series of three papers we focus on line emission from carbon monoxide (CO), the most commonly used tracer of molecular gas. Transitions between rotational states of the molecule, denoted by integer values $J$, lead to emission at discrete frequencies or wavelengths. The transition from $J=1$ to $J=0$ (the latter being the non-rotating state), denoted CO(1-0), produces emission at $\nu_0=115$ GHz or $\lambda_0=0.26$ cm. Emission from the $J=2$ to $J=1$ transition, denoted CO(2-1), produces emission at $\lambda_1=0.13$ cm, and so forth. The line luminosity is denoted by \LcoOneZeroSolarLum, with units of solar luminosity $L_\odot$ (where $L_\odot=3.828\times10^{33}\,\ergs$), or by \LcoOneZero$\equiv\int\int T_b\, dv\,dA$, with units $K\,\kms\,\pc^2$, the units favored by observers; $T_b$ is the brightness temperature in Kelvin, $v$ is the line-of-sight velocity across the line profile (in $\kms$), and $A$ is the area of the emitting region, in square parsecs. 

The critical element for producing mock maps for blind CO LIM surveys is the determination of the functional relationship between the CO emission and the dark matter field in N-body dark matter simulations, i.e. the response function (which is a distribution function) $g$, which we present in \citetalias{Paper3}. In such a blind survey, the model for the response function is $L_{CO}=g(M_{\rm halo};z)$. For simplicity, in our analytic fits we will assume that the distribution $g$ is log-normal, i.e., it is characterized by a mean of the log luminosity $\mu_{\log L_{\mathrm CO}}$ and a standard deviation of the log luminosity $\sigma_{\log L_{\mathrm CO}}$.\footnote{We show in \citetalias{Paper3} that the $\LcoOneZero$ distributions we predict are not always well described as log-normal, depending on which variables are used to construct them.}


\begin{figure*}
    \centering
    \includegraphics[width=1\linewidth]{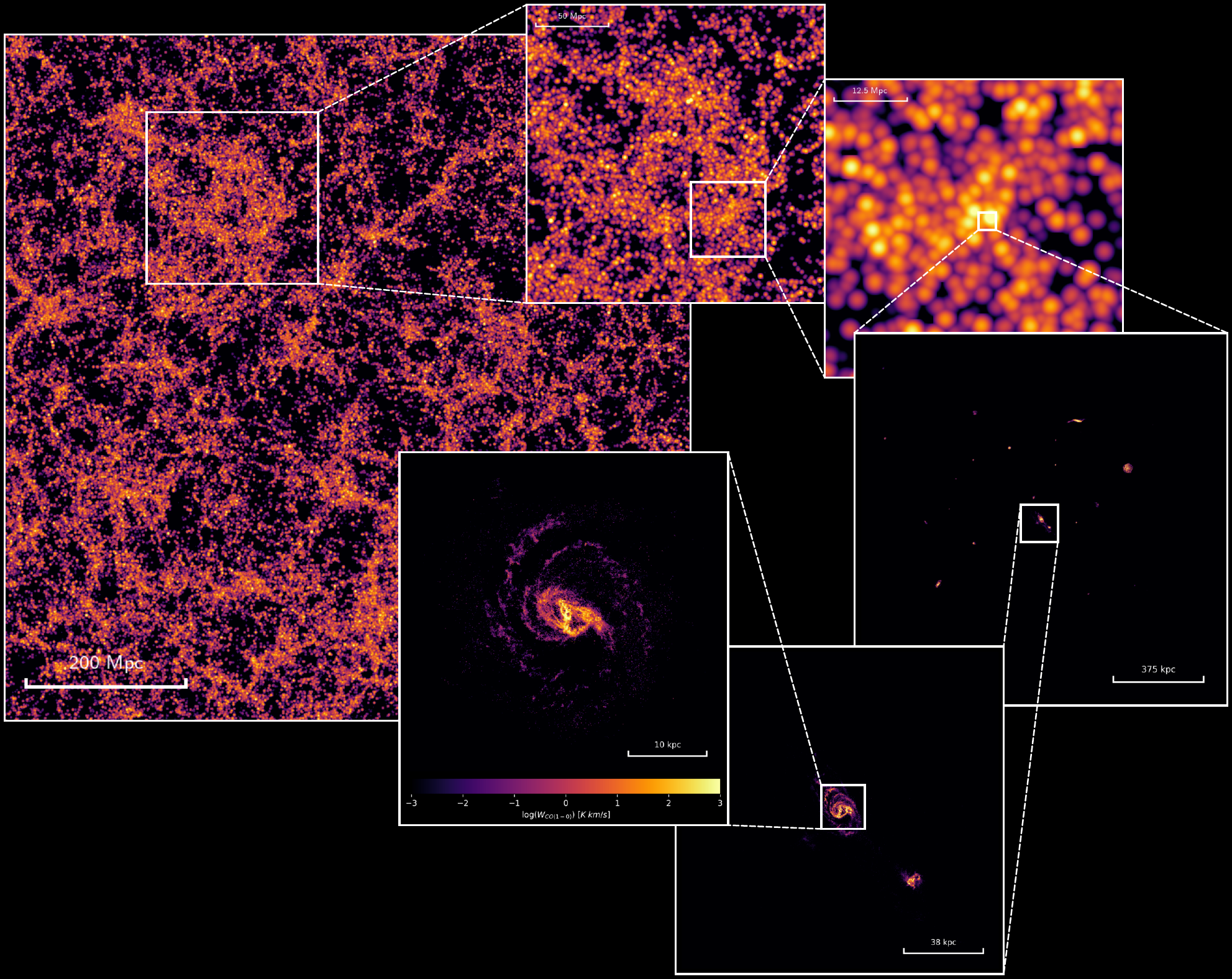}
    \caption{Illustration of a line intensity map produced using the response function formalism. We refer to the six panels by their scale, represented by the white bars, which range from 200 Mpc down to 10 kpc. The first three panels show progressively zoomed-in versions of the large volume (dark matter only) \websky\ maps converted to a CO LIM using $g(\Mhalo; z=0)$ (see Table~1 in Paper~III \citep{Paper3}) convolved with a Gaussian beam with $\sigma\sim0.5~\mathrm{Mpc}$ scales of 200, 50, and 12.5 Mpc). The next two panels show the zoomed-in version of one of the central halos in the \texttt{FIREbox} simulation, again using $g(\Mgas; z=0)$ (tabulated in Table~1 in \citetalias{Paper3}). Because the pixels shown are much smaller than the galaxy-scale systems used to calibrate the response function, these two images are intended as an artistic illustration rather than a quantitative prediction. The last panel is the face-on projection of the central halo in a \texttt{FIREbox} simulation where the map is generated from the CLOUDY-predicted \LcoOneZero\ luminosity of individual gas cells; this image is a quantified testable prediction. Together, these panels demonstrate how response functions can connect large-volume cosmological simulations to line-emission predictions across progressively smaller spatial scales.} 
    \label{fig:peakpatch_firebox_zoom_stacked_6panel}
\end{figure*}

Our goal is to produce mock CO intensity maps over the volumes that LIM experiments actually poise to observe and Figure~\ref{fig:peakpatch_firebox_zoom_stacked_6panel} shows that chain end to end. The first three panels are a mock of the kind a CO survey would measure. To produce this image, we paint the large-volume dark matter only \websky\ maps \citep{Stein_bond_webskyProject} with the $g(\Mhalo; z=0)$ response function of \citetalias{Paper3} and convolved with a LIM-like beam, on scales of 200, 50 and 12.5 Mpc. Instead of explicitly modelling the hydrodynamical gas physics within each halo (at high computational expense), the \websky\ simulations use the FIRE-informed response functions from \citetalias{Paper3} to efficiently map the observable CO line emission onto Peak Patch halos. The next two panels apply the same painting using $g(\Mgas; z=0)$ of \citetalias{Paper3} at giant molecular cloud scales ($\sim50\,\pc$) and are illustrative only, for the same reasons in the caption. The final panel is where this paper discuss. \LcoOneZero\ computed for each gas cell with the radiative post processing (see Section~\ref{sec:post_processing_fire} for more details). It is the generalization of this pipeline to many galaxies that calibrates the response functions used in the first three panel. One can of course do the same for emission lines other than CO. For example, a case study for the mock $[\mathrm{C\,II}]\,158\,\mu\mathrm{m}$ LIM is presented in \cite{horlaville2024_cii_peakPatch, nate_2025} or for rest-frame optical and far-infrared fine structure emission lines presented in \citet{Kannan2022_Thesan_LIMPredictions_multiTracer}.

In much of the current literature response functions are calculated using semi-analytical modeling, e.g. \citet{Gong_2011_coModelling, Sun_2019_coModelling, Yang_2022_coModelling, masRibas2023_LIMfast1, jasonSun2023_LIMfast2_coModelling}, or by using empirical relationships between line emission and galactic observables that are strongly correlated \citep{Lidz_2011_coModelling, pullen_2013_coModelling, Li_2016_coModelling, Keating_2020_coModelling, dongwoo_2022_comap_earlyscience5}. They are vulnerable to systematic issues that introduce substantial uncertainties, particularly in high-redshift studies. Many uncertainties remain regarding the modeling of CO(1-0) power spectrum \citep{Breysse_2016_highredshift_COmocks, dongwoo_2024_COMAP_season2} especially at high redshifts. The discrepancy between the predicted \LcoOneZero\ of different models range over two or three orders of magnitude. Current CO LIM experiments, such as CO Mapping Array Project (COMAP, \cite{Cleary_2022_comap_earlyScience1}), have so far found only upper limits. These limits eliminate some of the most luminous CO models, but other models are still viable. 

Cosmological simulations offer an alternative to the greatly simplified models used so far to estimate \LcoOneZero. We use simulated galaxies from the \fireTwo\ \citep{Hopkins_2018_fire2} project to generate response functions by finding the correlations between \LcoOneZero (or higher transitions)  and galactic observables as well as (unobserved) halo masses. \texttt{FIREbox} \citep{Feldmann_2023}, a full cosmological volume using the same FIRE-2 physics, simulates galaxies down to dwarf masses, so the population that surveys cannot currently reach is present in our sample by construction. Future work will use these response functions to populate dark matter halos with \LcoOneZero\ to make mock maps similar to \cite{nate_2025}.

\subsection{CO as a tracer of molecular gas \label{sec:intro_xco}}

Molecular gas masses are not observed directly, but they are often inferred from CO line emission through an assumed conversion factor. Both observations we compare our molecular gas predictions in Section~\ref{sec:comparsion_mh2_mco_xco_with_obs} assume a conversion factor to infer the molecular gasses of observed galaxies.

The relation between CO emission and $\Sigma_{\rm mol}$ or molecular hydrogen column number densities $\rm N(\rm H_2)=\Sigma_{\rm mol}/(2\mu m_p)$, (with units $\cm^{-2}$) where $m_p$ is the proton mass and $\mu=1.37$ is the mean molecular mass accounting for the presence of Helium, was first derived from Milky Way (MW) spatially resolved observations of giant molecular clouds, e.g., \citet{1984ApJ...276..182S, 1991ARA&A..29..581Y}. Observers characterize the CO emission line flux by the line brightness temperature integrated over the line profile, $W_{CO}\equiv \int T_b dv$ (also denoted $I_{CO}$), which observers report in units of ${\rm K}\kms$.\footnote{In cgs units the line flux (integrated over velocity) is $F_{CO}=10^5\cdot 2k_bW_{CO}/\lambda_1^3\,\ergs\,\cm^{-2}$, where the factor $10^5$ comes from converting $\kms$ to $\cm\,\s^{-1}$.} The relation between $W_{CO}$ and $N_{\rm H_2}$ is then
\begin{equation}
    \rm N(H_2) = X_{CO}\,W_{CO},
    \label{eq:Xco_theoretical_formulation}
\end{equation}
where the Milky Way (MW) value of $X_{\rm CO}$ is 
\begin{equation}
X_{\rm CO_{MW}}\approx2\times 10^{20}\,\cm^{-2}\left({\rm K}\,{\rm \kms}\right)^{-1}
\end{equation}
using observer units 
\citep{1991ARA&A..29..581Y, Leroy_2011_constantXco, Tacconi_2020_annualReviews}.

For spatially unresolved observations, which are common in high redshift surveys or for LIM generally, one measures luminosities rather than fluxes, which involves multiplying the flux (or $W_{CO}$) by the solid angle of the source (or beam, for LIM) and the square of the angular diameter distance, $\LcoOneZero=W_{\rm CO}\Omega_{\rm s} D_{\rm A}^2$ \citep{Solomon_1997_ULIRGs_Xco}. The analog of Equation~(\ref{eq:Xco_theoretical_formulation}) will convert the luminosity to the total number of $H_2$ molecules. More commonly, measured luminosities are converted to molecular gas mass, including the contribution from Helium, using
\be
\label{eq:alpha_co_definition}
M_{mol}=\alpha_{CO}\LcoOneZero,
\ee
where
\be
\alpha_{CO}={2\mu m_p}X_{CO}\approx4.3\,M_\odot\,(\obsunits)^{-1},
\ee
where the numerical value is valid for Milky Way-like galaxies.

While this value is known to be accurate for nearby Milky Way-like disk galaxies, e.g., \citet{Leroy_2011_constantXco}, $X_{\rm CO}$ and $\alpha_{\rm CO}$ are observed to be smaller by a factor of a few in ultra-luminous infrared galaxies, and much larger (by factors up to one hundred) in metal poor galaxies, e.g., figure 9 in  \citet{Bolatto_2013} (also see \cite{Wolfire_2010_coDarkGas, Feldmann2012_Xco_paperI, Madden_2020, bisbas_2015_co_dark_clouds}). The rapid increase of $\alpha_{\rm CO}$ with decreasing metallicity $Z$ is likely due to a rapid reduction in $L_{CO}$ at a fixed gas mass and star formation rate when $Z$ is decreased, a point we return to below. $L_{CO}$ is also observed to vary with stellar mass, e.g. \citet{Cicone_2017_allsmog}.

Two simulation based modeling approaches to model the change in the conversion factor. \citet{Feldmann2012_Xco_paperI} in post-processing of cosmological simulations, put a sub-grid model calibrated on small-scale magnetohydrodynamic simulation of the ISM, reaching to an effective resolution of $\sim0.1\,\pc$. On the scale of giant molecular clouds they found that $X_{\rm CO}$ is a strong function of metallicity, column density and interstellar UV flux; the impact of latter two weakens at $\sim\rm kpc$ scales. In contrast, $X_{\rm CO}$ remains a strong function of metallicity, with $X_{\rm CO}\propto Z^{-\gamma}$ for $\gamma\approx0.5-0.8$. \citet{Feldmann2012_Xco_paperII} extend this analysis to the molecular gas - SFR relation, and argue that on super-kpc scales spatial variations of the conversion factor account for most of the measured $M_{\rm H_2}-\rm SFR$ scatter. \citet{Narayanan_2012} post processed hydrodynamic simulations of isolated discs and galaxy mergers with a subgrid giant molecular cloud model in which the CO line intensities are obtained from an escape probability formalism. They created a fitting formula for $X_{\rm CO}$ in terms of CO intensity, metallicity and gas column. Their results show that $X_{\rm CO}$ rises at low metallicity due to the photodissociation of CO molecules. Conversely, $X_{\rm CO}$ falls in high surface density systems where larger gas temperatures and velocity dispersions raise $W_{\rm CO}$ at fixed $\Sigma_{\rm H_2}$.

\subsection{Predicting CO emission from cosmological galaxy formation models \label{sec:intro_prior_work_CO_from_cosmological_gal_formation}}

Predicting CO emission for a simulated galaxy formation simulation requires pairing a cosmological model with a treatment of the line emission prescription that could be either a radiative transfer code or semi-analytical approximations. For example, \citet{Lagos2012_Lco_simulatedGalaxies} coupled the \texttt{GALFORM} semi-analytical model to the \texttt{UCL\_PDR} code to predict CO rotational lines from CO(1-0) through CO(10-9) transitions from $z=0$ to 6. They use the pressure-based empirical star formation relation of \citet{Blitz_Rosolowsky_2006_pressureH2_relation} to predict the molecular gas content while using \texttt{UCL\_PDR} model to obtain the CO line emissions for each simulated galaxy. They recover the local CO(1-0) luminosity function and the CO-to-infrared luminosity ratios, and attribute departures from the Milky Way conversion factor mainly to metallicity, and secondarily to the mean gas and star formation rate surface densities.

\citet{Inoue2020_COUniverse} instead post-processed the cosmological \texttt{IllustrisTNG} simulation and assumed a subgrid prescription to model molecular clouds for which they compute the line luminosities using \texttt{DESPOTIC}. They were able to produce the observed CO luminosity function at $z=0$, and find that roughly ten per cent of the molecular gas in the local Universe resides in dwarfs with $\rm M_\star<10^9\,M_\odot$ that are largely CO-dark and have high $\alpha_{\rm CO}$.

\citet{Garcia2024_SLICK} similarly applied \texttt{DESPOTIC} to a sub-resolution cloud model (assuming a power-law density profile) for gas particles in the \texttt{SIMBA} simulation, and trained a random forest on a subset of clouds and galaxies to predict CO, [\ion{C}{1}] and [\ion{C}{2}] luminosities for the rest, which they used to build light cones out to $z=10$.

In contrast to both of these works we do not assign the molecular gas through an empirical relation, nor do we assume a subgrid structure for the unresolved scales. We calculate $\rm H_2$ and $\rm CO$ abundances and line luminosities of each simulated gas cell as a function of their simulated properties.

Closest to the present work, two studies have post-processed Feedback in Realistic Environments (\fireTwo, \citealt{Hopkins_2018_fire2}) galaxies to predict CO emission using the same set of codes. Both \citet{Keating_2020} and \citet{Thompson2024_FIRE_CO} take the local radiation field from the \texttt{LEBRON} \citep{Hopkins2020_Lebron} that is also used in the \fire\ simulation. They then use the \texttt{CHIMES} chemistry solver \citep{Richings_2014_CHIMESa, Richings_2014_CHIMESb} to calculate the CO and $\rm H_2$ abundances, and \texttt{RADMC-3D} \citep{Dullemond_2012_RADMC3D} to perform the line radiative transfer in the galaxy.

The two studies differ in their cosmology and chemistry assumptions. \citet{Keating_2020} ran \texttt{CHIMES} in equilibrium on Latte suite of \fireTwo\ cosmological zoom-in simulations of MW-mass galaxies \citep{Wetzel_2016, Hopkins_2018_fire2} at $z=0$, and found that the predicted CO(1-0) emission and $\rm X_{CO}$ (see Equation~\ref{eq:Xco_theoretical_formulation}) depend strongly on the assumed shielding length. Common choices, such as the Jeans length or the gas smoothing length, produce CO abundances that are too high at a given $\rm H_2$ abundance; only an order of magnitude decrease in shielding length, with a median of $\sim3\,\pc$ in cold ($T<300\,\mathrm{K}$) gas, reproduced the observed CO(1-0) luminosity and the inferred MW $X_{\rm CO}$ at a given SFR. They argue that the length of the shielding could be thought of as a subgrid parameter controlling how much dissociating radiation penetrates giant molecular clouds.

\citet{Thompson2024_FIRE_CO} instead ran \texttt{CHIMES} on the fly, allowing for possible out of equilibrium chemistry, on isolated disc galaxies ranging from dwarfs to MW-mass systems, making possible a direct comparison against equilibrium abundances. Their non-equulibrium $X_{\rm CO}$ differs from the equilibrium prediction by less than $5\%$ in massive galaxies, but the $\rm H_2$ masses and CO luminosities change by $\approx25\%$. Both works predict excess CO luminosity at a given star formation rate for $\rm SFR > 1 \,M_\odot\,{\rm yr}^{-1}$.

In this work we use a different set of codes. We calculate the radiation field with \skirt\ \citep{Baes_2011, Baes_2015, Camps_2015, Camps_and_baes_2020_skirt9}, so that dust radiative transfer is done by accounting for realistic scattering and absorption processes. Dust distribution attenuates the dissociating photons before we solve for chemistry, and we then use \cloudy\ \citep{2013RMxAA..49..137F, Ferland_2017, cloudy_version_c25_release} rather than \texttt{CHIMES} to predict CO luminosities. The caveat is that we do not transport the line photons through the galaxy, as \texttt{RADMC-3D} does. We return to this caveat in Section~\ref{sec:discussion_caveats}.

\subsection{This work and this series \label{sec:intro_this_series}}
In this paper, the first of three, we present the post-processing pipeline that produces the last panel of Figure~\ref{fig:peakpatch_firebox_zoom_stacked_6panel} (with a scale bar of 10 kpc), and we test the molecular gas content it predicts. We use the three-dimensional Monte Carlo dust radiative transfer code \skirt\ \citep{Baes_2011, Camps_2015} to compute the interstellar radiation field incident on each gas cell of a \fireTwo\ galaxy \citep{Hopkins_2018_fire2}. We then use the spectral synthesis code \cloudy\ \citep{cloudy_version_c25_release} to compute the resulting line emission and species abundances, given the local density, metallicity, turbulent velocity, radiation field, and cloud size.

We validate the results in three ways: first, we calculate the CO luminosities given the properties of individual Milky Way (MW) molecular clouds, and compare the predictions with the observed luminosity of each of those clouds; second, we calculate CO column densities and compare them to those measured along galactic translucent sight lines;  and third, we compare our predictions to inferred molecular gas masses of observed galaxies.

Paper~II \citep{Paper2} applies the pipeline to the full simulated sample, comparing predicted \LcoOneZero\ and \LcoTwoOne\ against observed CO scaling relations and luminosity functions at $z=0$, 1, 2 and 3, and examining how the emission depends on galactic properties and redshift. Paper~III \citep{Paper3} builds the response functions themselves, fitting \LcoOneZero\ as a function of halo mass, stellar mass, gas mass, metallicity, star formation rate (SFR) and infrared luminosity. We identify which combinations of variables best determine the CO luminosity, and compare the resulting models with those currently used in the LIM literature. The response functions are used to produce the top three panels of Figure~\ref{fig:peakpatch_firebox_zoom_stacked_6panel}. We also discuss how reliable the models are likely to be.

This paper is organized as follows. In Section~\ref{sec:Simulation_Methodology} we describe the \fireTwo\ simulations we use.  In Section~\ref{sec:post_processing_fire} we describe the post-processing of the \fireTwo\ snapshots, covering the \skirt\ radiation field calculation, the setup of the \cloudy\ runs, and the interpolation from \cloudy\ look-up table. In Section~\ref{sec:cloudy_runs_for_mw_molecular_clouds} we benchmark the calculation against MW molecular clouds, and in Section~\ref{section:Nco_Nh2_comparison} we compare the predicted CO column density ($\rm N_{CO}$) as a function of the molecular hydrogen column density ($\rm N_{H_2}$) with absorption measurements along Galactic translucent sight lines. In Section~\ref{sec:comparsion_mh2_mco_xco_with_obs} we compare our predicted $M_{H_2}$ and $X_{\rm CO}$ with the available galaxy observations. In Section~\ref{sec:discussion_caveats} we discuss the caveats when using our pipeline. We conclude in Section~\ref{sec:conclusion}.

\section{Simulation Methodology \label{sec:Simulation_Methodology}}

We use simulations from the Feedback in Realistic Environments (FIRE) project\footnote{\url{http://fire.northwestern.edu}} using {\sc GIZMO}\footnote{\url{http://www.tapir.caltech.edu/~phopkins/Site/GIZMO.html}} \citep{Hopkins_2015}, which employs an improved version of the {\sc gadget-3} \citep{Springel2005Gadget2} Tree-PM gravity solver with fully adaptive gravitational force softening.

We employ several suites of \fireTwo\ simulations \citep{Hopkins_2018_fire2}. We use the \core\ suite of zoom-in simulations of MW- and LMC-mass galaxies run to $z=0$ (see \cite{Wetzel_2023, Wetzel2025_secondDataRelease_FIRE2}), also including the higher-resolution resimulation of m12i as part of the new Triple Latte suite \citep{Wetzel_tripleLatte}. We also use the MassiveHalo (\massivefire) zoom-in suite run to $z=1$ \citep{Feldmann_2017_massiveFIRE, AnglesAlcazar2017_massiveFIRE_FIRE2}. Finally, we use the \firebox\ suite, which simulates the cosmological volume with baryonic physics rather than focusing on individual zoom-in regions \citep{Feldmann_2023}. All simulations use the FIRE-2 physics model \citep{Hopkins_2018_fire2} and the meshless finite-mass (MFM) hydrodynamic solver \citep{Hopkins_2015, Wetzel_2023}. All of our simulations adopt a (quasi-)Lagrangian hydrodynamic method, so the gas kernel (smoothing) lengths adapt to the local gas density; the gravitational softening of the gas is therefore fully adaptive and set equal to the hydrodynamic smoothing length.

The simulations self-consistently compute a wide range of radiative heating and cooling processes and explicitly track 11 elements. Simulated gas cells are converted into star particles when they meet several strict criteria: they must exceed a local density threshold, be predominantly molecular, and be locally self-gravitating and Jeans unstable. In practice, the self-gravitating criterion is generally the most restrictive requirement for star formation \citep{Hopkins_2018_fire2}.

\bgroup
\renewcommand{\arraystretch}{0.85}
\setlength{\tabcolsep}{4pt}
\small
\setlength{\LTcapwidth}{\textwidth}   

\startlongtable   
\begin{longtable}{l c c c c c c}
\caption{Individual galaxy properties from the zoom-in \fireTwo\ simulations used in this paper. The large number of \texttt{FIREbox} galaxies precludes our listing their properties here. We instead show the galactic properties of \texttt{FIREbox} snapshots in Figure~\ref{fig:galactic_properties_histogram_plots_mass_FIREbox}. From left to right we list the simulation name, redshift ($z$), star formation rate (\sfrHundredMyr, averaged over 100\Myr), mass-weighted gas metallicity within the half-light radius (\ZgasHl), total gas mass (\Mgas), total stellar mass (\Mstar), and halo mass of the \core\ \fireTwo, \tripleLatte\ resimulation of m12i galaxy at $z=0$, and \massivefire\ snapshots. The star formation rate, gas mass, and stellar mass are computed within 20\,\kpc\ of the galaxy center. For galaxies at $z=0$ with $\Mhalo > 3\times10^{11}\,M_\odot$ (the m12 galaxies), only gas cells within a perpendicular distance $d_\perp<5\,\kpc$ of the disc plane are used, whereas for all other galaxies we use all gas cells within 20\,\kpc\ (see Section~\ref{sec:Simulation_Methodology}). The halo masses of the galaxies listed here are the virial masses ($M_{\rm vir}$) from the \textsc{rockstar} halo finder \citep{Behroozi_2013}. We highlight that they are lower than the $M_{\rm 200m}$ values reported by \citet{Wetzel_2023} and \citet{Wetzel2025_secondDataRelease_FIRE2} for the same galaxies. We only use the m12i galaxy from the \tripleLatte\ suite at $z=0$. It has the same initial conditions as the \core\ Latte m12i galaxy tabulated here, but 8 times better baryonic mass resolution ($875$ versus $\rm 7100\,M_\odot$) and twice the force resolution, making it the highest-resolution galaxy analyzed in this work. The metallicities listed here are mass-weighted averages over the gas cells within the half-light radius, \RvisibleHl, which we measure from the face-on \skirt\ image of each galaxy. We sum the visible-band ($400$--$700\,{\rm nm}$) luminosity of the image pixels in concentric annuli around the galaxy center and define \RvisibleHl\ as the projected radius that encloses half of the total luminosity in the visible band of the image. The \massivefire\ sample includes four galaxies (A1, A2, A4, and A8) at $z=2$ and 3. These galaxies have a baryonic mass resolution of $\rm 3.2\times10^{4}\,M_\odot$. They reside in the most massive halos in our sample ($\rm \Mhalo>2\times10^{12}\,M_\odot$ at $z=2$ and $6.5\times10^{11}$--$\rm 1.6\times10^{12}\,M_\odot$ at $z=3$), extending the halo mass range of our sample. Gas masses are reported in units of $\rm 10^9\,M_\odot$, stellar masses in units of $\rm 10^{10}\,M_\odot$, and halo masses in units of $\rm 10^{11}\,M_\odot$. Gas metallicities are expressed in units of the solar metallicity, $Z_\odot$, for which we adopt a mass fraction of $Z_\odot=0.0134$ \citep{Grevesse2010_solarMetallicity}. Across the zoom-in \fireTwo\ galaxies, the mass-weighted gas metallicity evolves by more than an order of magnitude between $z=3$ and 0.}
\label{tab:galaxy_properties_zoomIn_tripleLatte} \\

\toprule
\textbf{Name} &
\textbf{$z$} &
\textbf{\sfrHundredMyr [$M_\odot\,\mathrm{yr}^{-1}$]} &
\textbf{\ZgasHl\ [$Z_\odot$]} &
\textbf{\Mgas\ [$10^9\,M_\odot$]} &
\textbf{\Mstar\ [$10^{10}\,M_\odot$]} &
\textbf{\Mhalo\ [$10^{11}\,M_\odot$]} \\
\midrule
\endfirsthead

\multicolumn{7}{c}{{\tablename\ \thetable{} -- continued from previous page}} \\
\toprule
\textbf{Name} &
\textbf{$z$} &
\textbf{\sfrHundredMyr [$M_\odot\,\mathrm{yr}^{-1}$]} &
\textbf{\ZgasHl\ [$Z_\odot$]} &
\textbf{\Mgas\ [$10^9\,M_\odot$]} &
\textbf{\Mstar\ [$10^{10}\,M_\odot$]} &
\textbf{\Mhalo\ [$10^{11}\,M_\odot$]} \\
\midrule
\endhead

\midrule
\multicolumn{7}{r}{{\emph{Continued on next page}}} \\
\endfoot

\bottomrule
\endlastfoot

m11d & 0.0 & 0.0182 & 0.892 & 3.95 & 0.428 & 2.62 \\
m11d & 1.0 & 0.0758 & 0.202 & 1.44 & 0.0291 & 0.579 \\
m11d & 2.0 & 0.208 & 0.164 & 1.93 & 0.0219 & 0.291 \\
m11d & 3.0 & 0.164 & 0.0673 & 2.20 & 0.00871 & 0.248 \\
\midrule
m11e & 0.0 & 0.433 & 0.706 & 2.17 & 0.152 & 1.40 \\
m11e & 1.0 & 0.113 & 0.261 & 1.73 & 0.0617 & 0.573 \\
m11e & 2.0 & 0.151 & 0.0731 & 2.65 & 0.0126 & 0.271 \\
m11e & 3.0 & 0.0110 & 0.0417 & 1.55 & 0.00414 & 0.134 \\
\midrule
m11h & 0.0 & 0.525 & 1.34 & 4.27 & 0.399 & 1.74 \\
m11h & 1.0 & 0.359 & 0.422 & 4.23 & 0.126 & 0.983 \\
m11h & 2.0 & 0.0611 & 0.132 & 4.01 & 0.0386 & 0.779 \\
m11h & 3.0 & 0.197 & 0.115 & 2.56 & 0.0171 & 0.253 \\
\midrule
m11i & 0.0 & 0.0794 & 0.785 & 1.61 & 0.102 & 0.671 \\
m11i & 1.0 & 0.0278 & 0.114 & 1.63 & 0.00939 & 0.268 \\
m11i & 2.0 & 0.0867 & 0.0594 & 1.08 & 0.00309 & 0.178 \\
m11i & 3.0 & 0.00105 & 0.00641 & 0.769 & 0.000787 & 0.123 \\
\midrule
m11q & 0.0 & 0.0710 & 0.585 & 1.53 & 0.0677 & 1.40 \\
m11q & 1.0 & 0.00527 & 0.143 & 0.809 & 0.0368 & 0.865 \\
m11q & 2.0 & 0.0198 & 0.125 & 0.856 & 0.0172 & 0.345 \\
m11q & 3.0 & 0.0337 & 0.0448 & 1.26 & 0.0133 & 0.264 \\
\midrule
m12b & 0.0 & 7.01 & 3.96 & 20.6 & 9.12 & 10.9 \\
m12b & 2.0 & 3.00 & 0.427 & 13.7 & 0.687 & 3.92 \\
m12b & 3.0 & 0.893 & 0.204 & 7.81 & 0.113 & 1.21 \\
\midrule
m12c & 0.0 & 6.79 & 3.65 & 19.1 & 6.30 & 10.6 \\
m12c & 1.0 & 9.37 & 0.821 & 16.1 & 1.53 & 4.88 \\
m12c & 2.0 & 0.821 & 0.272 & 4.11 & 0.179 & 1.13 \\
m12c & 3.0 & 0.469 & 0.126 & 5.33 & 0.0798 & 0.807 \\
\midrule
m12f & 0.0 & 7.78 & 3.48 & 27.7 & 8.53 & 12.7 \\
m12f & 1.0 & 6.77 & 1.60 & 11.1 & 2.67 & 6.14 \\
m12f & 2.0 & 7.36 & 0.520 & 16.5 & 0.882 & 4.10 \\
m12f & 3.0 & 3.04 & 0.179 & 9.45 & 0.159 & 1.46 \\
\midrule
m12i {\rm (triple latte)} & 0.0 & 2.53 & 5.39 & 8.98 & 6.97 & 9.13 \\
\midrule
m12i & 0.0 & 5.38 & 3.88 & 21.8 & 6.78 & 9.04 \\
m12i & 1.0 & 8.25 & 0.872 & 11.7 & 1.98 & 7.07 \\
m12i & 2.0 & 2.75 & 0.345 & 13.8 & 0.403 & 2.46 \\
m12i & 3.0 & 4.17 & 0.207 & 12.8 & 0.118 & 1.23 \\
\midrule
m12m & 0.0 & 8.78 & 5.12 & 26.4 & 12.4 & 11.6 \\
m12m & 1.0 & 13.1 & 1.09 & 31.5 & 2.59 & 6.17 \\
m12m & 2.0 & 1.10 & 0.246 & 12.1 & 0.328 & 4.47 \\
\midrule
m12r & 0.0 & 1.92 & 1.82 & 11.9 & 1.79 & 8.95 \\
m12r & 1.0 & 0.737 & 0.777 & 3.05 & 0.462 & 2.63 \\
m12r & 2.0 & 1.21 & 0.365 & 4.60 & 0.270 & 1.59 \\
m12r & 3.0 & 0.143 & 0.146 & 5.30 & 0.126 & 1.12 \\
\midrule
m12w & 0.0 & 8.60 & 4.89 & 6.04 & 6.05 & 8.30 \\
m12w & 1.0 & 3.96 & 0.992 & 13.8 & 0.948 & 3.43 \\
m12w & 2.0 & 2.36 & 0.282 & 14.6 & 0.180 & 1.91 \\
m12w & 3.0 & 0.356 & 0.154 & 5.28 & 0.0501 & 1.14 \\
\midrule
A1 & 2.0 & 85.9 & 3.57 & 23.5 & 17.6 & 22.7 \\
A1 & 3.0 & 89.8 & 3.29 & 22.6 & 9.83 & 13.6 \\
\midrule
A2 & 2.0 & 118 & 3.90 & 38.5 & 29.1 & 27.3 \\
A2 & 3.0 & 149 & 2.11 & 58.5 & 13.0 & 16.1 \\
\midrule
A4 & 2.0 & 53.3 & 2.79 & 26.0 & 12.8 & 24.6 \\
A4 & 3.0 & 62.4 & 0.658 & 44.2 & 2.25 & 7.77 \\
\midrule
A8 & 2.0 & 105 & 3.55 & 69.8 & 5.22 & 28.9 \\
A8 & 3.0 & 35.0 & 0.558 & 43.0 & 1.06 & 6.51 \\

\end{longtable}
\egroup

\begin{figure}[h!]
	\includegraphics[width=\columnwidth]{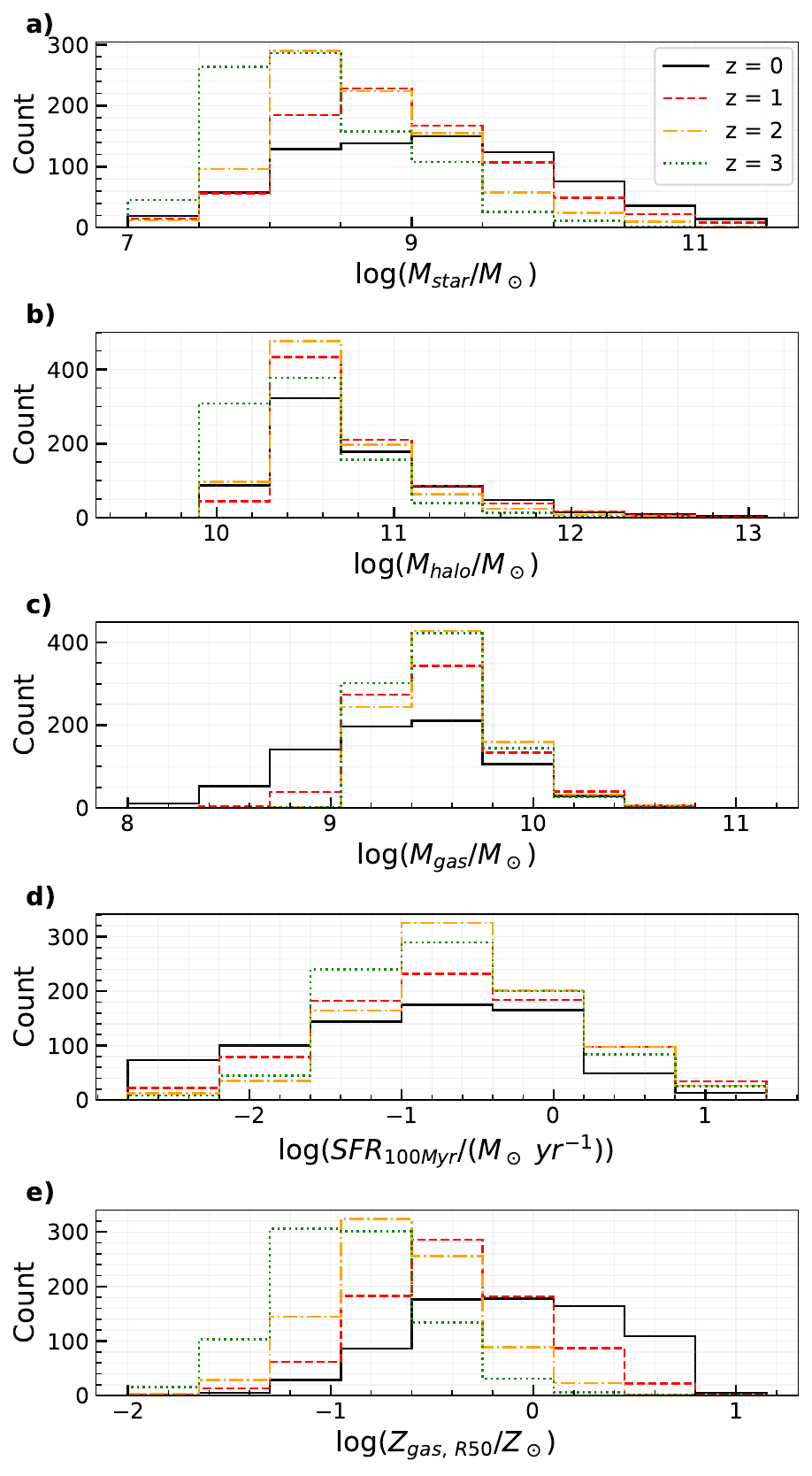}
    \caption{Histograms of the properties of the \firebox\ galaxies: (a) stellar mass \Mstar, (b) halo mass \Mhalo, (c) gas mass \Mgas, (d) star formation rate \sfrHundredMyr, and (e) mass-weighted gas metallicity \ZgasHl\ within the half-light radius. Quantities are defined as in Table~\ref{tab:galaxy_properties_zoomIn_tripleLatte}, which lists the zoom-in galaxies. The halo masses are taken from the virial masses reported by the \texttt{AMIGA Halo Finder} \citep{Gill_2004, Knollmann_2009}. Distributions are shown at $z=0$ (black solid), $z=1$ (red dashed), $z=2$ (orange dash-dotted), and $z=3$ (green dotted). \Mstar\ spans about 5 dex and \ZgasHl\ nearly 3 dex.}
    \label{fig:galactic_properties_histogram_plots_mass_FIREbox}
\end{figure}

Star particles influence the surrounding gas cells through various feedback mechanisms, including photoionization, photoelectric heating, radiation pressure, stellar winds, and supernovae. The momentum and energy injected by the feedback are determined by assuming a single stellar population model with a Kroupa initial mass function \citep{Kroupa_2001}, coupled with the STARBURST99 stellar population synthesis model \citep{Leitherer_1999}. None of our simulated galaxies include feedback from active galactic nuclei (AGN). The adopted cosmological parameters differ slightly between suites. The zoom-in halos use either the AGORA cosmology \citep{Kim2014_agoraCosmology_FIRE} ($\Omega_{m} = 0.272$, $\Omega_{\Lambda} = 0.728$, $\Omega_{b} = 0.0455$, $h = 0.702$, $\sigma_{8} = 0.807$, $n_{s} = 0.961$; m11q, m12i, m12f, m12m, m12b, m12c) or a Planck 2013 cosmology \citep{Planck_2014} ($\Omega_{m} = 0.31$, $\Omega_{\Lambda} = 0.69$, $\Omega_{b} = 0.048$, $h = 0.68$, $\sigma_{8} = 0.82$, $n_{s} = 0.97$; m11d, m11e, m11h, m11i, m12r, m12w) \citep{Hopkins_2018_fire2, ElBadry2018_m11halosPlanck}. FIREbox adopts \cite{Planck_2016} parameters ($\Omega_{m} = 0.3089$, $\Omega_{\Lambda} = 0.6911$, $\Omega_{b} = 0.0486$, $h = 0.6774$, $\sigma_{8} = 0.8159$, $n_{s} = 0.9667$), while the MassiveFIRE \citep{Feldmann2016_massiveFIRE_FIRE1, Feldmann_2017_massiveFIRE, AnglesAlcazar2017_massiveFIRE_FIRE2} halos adopt $\Omega_{m} = 0.2821$, $\Omega_{\Lambda} = 0.7179$, $\Omega_{b} = 0.0461$, $h = 0.697$, $\sigma_{8} = 0.817$ and $n_{s} = 0.9646$ \citep{Feldmann_2023}. All runs assume a flat $\Lambda$CDM cosmology with $w = -1$.

We begin our post-processing by centering each galaxy in every snapshot using halo catalogs generated by the \fireTwo\ collaboration using the ROCKSTAR halo finder \citep{Behroozi_2013} for the \core\ \fireTwo, \tripleLatte\ and \massivefire\ simulations and the AMIGA Halo Finder (AHF; \citealt{Gill_2004, Knollmann_2009}) for the \texttt{FIREbox} simulations. We started with the 1000 most massive halos in the \texttt{FIREbox} simulation. Then we restricted our analysis to only central halos.

This selection yields a final sample of 848 central halos at $z=0$, 866 at $z=1$, 913 at $z=2$, and 956 at $z=3$. The large number of \firebox\ galaxies dominate our sample and therefore our response functions (\citetalias{Paper3}); including high-resolution \fireTwo\ galaxies (zoom-in and triple latte) and \massivefire\ allows us to test the resolution dependency of the predicted emission. The \firebox\ galaxies are supplemented by a set of \core\ \fireTwo\ galaxies with a baryonic mass resolution of $7100\,M_\odot$ (listed in Table~\ref{tab:galaxy_properties_zoomIn_tripleLatte}): eleven galaxies at $z=1$ and $z=3$, and twelve at $z=0$ and $z=2$. We also include the m12i galaxy resimulated at \tripleLatte\ resolution ($880\,M_\odot$ baryonic mass resolution) at $z=0$, and eight \massivefire\ snapshots at $z=2$ and $z=3$ ($3.3\times10^{4}\,M_\odot$ baryonic mass resolution).

A limitation of the rather small \firebox\ cosmological volume $(22.1\, \rm cMpc)^3$ is that its maximum halo mass only reaches to about $\approx 10^{12}\,M_\odot$, falling well short of the most massive observed halos. The most massive halos are rare, and the ratio of molecular gas mass to their total baryonic mass is smaller than that in $10^{12}\,M_\odot$ halos, so their contribution to the CO line intensity is believed to be small, e.g., \citet{Li_2016_coModelling}. Crucially for our purposes, the sample of \firebox\ halos does include a few halos at or just above $10^{12}\,M_\odot$ at $z>1$, a mass range which does contribute, possibly significantly, to the LIM CO signal. To further account for such galaxies, we incorporated \massivefire\ galaxies at $z=2$ and $z=3$. We caution, however, that even including the most massive of these halos ($\approx 1.6 \times 10^{12}\,M_\odot$ at $z=3$) will only provide a lower limit on the most massive halos typical in the Websky simulation \citep{Stein_bond_webskyProject}.

For each galaxy, we restricted our analysis to particles within 20 kpc of the center of the main halo. We rotated each galaxy so that the $z$-axis aligns with the net angular momentum vector of its gas, which defines the disc plane, and denote by $d_\perp$ the perpendicular distance of a gas cell from this plane. Most MW-mass galaxies at $z=0$ formed well-defined discs, whereas lower-mass galaxies, specifically those with $\Mhalo < 3\times10^{11}\,M_\odot$, tended not to. Therefore, for galaxies at $z=0$ with $\Mhalo > 3\times10^{11}\,M_\odot$ we further restricted the sample to gas cells with $d_\perp < 5\,$kpc, and for all other galaxies we kept all gas cells within 20 kpc. At higher redshifts, we likewise kept all gas cells within 20 kpc.

We generally work with logarithmic variables, so galaxies that have no star formation within 100 million years or sooner or no gas inside 20 $\kpc$ or half light radius are problematic. To simplify our analysis, we exclude such galaxies. After applying these cuts, the analysis includes 765 galaxies at $z=0$, 850 at $z=1$, 887 at $z=2$, and 915 at $z=3$. These exclusions mean that our response functions will not include the $\approx5\%$ of galaxies that lack either star formation within 100 million years or gas. One could use survival analysis to account for these excluded galaxies; that is beyond the scope of this work. 

We present some relevant galactic properties of our \firebox\ sample in Figure~\ref{fig:galactic_properties_histogram_plots_mass_FIREbox}. The \tripleLatte, \core\ \fireTwo\ and \massivefire\ galaxies properties are listed in Table~\ref{tab:galaxy_properties_zoomIn_tripleLatte}. In both the figure and the table we give values at four different redshifts. The catalog spans a wide dynamic range in galactic properties: 5.6 orders of magnitude in stellar mass ($6 < \log(\Mstar/M_\odot) < 11.6$), 3.1 orders of magnitude in halo mass ($10.1 < \log(\Mhalo/M_\odot) < 13.2$), 2.9 orders of magnitude in gas mass ($7.9 < \log(\Mgas/M_\odot) < 10.8$), 5.5 orders of magnitude in star formation rate ($-3.3 < \log(\sfrHundredMyr/(M_\odot,\mathrm{yr}^{-1})) < 2.2$), and 3.1 orders of magnitude in gas-phase metallicity calculated within the half-light radius ($-2.2 < \log(\ZgasHl/Z_\odot) < 0.9$). Here it is important to note the gas cell metallicities in \fireTwo\ are stored as mass fractions. We normalized these to solar metallicity using a fiducial value of 0.0134 \citep{Grevesse2010_solarMetallicity}. Later, we use these metallicities to scale the ISM abundances in \cloudy\ in Section~\ref{sec:cloudy}.

\subsection{CORE FIRE-2 simulations \label{sec:fireLatte_runs_description}}

The FIRE-2 simulations used here are cosmological zoom-in simulations from the \core\ suite comprising seven Milky Way-mass halos from the Latte suite: m12b \citep{Garrison-Kimmel2019_m12b_m12c}, m12c \citep{Garrison-Kimmel2019_m12b_m12c}, m12f \citep{Garrison-Kimmel2017_m12f}, m12i \citep{Wetzel_2016}, m12m \citep{Hopkins_2018_fire2}, m12r \citep{Samuel2020_m12w_m12r}, m12w \citep{Samuel2020_m12w_m12r} together with five lower masss halos: m11d, m11e, m11h, m11i
\citep{ElBadry2018_m11halosPlanck} and m11q. Each was generated by first evolving a dark-matter-only periodic cosmological box to $z=0$, identifying haloes with \textsc{rockstar} \citep{Behroozi_2013}, selecting a target halo based on its mass at $z=0$ and an isolation criterion, and re-simulating the Lagrangian region around it with baryons at much higher resolution, with initial conditions generated using \textsc{music} \citep{Hahn_2011}. The initial conditions for m12b, m12c, m12f, m12i, m12m and m11q derive from the $(85.5\,\mathrm{Mpc})^3$ AGORA box \citep{Kim2014_agoraCosmology_FIRE}. The remaining simulations were generated within periodic boxes of side $85.5-172\,\mathrm{Mpc}$ \citep{Samuel2020_m12w_m12r, Wetzel_2016}. m12r and m12w \citep{Samuel2020_m12w_m12r} in particular were added to the Latte suite later, selected to host an LMC-mass satellite at $z\approx0$, and adopt updated \cite{planck_2018} cosmology.

All of the halos analyzed here have an initial baryonic mass of $7100\,M_\odot$, with a dark matter particle mass of $3.5\times10^{4}$ or $3.9\times10^{4}\,M_\odot$ (for the AGORA and Planck cosmologies, respectively), i.e. $\approx 5$ times the baryonic mass. The gas smoothing lengths shrinks to a minimum of $1\,\mathrm{pc}$ in the densest regions, while a typical gas cell at $n\sim1\,\mathrm{cm^{-3}}$ has a softening length of $\sim40\,\mathrm{pc}$. In contrast, the gravitational softening lengths of the collisionless particles are fixed at $4\,\mathrm{pc}$ for stars and $40\,\mathrm{pc}$ for dark matter.

\subsection{Triple Latte version of m12i}

In addition to the fiducial version of m12i, we also include the new higher-resolution version of m12i as part of the new \tripleLatte\ suite (Wetzel et al., in prep.).
Compared with the fiducial m12i, it has 8 times better mass resolution for baryonic particles, at 880 $M_\odot$, with all force softenings a factor of two smaller. We use this simulation to assess the effect of resolution on our calculations. Some of the properties of m12i \tripleLatte\ are in Table~\ref{tab:galaxy_properties_zoomIn_tripleLatte}, and Wetzel et al., in prep. will provide more details.

\subsection{\massivefire}

The \massivefire\ simulations \citep{Feldmann2016_massiveFIRE_FIRE1, Feldmann_2017_massiveFIRE, AnglesAlcazar2017_massiveFIRE_FIRE2} are also zoom-in simulations. They closely resemble our \core\ \fireTwo\ simulations (see Section~\ref{sec:fireLatte_runs_description}).  The position of the massive halos are initially drawn from a low-resolution dark matter only simulation within a 144 cMpc box. The initial conditions for the high resolution gas cells are defined with multiscale initial conditions \texttt{MUSIC} \citep{Hahn_2011}. Then these identified halos are re-simulated, with increased resolution, using the GIZMO code \cite{Hopkins_2014} with FIRE-2 physics.

 The baryonic particles have a mass resolution of $\rm 3.3\times10^{4}\,M_\odot$ and dark matter particles are of masses $\rm 1.7\times10^{5}\,M_\odot$. For all \massivefire\ simulations, the gravitational softening lengths are fixed at 143 pc for dark matter particles and 21 pc for stellar particles. The gravitational softening lengths for gas cells are fully adaptive, reaching a minimum value of 9 pc within the dense interstellar medium.

They were designed to target massive halos at $z\sim2$. They provide an important set of galaxies for our analysis at high redshifts. These halos are the most massive we include at $z=3$ ($\sim10^{12}\,M_\odot$), but are still less massive than the extreme high-mass tail of observed galaxies at that redshift.



\subsection{FIREbox}

\begin{figure*}[t]
	\includegraphics[width=\linewidth]{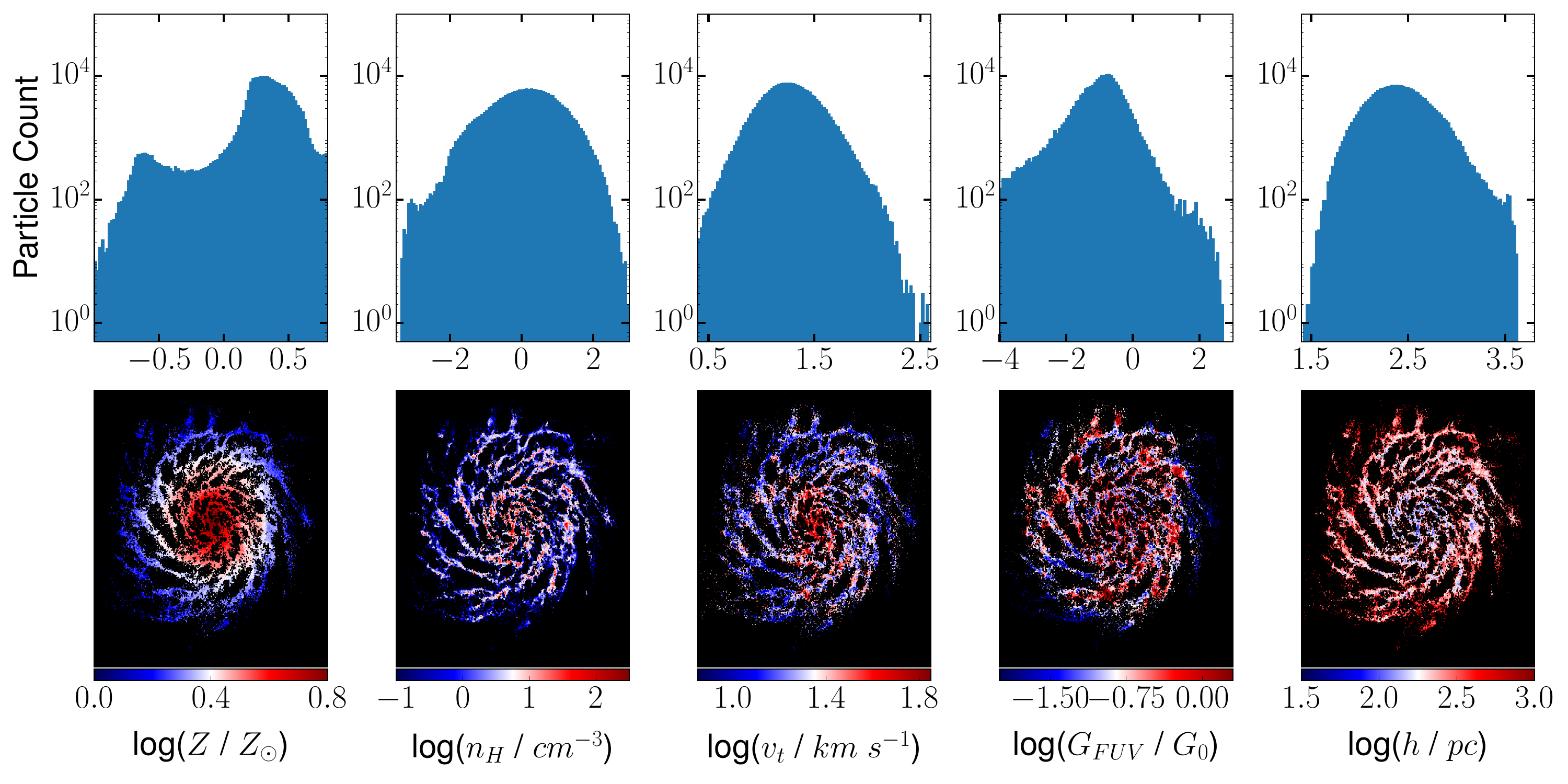}
    \caption{The distribution of various gas cell properties of one of the central halo in \firebox\ suite at redshift zero, for particles within $20\,\kpc$ of the center of the host halo, and within $5\,\kpc$ of the plane of the galactic disk ($|z|<5\,\kpc$). The panels, from left to right, illustrate the gas phase metallicity ($Z$), hydrogen number densities ($n_H$), turbulent velocity ($v_t$), interstellar radiation field ($G_{\rm FUV}$) normalized to \cite{Draine_1978_ISRF_MW_value_habingUnits}, and shielding length ($h$). We calculated the shielding of the gas cell using its smoothing length. The top row presents 1D histograms, showing the distribution of the gas cell properties, whereas the bottom row presents 2D maps, showing the spatial two-dimensional (1024 X 1024 bin, 20$\kpc$ on a side) histograms of average gas cell properties of galaxies viewed face-on.}  \label{fig:histogram_plots_of_cloudy_parameters_firebox}
\end{figure*}

The \texttt{FIREbox} simulation \citep{Feldmann_2023} employs a cubic cosmological volume with a side length of $15\,{\rm cMpc}\,h^{-1}\sim 22.1\,{\rm cMpc}$, and periodic boundary conditions, with a baryonic mass of $\rm 6.26\times10^4\,M_\odot$, i.e., a mass resolution 8 times lower and force resolution twice lower than in \core\ \fireTwo\ simulations. The softening lengths of star and dark matter particles are fixed to 12 and $80\,\pc$, respectively. \texttt{FIREbox} uses standard FIRE-2 physics.

Figure~\ref{fig:histogram_plots_of_cloudy_parameters_firebox} shows the spatial distribution of various properties of the gas cells in the one of the central halo at $z=0$, in a face-on view. We display their metallicity ($Z$) normalized to \cite{Grevesse2010_solarMetallicity} solar metal mass ratio of 0.0134, number density ($n_H$), turbulent velocity ($v_t$), radiation field in the FUV band (6-13.6 eV) normalized to \cite{Draine_1978_ISRF_MW_value_habingUnits} ($G_{\rm FUV}$) computed with SKIRT (see Section~\ref{sec:skirt}), and the gas cell smoothing length ($h$). We use these gas cell properties to compute the line luminosities for each gas cell, as described in section below.

\section{Calculating CO Emission, Molecular Gas Masses, and CO Abundances from FIRE Galaxy Snapshots \label{sec:post_processing_fire}}
We post-process the \fireTwo\ simulation snapshots using a combination of \skirt\ (to calculate the local radiation field) and \cloudy\ (to calculate the line emission and abundances given the local radiation and density fields). We describe \skirt\ and \cloudy\ below.

\subsection{Calculation of Radiation Field that Each Gas Cell is Subject to Using SKIRT} \label{sec:skirt}

We used the open-source 3D Monte Carlo dust radiative transfer code, \skirt\ (version 9, \cite{Baes_2011, Baes_2015, Camps_2015, Camps_and_baes_2020_skirt9}) to determine the interstellar radiation field (ISRF) that gas cells are exposed to. \skirt\ simulates the full radiative transfer,  self-consistently computing absorption, scattering and thermal emission from stars and dust. The \skirt\ code simulates complex geometries using adaptive spatial grids, and computes the radiation field in each \skirt\ simulation cell. The code can generate galactic spectral energy distributions (SED) and images.

\skirt\ can accept both grid-based and particle based input files, among others. FIRE-2 uses a Lagrangian (MFM) hydrodynamic solver. Accordingly, we used the \skirt\ ``ParticleSnapshot'' routine to input the particles into the \skirt\ simulation. In both \fireTwo\ and \skirt\ codes star particles are assumed to be a single stellar population, with emission given by STARBURST99 \citep{Leitherer_1999} as a function of age, metallicity, and initial mass of the star (cluster) particles. 

When running SKIRT, we used $10^7$ photon packages for each of 350 points in the equally spaced logarithmic wavelength grid spanning $\lambda = 0.03 - 2000\,\rm\mu m$. We note that both \textsc{Gizmo} and \skirt\, when using the ``ParticleSnapshot'' input, employ a cubic spline to calculate smooth properties (like density) from the (discrete) gas cells and their properties.

We assume a constant dust-to-metal ratio $\delta_{dzr} = 0.4$, i.e., $40\%$ of the metals are in the form of dust, consistent with the ratio used in \fireTwo\ simulations \citep{Hopkins_2018_fire2}, except when the gas temperature was higher than $10^6\,{\rm K}$, in which case we assumed that the gas was dust free as the result of thermal sputtering \citep{Tielens_1994}. We used the \cite{Weingartner_2001} dust mix to set the size distributions of the carbonaceous and silicate grain populations to MW observed silicates and PAH sizes, $R_v = 3.1$. 

When running \skirt\, we gridded the \fireTwo\ galaxies using $10^5$ Voronoi tessellation cells. We ran a number of test runs using $10^6$ cells, but found that the predicted CO luminosities changed by less than 7\%. 

We use \skirt\ to compute the local interstellar radiation field (ISRF) for each gas cell. The commonly used band to compute the radiation field is the FUV band, defined between 6 - 13.6 eV. Photons in this energy range are particularly important for neutral ISM chemistry because because they are the main heating source of the molecular region, through the photoelectric effect. They are also critical for photodissociation of CO molecules \citep{Draine_2011}. We computed the FUV energy density by integrating the SKIRT-calculated mean intensity per spatial cell $(J_\lambda)$, with units of $\rm W m^{-2} \mu m^{-1} sr^{-1}$, over the wavelength range of 6 - 13.6 eV and solid angle ($\Omega$) (using the Simpson integration routine from the SCIPY PYTHON library \cite{scipy_2020}):

\begin{equation}
    \label{eq:radiation_field_calculation_fuv_habing}
    u_{6-13.6\, {\rm eV}} = \frac{4\pi}{c} \int_{91\,{\rm nm}}^{206\,{\rm nm}} J_{\lambda} \,d\lambda,
\end{equation}
where $u_{6-13.6\, {\rm eV}}$ is the energy density normalized to the Habing radiation field \citep{Habing1968_radiationFieldMeasurement, Rybicki_Lightman_1986, Draine_2011} and to the MW value observed by \cite{Draine_1978_ISRF_MW_value_habingUnits}. We calculate the interstellar radiation field (ISRF) for every \skirt\ cell in the unit of MW measured interstellar radiation field in the FUV band, $\rm G_{FUV}$, in the unit of $\rm G_0$ using 
\begin{equation}
    \label{eq:normalizing_radiation_field_calculation}
    G_{FUV}/G_0 = \frac{u_{6-13.6\, {\rm eV}}}{5.29 \times 10^{-14}\,{\rm erg}\,{\rm c}m^{-3}} \times \frac{1}{1.69} ,
\end{equation}
where the factor $1.69$ is from \cite{Draine_1978_ISRF_MW_value_habingUnits}.

\section{\LFIR\ calculation for Each Galaxy \label{sec:Lfir_calculation_for_each_galaxy}}
We calculated the bolometric far-infrared (FIR) luminosity ($\LFIR$) of each \fireTwo\ galaxy using \skirt. For each galaxy, a synthetic observation instrument was placed at a luminosity distance of $10~\mathrm{Mpc}$ ($D_{L,\mathrm{skirt}}=10~\mathrm{Mpc}$). We defined a logarithmic wavelength grid spanning $0.03~\mathrm{\mu m}$ to $2000~\mathrm{\mu m}$ with 350 bins to measure the face-on spectral energy distribution (SED), $F_{\lambda}$. The total FIR luminosity was then computed via:
\begin{equation}
    \label{eq:Lfir_calculation_from_skirt_sed}
    \LFIR / (\ergs) = 4 \pi D_{L,\mathrm{skirt}}^2 \int_{0.03 \mu m}^{1000 \mu m} F_\lambda \, d\lambda  
\end{equation}

\subsection{Setup of the CLOUDY Simulations}
\label{sec:cloudy}

We calculate the line luminosities ($L_\mathrm{CO(1-0)}$, $L_\mathrm{CO(2-1)}$, $L_\mathrm{H\alpha}$, $L_\mathrm{H\beta}$, $L_\mathrm{[OII]}$, $L_\mathrm{[OIII]}$ and $L_\mathrm{[NII]}$) and species abundances ($M_{H_2}$ and $M_{CO}$) using version 25.00 of \cloudy\ last described by \cite{cloudy_version_c25_release}. The computed [\ion{O}{2}] lines are the $\lambda\lambda\,3727.09,\,3729.88\,\mathrm{\AA}$ doublet, the computed [\ion{O}{3}] lines are $\lambda\lambda\lambda\,4364.44,\,4960.30,\,5008.24\,\mathrm{\AA}$ and $88.36\,\mu\mathrm{m}$, and the computed [\ion{N}{2}] lines are $\lambda\lambda\,5756.19,\,6585.27\,\mathrm{\AA}$, where all wavelengths are given in vacuum. The lines other than CO will be used to infer the physical quantities of the galaxies similar to observations in \citetalias{Paper2} and \citetalias{Paper3}.

\cloudy\ is a non-local spectral synthesis and plasma simulation code that calculates spectra for a range of physical environments. Performing a \cloudy\ run for every gas cell in each of our galaxies was prohibitively expensive. Instead, we determined the range of five physical quantities (metallicity, number density, turbulent velocity, radiation field, and physical size or stopping length) for each of our particles, and created a lookup table covering that range. We then interpolated between \cloudy\ runs to estimate the CO and other line emissions for each gas cell in our \fireTwo\ simulations. Although the table also contains the other lines listed above, this paper concentrates on CO and molecular gas, and we leave the analysis of the remaining lines to the companion papers.

\cloudy\ includes $\approx$ 1000 reactions for 68 molecules  \citep{Abel_2005} which integrates chemistry for both light and heavier elements. \cloudy\ calculates the cloud's non-equilibrium ionization, thermal, and chemical state based on the pre-determined radiation field. The level populations and line emissions for atoms and molecules are predicted using the Leiden Atomic and Molecular Database (LAMDA, \cite{Schoier_2005}).

Energetic particles such as cosmic rays and suprathermal electrons can change the physical conditions inside the molecular clouds \citep{Spitzer_1968}. They are implemented in our \cloudy\ simulation as described by \cite{Ferland_1984} and \cite{Ferland_2009}. We used the ``cosmic rays background 1.000 linear" option to set the cosmic ray background to the average MW value of $2 \times 10^{-16} s^{-1}$. 



We model the gas as a plane-parallel slab of constant density, illuminated on one face by an external radiation field (an open geometry in the CLOUDY terminology). These choices are motivated by the physical configuration of the CO-emitting gas in real galaxies. A closed geometry, in which the gas fully covers a central source is an appropriate representation for an H\textsc{ii} region surrounding a young cluster, where the ionization and heating source originate inside the cloud as a result of recent star formation. However, in the Milky Way only a small fraction of molecular clouds are actively star forming at any given time \citep{Miville-Deschenes_2017_MW_paper}. The bulk of the CO emission arises from clouds that are heated by the ambient interstellar radiation field rather than by embedded sources. An open geometry, in which the illuminated face receives the external field is therefore a more accurate representation of the CO emitting gas in real galaxies.

We computed the hydrogen number density $n_H$ of each gas cell as
\begin{equation}
    n_H = \frac{\rho}{\mu_{\rm sim} \, m_p},
\end{equation}
where $\rho$ is the simulated gas density, $m_p$ is the proton mass, and $\mu_{\rm sim}$ is the mean molecular mass of the cell. We derived $\mu_{\rm sim}$ from the helium and electron abundances in the \fireTwo\ simulation:
\begin{equation}
    \label{eq:mean_molecular_mass_per_H_muH_def}
    \mu_{\rm sim} = \frac{1 + 4Y}{1 + Y + n_e},
\end{equation}
where $n_e$ is the electron abundance and $Y \equiv n_{\rm He} / \left[4(1 - n_{\rm He})\right]$.

We used CLOUDY's pre-defined input spectral shape corresponding to the Milky Way radiation field. Then we scaled the MW radiation field using the ``table ism" flag by the appropriate amount, with a range that covered the radiation fields produced by the \skirt\ runs. For all of our \cloudy\ runs we included the Cosmic Microwave Background (CMB) radiation to the incident continuum, evaluated at the redshift of each snapshot.

As noted above, we converted the \fireTwo\ metal mass fractions to solar metallicities by assuming $Z_\odot=0.0134$ \citep{Grevesse2010_solarMetallicity} as the fiducial value for metal mass fraction. Then we used  the ``abundances ISM" option, an empirically determined set of gas phase abundances and grains. We scaled the metallicities in our set of \cloudy\ simulations to cover the range produced by the \fireTwo\ simulations. Thus, our \cloudy\ calculations do not use the cell-by-cell carbon and oxygen abundances predicted by \fireTwo; instead, the \fireTwo\ metal mass fractions are used only to set the overall metallicity relative to solar. We note that predicted CO luminosities are sensitive to the adopted value of $Z_\odot$. We assumed that the grain/metal ratio in \cloudy\ simulation does not change with the metallicity, but that the grain-to-gas ratio scales linearly with metallicity.

In molecular clouds, the internal (possibly turbulent) motions of the gas are typically highly supersonic \citep{Miville-Deschenes_2017_MW_paper}. The Mach numbers can easily reach to $\sim 10-50$. These high velocities can change the effective optical depth of the lines \citep{Goldreich_and_kwan_1974_turbulence}. We calculate the ``turbulent'' velocities of gas cells in the FIRE simulations using Equation~\ref{eq:turbulence}. Our definition of the turbulent velocities include rotation, shear, shocks and bulk motions rather than only microturbulence.
\begin{equation}
    \label{eq:turbulence}
    v_{t} = \sqrt{\sum_{i=x,y,z}(v^{i}_{gas} - v^{i}_{CM})^2},
\end{equation}
where $v^i_{gas}$ is the velocity of an individual gas cell. $v^i_{CM}$ is the center of mass velocity of the gas around cell $i$, weighted by the cubic spline kernel for particles inside the region, calculated using \texttt{Meshoid}\footnote{\url{https://github.com/mikegrudic/meshoid}} package by:

\begin{equation}
    \label{eq:turbulence_CM}
    v^{i}_{CM} = \frac{\sum_{j=1}^{k} W(r_{j},h) \times v^{i}_{j} \times M_{j}}{\sum_{j=1}^{k} W(r_j,h) \times M_{j}}
\end{equation}

In this expression, W is the cubic spline kernel defined in \texttt{Meshoid}, h is the smoothing length of the chosen gas cell, $r_{j}$ is the distance between that gas cell and its $j^{th}$ neighbor gas cell, and k is the number of gas cells inside the two smoothing lengths. We then set a range of $v_t$ to be used in our \cloudy\ simulations that would cover the turbulent velocities of most of our gas cells. 


We ran our set of \cloudy\ simulations until they reached a pre-defined stopping length $R_{stop}$, chosen to cover the range determined from the smoothing length of our gas cells is \fireTwo\ snapshots. When interpolating, we set $R_{stop}$ equal to the smoothing length for each gas cell in the snapshot. 

The CO lines can reach to high optical depths deep inside the molecular clouds. If they do, multiple iterations are needed for the emission calculation to converge. Therefore, we iterated our \cloudy\ simulations until convergence  using the ``iterate to converge" command.

\subsection{CLOUDY Line Luminosity Per Mass Calculation \label{sec:cloudy_lineLuminosityPerMass}}

In this section, we describe our methodology for calculating the $\mathrm{CO(1-0)}$, $\mathrm{CO(2-1)}$, $\mathrm{H\alpha}$, $\mathrm{H\beta}$, $\mathrm{[OII]}$, $\mathrm{[OIII]}$, and $\mathrm{[NII]}$ luminosities normalized by gas mass. As described above, we use \cloudy\ calculations assuming a one-dimensional, plane-parallel geometry. However, the gas cells in the \fireTwo\ simulations are inherently three-dimensional, and their boundaries are not uniquely defined. Mapping the one-dimensional \cloudy\ calculations onto the three-dimensional \fireTwo\ gas distribution therefore introduces an additional source of uncertainty.

The \cloudy\ code divides the simulated region into small zones, and for each zone prints out the angle integrated line (also integrated over frequency or velocity) emissivity
\begin{equation}
    \epsilon_{line} = 4 \pi j_{line}\ \ergs\,\cm^{-3}.
\end{equation}
The non-angle integrated emissivity for each line emission ($L_
\mathrm{CO(1-0)}$, $L_\mathrm{CO(2-1)}$, $L_\mathrm{H\alpha}$, $L_\mathrm{H\beta}$, $L_\mathrm{[OII]}$, $L_\mathrm{[OIII]}$ and $L_\mathrm{[NII]}$) is calculated by dividing $\epsilon_{line}$ by $4\pi$. Then the line intensity is calculated by following \cite{Rybicki_Lightman_1986}:
\begin{equation}
    I_{\mathrm{line}} =
    \int_{0}^\Rs ds \;
    j_{\mathrm{line}}\,\ergs\,\cm^{-2} sr^{-1},
    \label{eq:intensity_from_emissivity_cloudy}
\end{equation}
where $R_{\mathrm{stop}}$ is the stopping length of the \cloudy\ simulation, which in our case will be interpolated to the smoothing length for each gas cell. As already noted, the range of thicknesses $R_{stop}$ is chosen to span the distribution of gas cell smoothing lengths ($h$) in the inner $20\,\kpc$ of our FIRE simulations. We show the distribution of $h$ in the central \firebox\ halo in Figure~\ref{fig:histogram_plots_of_cloudy_parameters_firebox} as an example.

The flux at the cloud surface is obtained by integrating the intensity over solid angle, 
\begin{align}
    F_{\mathrm{line}}(\Rs) &=
    \int d\Omega \;  I_{\mathrm{line}} \cos\theta\\
    &= \pi I_{\mathrm{line}} \, \ergs\,\cm^{-2},
    \label{eq:cloudy_flux_calculation}
\end{align}
where we have assumed a circular beam, with beam radius at the location of the cloud matched to the cloud radius, i.e., $R=R_{\mathrm{stop}}$.

The line luminosity is
\begin{align}
  L_{line}&=4\pi D^2 F_{line}(D)\\
  &=4\pi D^2\left(\frac{\Rs}{D}\right)^2\,\pi\int_0^{\Rs}ds\, j_{line}\,\ergs,
\end{align}
or
\be
L_{line}=4\pi R_{{\mathrm stop}}^2\,\pi\int_0^\Rs ds\, j_{line}\,\ergs.
\ee

The mass of the cylinder in the beam is
\begin{equation}
    M(R_s) = \pi R^3_{stop} \, \mu \, m_p n_H\ \g.    \label{eq:column_density_calculation_spherical}
\end{equation}
We remind the reader that $\mu = 1.37$ corrects the molecular mass for the presence of Helium.

The luminosity per mass is then
\begin{equation}
    \frac{L_{\mathrm{line}}}{M}
    =
    \frac{4\pi}{\rho \Rs} \int_{0}^{R_{\mathrm{stop}}}    j_{\mathrm{line}}(s)\, ds \, (\ergs M_\odot^{-1})
    \label{eq:luminosity_per_mass_calculation a}
\end{equation}
where $\rho$ is the density, assumed independent of depth $s$ for simplicity, for each \cloudy\ run. 

We then use the resulting luminosity-per-mass values in the units of $erg \, s^{-1} /M_\odot$ to interpolate the line luminosities assigned to each gas cell in the \fireTwo\ simulation, as described in Section~\ref{sec:interpolating_between_cloudy_runs}.

We created a table of \cloudy-predicted line luminosity per mass characterized by five distinct parameters: metallicity $Z_{\rm gas}$ over the range $-3 \leq \log(Z_{\rm gas}/Z_\odot) \leq 1$, number density $n_{\rm H}$ over $-4 \leq \log(n_{\rm H}/{\rm cm^{-3}})\leq 3.5$, turbulent velocity $0 \leq \log(v_{t}/\kms)\leq 2$, ISRF over $-5 \leq \log(G_{\rm FUV}/G_0)\leq 5$, and stopping length $R_{\rm stop}$ over $1 \leq \log(h/{\rm pc})\leq 3.5$. The grid spacing is 0.5 dex for $Z_{\rm gas}$, $n_{\rm H}$ and $R_{\rm stop}$, and 1 dex for the other two parameters. The hydrogen column density, $n_{\rm H}R_{\rm stop}$, of the gas cells in our \fireTwo\ simulations does not exceed $10^{5.5}\,{\rm cm^{-3}\,pc}$. We therefore ran \cloudy\ only for grid points with $n_{\rm H}R_{\rm stop} < 10^{5.6}\,{\rm cm^{-3}\,pc}$ (i.e., up to $10^{5.5}\,{\rm cm^{-3}\,pc}$ on our grid), which leaves 27,324 parameter combinations and makes the table considerably cheaper to compute.

Because the CMB temperature changes with redshift, we computed this table separately at each of the four redshifts we analyze, $z=0$, 1, 2 and 3. The grid of parameters is identical at every redshift; only the CMB temperature differs. When interpolating (Section~\ref{sec:interpolating_between_cloudy_runs}), every gas cell uses the table that corresponds to the redshift of its own snapshot, so no galaxy is evaluated with a CMB temperature other than its own.

In total we ran \cloudy\ for $4\times27{,}324 = 109{,}296$ models. Of these, $664$ ($0.61\%$) failed to complete successfully. Details of these problematic runs are presented in Appendix~\ref{sec:problematic_cloudy_runs}.

\subsection{CLOUDY Species Mass Fraction Calculation}
\label{sec:cloudy_mass_fraction_calculation}

We calculated the molecular hydrogen and CO mass fractions for each of our \cloudy\ runs. In each \cloudy\ simulation, the gas is modeled as a plane-parallel slab divided into many layers with widths $d_i$. For each layer, \cloudy\ outputs the relative abundance of molecular hydrogen $(n_{\mathrm{H}2}/n_{\mathrm{H}})_i$. We use this quantity to compute the molecular hydrogen column density in layer $i$ as
\begin{equation}
    \label{eq:column_denisty_h2_in_one_layer_a}
    \Sigma_{\rm H2} = 2 \left(\frac{n_{H2}}{n_H}\right)_i n_{H} m_p d_i,
\end{equation}
where the factor of 2 accounts for the two hydrogen nuclei in each $\rm{H_2}$ molecule, $n_{\rm H}$ is the pre-defined hydrogen number density and $m_{\rm p}$ is the proton mass. The corresponding total hydrogen surface density in the same layer is
\begin{equation}
    \label{eq:column_denisty_h_in_one_layer}
    \Sigma_{\rm H,i} = n_{\rm H,i}  m_{\rm p}  d_{\rm i}.
\end{equation}

The molecular hydrogen mass fraction ($f_{H_2}$) for each \cloudy\ slab is then obtained by summing over all layers:
\begin{equation}
    \label{eq:fh2_cloudy}
    f_{\rm H_2} = \frac{\sum_i\Sigma_{\rm H2,i}}{\sum_i\Sigma_{\rm H,i}}.
\end{equation}

We compute CO mass fraction in an analogous manner; for each layer, \cloudy\ outputs the fraction of carbon in CO, $(n_{\mathrm{CO}}/n_{\mathrm{C}})_i$. We first estimate the total carbon column density in layer $i$ as:
\begin{equation}
    \label{eq:column_denisty_c_in_one_layer}
    \Sigma_{\rm C,i} = \Sigma_{\rm H,i}
    \left(\frac{n_{\rm{C}}}{n_{\rm{H}}}\right)_{Z=Z_\odot}
    \left(\frac{m_{\rm{C}}}{m_{\rm{H}}}\right)
    Z,
\end{equation}
where $\Sigma_{\mathrm{\rm H},i}$ is given by Equation~\ref{eq:column_denisty_h_in_one_layer}. Here, $(n_{\rm{C}}/n_{\rm{H}})_{Z=Z_\odot}$ is the carbon-to-hydrogen number abundance, which is fixed to the ISM value $2.51 \times 10^{-4}$ \citep{Cowie_songaila_1986_ismAbundance, Savage_sembach_1996_ismAbundance} and not the carbon abundance of each gas cell. $m_{\mathrm{C}}/m_{\rm H}$ is the carbon-to-hydrogen atomic mass ratio. $Z$ is the metallicity of the \cloudy\ model, relative to solar, i.e., the factor $Z$ scales the carbon abundance relative to the adopted solar-neighborhood ISM abundance.

We then calculate the CO column density in each layer as
\begin{equation}
    \label{eq:column_denisty_co_in_one_layer}
    \Sigma_{{\rm CO},i} = \Sigma_{{\rm C},i}
    \left(\frac{n_{\rm CO}}{n_{\rm C}}\right)_i
    \left(\frac{m_{\rm CO}}{m_{\rm C}}\right),
\end{equation}
where $(n_{\mathrm{CO}}/n_{\mathrm{C}})_i$ is the CO-to-carbon number ratio output by \cloudy\ for layer $i$, and $m_{\mathrm{CO}}/m_{\mathrm{C}}$ is the carbon monoxide to carbon mass ratio ($28/12=2.33$). We then define the CO mass fraction of the slab as
\begin{equation}
    \label{eq:fco_cloudy}
    f_{\rm CO} = \frac{\sum_i\Sigma_{{\rm CO},i}}{\sum_i\Sigma_{{\rm H},i}}.
\end{equation}

\subsection{Interpolating between CLOUDY Runs} \label{sec:interpolating_between_cloudy_runs}

We used this table to interpolate line luminosities $(\mathrm{erg\,s^{-1}})$ and mass fractions for every gas cell in the host galaxy, defined as above ($r<20\,\kpc$ and, for galaxies with clear disks, $d_\perp<5\,\kpc$). We use the \texttt{RBFInterpolator} implemented in the \texttt{SCIPY} \texttt{Python} library \citep{scipy_2020}. The interpolator is trained using the 200 nearest neighbors in the parameter space of our \cloudy\ lookup table. 

A natural question is why we adopt \RBFInterpolator\ \citep{scipy_2020} rather than a simpler interpolation framework such as \LinearNDInterpolator. Both interpolators interpolates unstructured data in $d$ dimensions but the former is trained by finding the weights of the summed radial basis functions centered on table entities, plus a low-order polynomial, while \LinearNDInterpolator\ is a piecewise linear regression model.

We prefer this to piecewise linear interpolation for two reasons. First, \LinearNDInterpolator\ requires a Delaunay tessellation of the full table, and the total number of building blocks to construct the interpolator grows as $d^2$, and for our own $\rm d=5$ table this is expensive to build, whereas the cost of the radial basis function is set by the neighbor cost much less. Secondly, and more importantly, the linear interpolant requires significantly more table entries to start interpolating accurately for a training set with large curvature. On the other hand, \RBFInterpolator\ is constructed to represent the curvature that the linear scheme could not account for using the weighted sum of the radial basis functions, therefore could converge faster than \LinearNDInterpolator.

Yet, a faster converge comes with a caveat. Unlike piecewise-linear interpolation, the radial basis function interpolant can overshoot where the table varies sharply, typically near the table entries where density and metallicity thresholds at which CO starts to get photodissociated. We therefore create a test to validate the converge of our interpolators. We ran \cloudy\ explicitly for 3000 randomly selected gas cells within the $5\,\rm kpc$ from the galactic center in the \core\ \fireTwo\ \texttt{m12i} Latte galaxy with baryonic mass resolution of $7100\, \rm M_\odot$. Our \RBFInterpolator\ with 200 neighbors recovers the summed \LcoOneZero\ to $\sim 0.6\%$, whereas, a \LinearNDInterpolator\ trained on the same table underpredicts it by $\sim 28\%$. The accuracy of our \RBFInterpolator\ improves with neighbor count up to $\rm k \approx 200$ and degrades beyond it, we therefore adopt $\rm k = 200$ while interpolating for all gas cells. 

Using this trained model, we estimate the line luminosities of each gas cell in the central galaxy of our \fireTwo\ snapshots. In rare cases the \texttt{RBFInterpolator} must extrapolate outside the parameter space of the \cloudy\ table, which results in NaN values. For these particles we instead use the \texttt{NearestNDInterpolator} from the same library.  We then multiply the interpolated line-luminosity per mass by the mass of the gas cell to find \LcoOneZero\ and other line emissions (\LcoTwoOne, $L_\mathrm{H\alpha}$, $L_\mathrm{H\beta}$, $L_\mathrm{[OII]}$ , $L_\mathrm{[OIII]}$ and $L_\mathrm{[NII]}$) in units of $\ergs$. Since observers commonly report their CO results in the units of $\obsunits$, we converted the units of our CO line emission using \citep{Carilli_2013}
\begin{equation}
    \label{eq:co_emission_unitConversion}
    L'_{\rm CO} = \frac{L_{\rm CO}(\ergs)}{L_\odot} \times \frac{1}{3 \times 10^{-11} \times (\nu_{J}/{\rm GHz})^3}, 
\end{equation}
where $\nu_{J}$ is the rest frequency of the corresponding CO emission.

We interpolated the tabulated values of $f_{\mathrm{H}2}$ and $f_{\mathrm{CO}}$ onto the gas cells in our \fireTwo\ simulations. For each gas cell j, the molecular hydrogen mass is computed as
\begin{equation}
    \label{eq:H2_mass_calculation_gas_cell}
    M_{{\rm H_2},j} = f_{{\rm H_2},j} \times M_{{\rm gas},j}/\mu ,
\end{equation}
and the CO mass is computed as
\begin{equation}
    \label{eq:CO_mass_calculation_gas_cell}
    M_{{\rm CO},j} = f_{{\rm CO},j} \times M_{{\rm gas},j}/\mu ,
\end{equation}
where $\mu=1.37$. These quantities are used to calculate the total molecular hydrogen mass ($\rm M_{H_2}$) and total CO mass ($\rm M_{CO}$) of each galaxy by summing them over all gas cells associated with the galaxy.

Observational datasets often report molecular gas mass values that include the contribution from helium, e.g., \cite{Saintonge_2017, Freundlich_2019, boogaard_2019_alma_aspecs}. In this paper, we distinguish between $\rm M_{H_2}$, which refers to the molecular hydrogen mass, and $\rm M_{mol}$, which includes the helium contribution. We calculate the latter by multiplying $\rm M_{H_2}$ by $\mu$:
\begin{equation}
    \label{eq:molecular_mass_inc_Helium_calculation_gas_cell}
    M_{\rm mol} = M_{\rm H_2} \times \mu
\end{equation}

\section{Cloudy CO predictions for MW Molecular Clouds}
\label{sec:cloudy_runs_for_mw_molecular_clouds}

\cite{Guszejnov2020_GMCs_FIRE} and \cite{Benincasa2020_GMCs_FIRE2} showed that the giant molecular clouds (GMCs) in the \core\ \fireTwo\ MW mass galaxies and dwarfs have structural properties broadly similar to those observed in the MW and nearby galaxies \citep{Miville-Deschenes_2017_MW_paper, Colombo2014_M51_GMC_properties}. The simulated GMCs exhibit a sharp drop off in the number of GMCs below $10^5M_\odot$, near the numerical resolution limit, while the observed GMCs show a slow flattening, starting at $\approx3\times10^5M_\odot$. This flattening results from a combination of the limited spatial resolution and limited sensitivity of the Dame \citep{Dame2001_COdata_for_MivilleDeschenes} survey. However, neither simulation-based study predicted the \LcoOneZero\ of the GMCs in \fireTwo\ simulations. An agreement in cloud masses and sizes does not guarantee agreement in CO luminosity, as \LcoOneZero\ depends on the density structure, temperature, radiation field, and, most importantly, the metallicity and chemical state of the gas. 

Because the simulated clouds resemble observed ones in these structural properties, using MW clouds as input to our \LcoOneZero\ calculations provides an excellent test of our CO predictions, which we now proceed to administer.

The \citet{Miville-Deschenes_2017_MW_paper} catalog contains 8107 molecular clouds within $-5^{\circ} < b < 5^{\circ}$ and recovered approximately $98\%$ of the total \LcoOneZero\ in the MW. We have used all of the compiled molecular cloud data within 15 kpc from the galactic center, which corresponds to $\sim97\%$ of their total \LcoOneZero\ data. The catalog uses the observed velocity dispersion together with the expected relation between velocity dispersion, cloud size, and surface density to determine which of the two possible distances (for clouds inside the Solar circle) is more likely for each cloud. We adopt the distance favored by the catalog for each cloud. Appendix~\ref{sec:appendix_mw_cloud_catalog} describes the used MW catalog in more detail.

The catalog also provides cloud-scale quantities including molecular gas column density ($N_{\rm H_{2,cloud}}$), velocity dispersion, and size, which we use as inputs for our \cloudy\ models. For the remainder of this section, we drop the subscript ``cloud'' for clarity. The catalog does not provide the metallicity or ISRF for the clouds. We estimate the density of the molecular clouds as
\begin{equation}
    \label{eq:MW_clouds_hden_calculations}
    n_{\rm H}/{\rm cm^{-3}} = 2 \times N_{\rm H_2} / \rm R.
\end{equation}

We used three different Galactic-metallicity profiles to estimate the GMC metallicity. First, we adopted the Milky Way metallicity measured by \citet{rudolph_2006_MW_metallicity_profile} using optical data. In this prescription, the gas-phase metallicity at Galactocentric radius $\rm R_{gal}$ is given by
\begin{equation}
    (12 + \log({\rm O/H}))_{R_{\rm gal}} = 8.67 + (\rm R_{gal} - 8.5)(-0.06).
    \label{eq:rudolph_metallicity_profile}
\end{equation}
We then converted this oxygen abundance to a metallicity relative to solar by assuming a solar abundance of $\rm (12+\log(O/H))_\odot = 8.69$ \citep{asplund_2009_Mw_metallicity}:
\begin{equation}
    Z / Z_\odot = 10^{(12 + \log({\rm O/H}))_{R_{\rm gal}} \, - \, 8.69}.
\end{equation}

In addition to this radially varying metallicity profile, we also considered constant metallicity models with $Z/Z_\odot = 1$ and 3.

We adopted the MW ISRF value for all clouds. A deviation from this assumption is not expected to strongly alter our results, since our \cloudy\ runs show that the calculated \LcoOneZero\ is less sensitive to variations in the radiation field compared to the variations in density ($n_H$) and metallicity ($Z$).

The catalog does not directly report the CO luminosity, $\LcoOneZero$, for individual clouds. Instead, it provides the total integrated CO emission, $W_{\rm CO(1-0)}$, which represents the sum of the measurements across all pixels associated with a cloud rather than the spatial average. To determine the average integrated intensity, this total must be divided by the number of pixels on the sky, $\rm N_{pix}$. We therefore calculated the CO luminosity by multiplying this average intensity by the physical area of the cloud, $\rm S$:
\begin{equation}
    \LcoOneZero = \frac{\rm W_{\rm CO(1-0)}}{\rm N_{pix}} \times \rm S
\end{equation}

We first ran \cloudy\ directly for each of the MW GMCs in our sample, using their derived physical properties and each of the three metallicity assumptions. Thus, no interpolation was used in calculating the initial CO luminosity predictions. We calculated the \LcoOneZero\ per mass ($\LcoOneZero/\rm M$) following the method described in Section~\ref{sec:cloudy_lineLuminosityPerMass}, fixing the CMB temperature at $z=0$, and multiplied this value by the cloud mass to obtain the total predicted \LcoOneZero. For the cloud masses, we again adopted the distance favored by the catalog for each cloud. In Table~\ref{tab:mw_cloud_lco_comparison}, we show the resulting total CO luminosities for the three metallicity profiles, labeled `Rudolph et al. 2006'', `Z=1'', and `Z=3''. 

We also used this procedure to test the accuracy of our interpolation. We applied our \RBFInterpolator\ to the same clouds with \cite{rudolph_2006_MW_metallicity_profile} metallicity profile assumed. These interpolated results are listed as `Interpolated (with Rudolph+2006)'' in the table.

For reference, the summed \LcoOneZero\ of the MW GMCs in our sample, which is a subset of the clouds from \citet{Miville-Deschenes_2017_MW_paper}, is measured to be approximately $3.68\times10^8\,\obsunits$. The third column gives the ratio of the observed \LcoOneZero\ to the predicted value. The interpolated luminosity differs from the direct `Rudolph et al. 2006'' calculation by only $1.7\%$, indicating that interpolation introduces only a small uncertainty in the total \LcoOneZero\ estimate for the MW molecular clouds.

\begin{table}
    \caption{Comparison between the total CO(1-0) luminosity from \cite{Miville-Deschenes_2017_MW_paper} and the luminosities calculated using different model assumptions and the prediction by interpolating the \cloudy\ table. The ratio is the observed total divided by the model total. The measured total CO luminosity for all the molecular clouds within 15 kpc from the galactic center is $3.68 \times 10^{8}$ $\obsunits$.}
    \label{tab:mw_cloud_lco_comparison}
    \begin{tabular*}{\columnwidth}{@{\extracolsep{\fill}}lcc}
        \hline
        Model & 
        $L_{\mathrm{CO}(1-0)}^{\mathrm{'model}}$ & 
        Ratio \\
        & 
        \multicolumn{1}{c}{[K km s$^{-1}$ pc$^{2}$]} & 
        \\
        \hline
        Interpolated (with Rudolph+2006) & $1.14 \times 10^{8}$ & $3.24$ \\
        Rudolph et al. 2006 & $1.16 \times 10^{8}$ & $3.17$ \\
        $Z=1$ & $2.09 \times 10^{7}$ & $17.64$ \\
        $Z=3$ & $2.42 \times 10^{8}$ & $1.52$ \\
        \hline
    \end{tabular*}
\end{table}

The $\rm Z=3\,Z_\odot$ assumption produced nearly an order of magnitude higher total \LcoOneZero\ compared to the result when using a fixed, solar metallicity. In other words, a factor of three increase in the assumed metallicity increased the predicted \LcoOneZero\ by an order of magnitude. Therefore, any simulation that aims to accurately predict the CO LIM signal will have to track the metallicity to high precision, which will require accurate modeling of metal production and ejection. This can of course be turned around and used as a test of feedback models in simulations. 

The impact of the assumed metallicity is also evident in the comparison with the observed total \LcoOneZero. The $\rm Z=3\,Z_\odot$ profile predicts a value within $30\%$ of the observed luminosity, whereas the solar-metallicity profile underpredicts the observations by a factor of $\sim20$. Our ``Rudolph et al. 2006'' profile yields a total \LcoOneZero\ between the two constant-metallicity cases, but much closer to the $\rm Z=3\,Z_\odot$ prediction. This is expected because \cite{rudolph_2006_MW_metallicity_profile} profile assigns $\rm Z\sim3\,Z_\odot$ for clouds at $\rm R_{gal}<5.5\,kpc$, which contribute $\sim50\%$ of the total luminosity.

\begin{figure}[h!]
    \centering
    \includegraphics[width=0.99\linewidth]{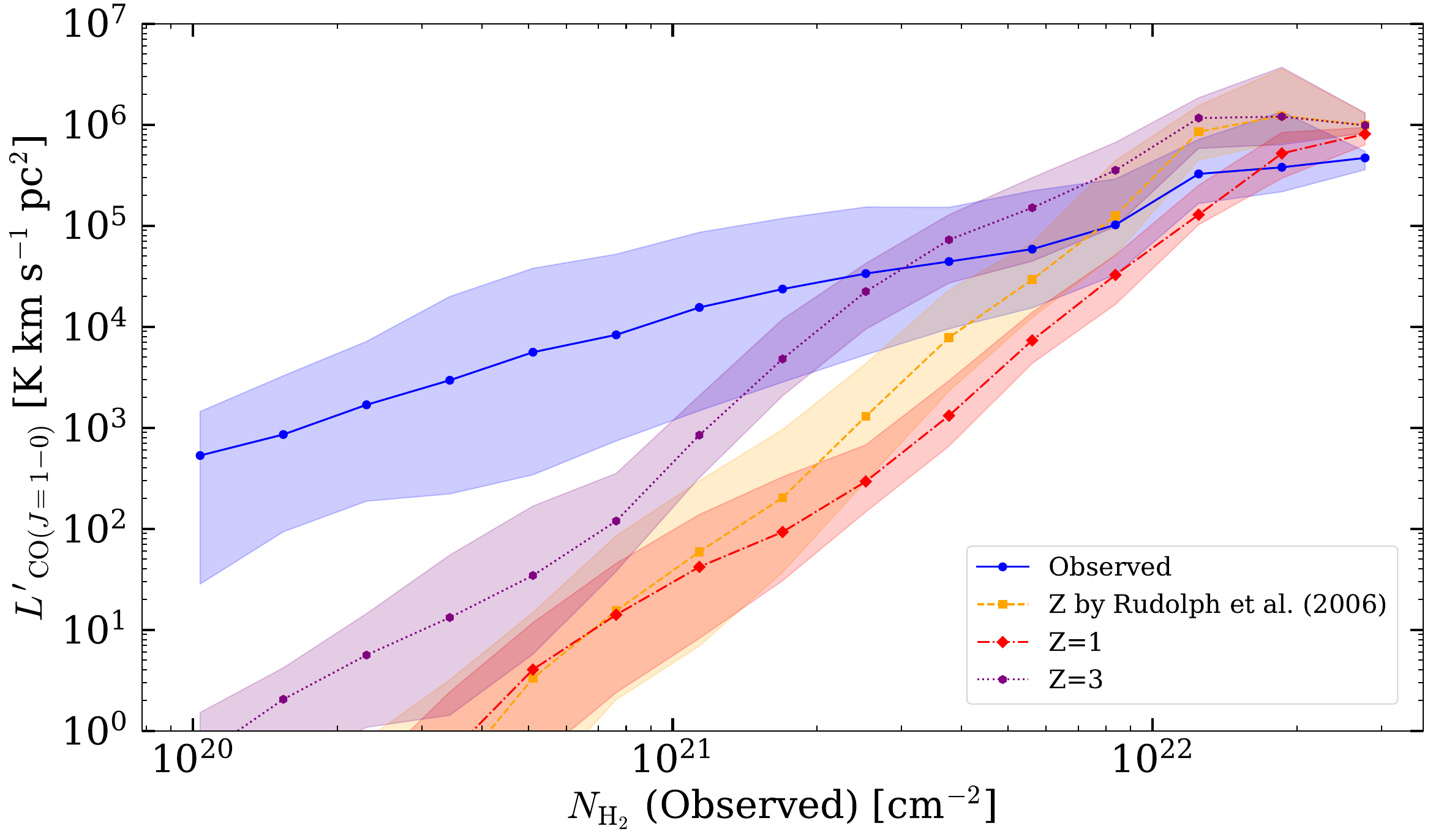}
    \caption{CO(1-0) line luminosity, \LcoOneZero, as a function of the observed H$_2$ column density, $N_{\mathrm{H_2}}$, for the Milky Way molecular clouds of \citet{Miville-Deschenes_2017_MW_paper}. The observed luminosities are shown in blue. The other curves show the \cloudy\ predictions for the same clouds, computed from each cloud's observed properties with three metallicity choices: the Galactocentric metallicity gradient of \citet{rudolph_2006_MW_metallicity_profile} (orange squares), and fixed metallicities of $Z = 1\,Z_\odot$ (red diamonds) and $Z = 3\,Z_\odot$ (purple hexagons). All samples share the same 15 logarithmically spaced bins for $8.4\times10^{19} \leq \rm N_{H_2} \leq 3.4 \times10^{22}$. Points mark the median $L'_{\mathrm{CO}}$ in each bin, placed at the geometric bin center, and the shaded bands show the 16th--84th percentile range.}
    \label{fig:lco_vs_nh2_mw_clouds}
\end{figure}

While encouraging, the convergence in total \LcoOneZero\ does not imply convergence in the cloud-by-cloud predictions. To assess the latter, in Figure~\ref{fig:lco_vs_nh2_mw_clouds} we show \LcoOneZero\ versus observationally determined $\rm N_{H_2}$ for the individual molecular clouds binned according to their $\rm N_{H_2}$. In this figure we plotted our \LcoOneZero\ predictions for different metallicity profiles against the observed $\rm N_{H_2}$, not the \cloudy\ predicted abundances. We choose to use observed $\rm N_{H_2}$ to be consistent with our assumption in Equation~\ref{eq:MW_clouds_hden_calculations}. In other words we assumed the hydrogen density of the molecular clouds as calculated from their CO inferred $\rm H_2$ column densities and their sizes.

For the $\rm Z=3\,Z_\odot$ metallicity profile, our \LcoOneZero\ predictions at a given $\rm N_{H_2}$  approach and then track the observed luminosities for $\rm N_{H_2}\geq10^{21}\,cm^{-2}$. At low $\rm N_{H_2}$ the predicted luminosity is well below the measured luminosities; we predict a steep increase in \LcoOneZero\ with increasing $\rm N_{H_2}$. A similar sharp increase emerges for the solar-metallicity profile, but at a column density approximately three times higher, $\rm N_{H_2}\sim3\times10^{21}\,cm^{-2}$. In other words, increasing the metallicity by a factor of three shifts the onset of this  increase in \LcoOneZero\ to a column density lower by approximately the same factor. Why does a factor of three increase in metallicity shift the characteristic column density for the onset of bright CO emission by a factor of three?

\begin{figure*}
    \centering
    \begin{minipage}{0.49\textwidth}
        \centering
        \includegraphics[width=\linewidth]{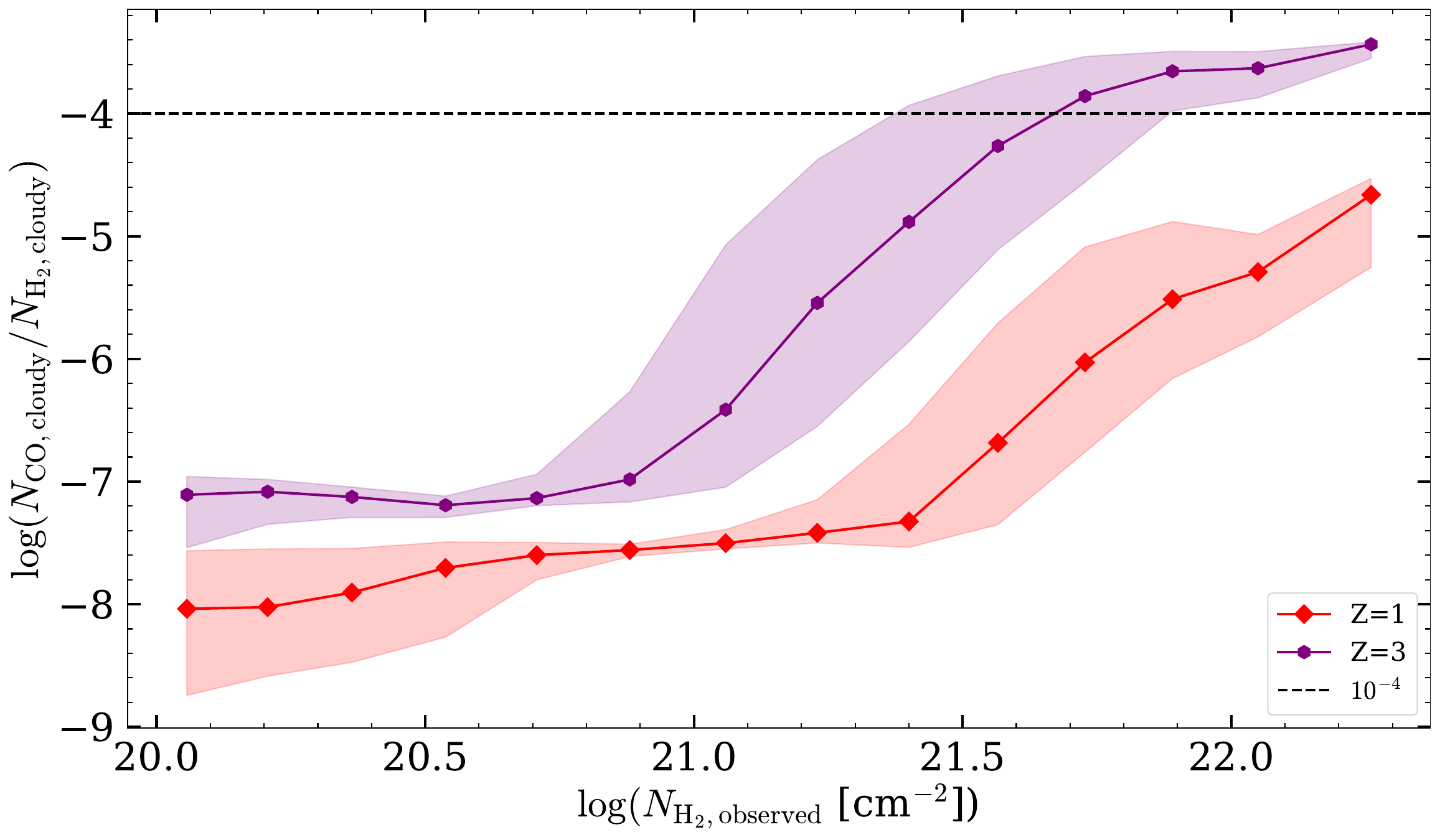}
    \end{minipage}
    \hfill
    \begin{minipage}{0.49\textwidth}
        \centering
        \includegraphics[width=\linewidth]{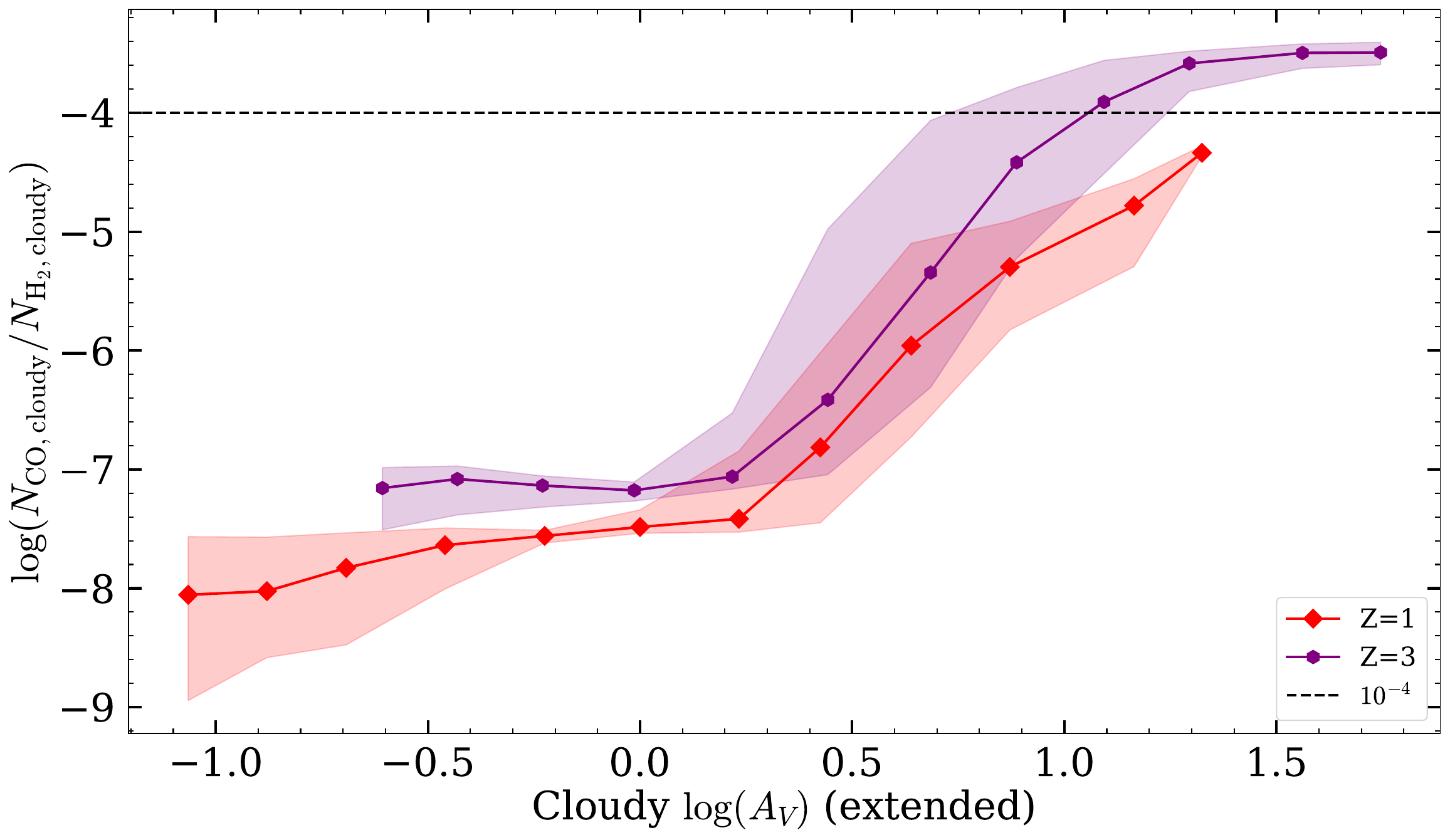}
    \end{minipage}
    \caption{CO-to-H$_2$ abundance ratio, $\rm N_{CO, cloudy}/N_{H_2, cloudy}$, predicted by \cloudy\ for the Milky Way molecular clouds of \citet{Miville-Deschenes_2017_MW_paper} at Galactocentric radii $R_{\mathrm{gal}} < 15$\,kpc. The abundance ratio is the number ratio derived from the CO and H$_2$ mass fractions of each cloud model. \textit{Left:} $\rm N_{CO}/N_{H_2}$ as a function of the observed H$_2$ column density, $N_{\mathrm{H_2}}$, in 15 bins of equal width in $\log N_{\mathrm{H_2}}$ spanning $8.4\times10^{19} < N_{\mathrm{H_2}}/\mathrm{cm^{-2}} < 3.4\times10^{22}$. \textit{Right:} $\rm N_{CO}/N_{H_2}$ as a function of the visual extinction, $\rm A_V$, computed by \cloudy\ for an extended source, in 15 bins of equal width in $\rm \log(A_V)$ spanning the combined range of both runs. Because the dust content scales with metallicity, the same cloud has a different $A_V$ in each run. In both panels, colors show two fixed metallicities, $Z = 1\,Z_\odot$ (red diamonds) and $Z = 3\,Z_\odot$ (purple hexagons). Points mark the median of each bin, placed at the bin's median $x$-value, and shaded bands show the 16th--84th percentile range. The dashed black line marks the canonical Milky Way GMC CO abundance, $N_{\mathrm{CO}}/N_{\mathrm{H_2}} = 10^{-4}$.}
    \label{fig:nco_over_nh2_mw_clouds}
\end{figure*}

To answer this question we have to look at the predicted CO abundances. Using the same \cloudy\ setup discussed above we predict the $\rm H_2$ mass fraction ($\rm f_{H_2}$) and CO mass fraction ($\rm f_{CO}$) as defined in Equations~\ref{eq:fh2_cloudy} and \ref{eq:fco_cloudy} respectively. We then multiplied the predicted mass fractions with the mass of individual species in the MW clouds. Finally we compute the number column densities for each cloud i by dividing their mass ratios and normalizing to atomic masses for each species by: 
\begin{equation}
    \left(\frac{\rm N_{CO}}{N_{H_2}}\right)_i = \rm \left(\frac{M_{CO}}{M_{H_{_2}}}\right)_i\,\frac{m_{H_2}}{m_{CO}}
\end{equation}
Where $\rm M_{CO}$ and $\rm M_{H_2}$ is the predicted CO and $\rm H_2$ masses for individual cloud and $\rm m_{H_2}/m_{CO}$ is equal to $2.016/28.01\sim0.07$. 

In Figure~\ref{fig:nco_over_nh2_mw_clouds} we plot the predicted CO-to-H$_2$ abundance ratio as a function of the observed H$_2$ column density, $N_{\mathrm{H_2}}$ on the left and extended visual extinction directly printed out by \cloudy\ on the right. Note that the two quantities in the figure on the left come from different sources. The abundance ratio, $\rm N_{CO, cloudy}/N_{H_2, cloudy}$, is computed entirely from our \cloudy\ predictions. The $\rm N_{H_2, observed}$ on the x-axis, however, is the observationally derived molecular hydrogen column density of \cite{Miville-Deschenes_2017_MW_paper} from the CO data by assuming a constant $\rm X_{CO}=2\times10^{20} \rm cm^{-2}(K \, km/s)^{-1}$.

The \cloudy\ values of $\rm N_{H_2, observed}$ are systematically lower than those of \cite{Miville-Deschenes_2017_MW_paper}. This is because we set the hydrogen density $n_{\mathrm{H}}$ of each \cloudy\ model directly from the $\rm N_{H_2, observed}$ using Equation~\ref{eq:MW_clouds_hden_calculations}, so the total hydrogen content is matched, but \cloudy\ does not place all of that hydrogen in molecular form. In clouds with low column density ($A_V < 1$), a significant fraction of the hydrogen remains atomic or ionized because the cloud provides insufficient shielding to form H$_2$.

For our $Z=3\,Z_\odot$ profile, the CO abundance rises sharply at $N_{\mathrm{H_2,\,observed}} \gtrsim 10^{21}\,\mathrm{cm^{-2}}$, consistent with the onset of bright CO emission in these clouds. Decreasing the metallicity by a factor of three, to $Z=1\,Z_\odot$, shifts this rise to a higher column density of $N_{\mathrm{H_2,\,observed}} \sim 4\times10^{21}\,\mathrm{cm^{-2}}$, a factor of $\sim3$--4 higher. The onset of bright CO emission is therefore set by how well CO is protected from photodissociation by the far-UV radiation field, which depends on the dust shielding of the cloud. The size of the shift is not a coincidence: the visual extinction scales as $A_V \propto Z\,N_{\mathrm{H}}$, so the column density at which $A_V$ reaches the critical value of 1 shifts by the same factor as the metallicity.

This is in line with the behavior predicted by \citet{2011MNRAS.412..337G}.

Our predicted $N_{\mathrm{CO,\,cloudy}}/N_{\mathrm{H_2,\,cloudy}}$ as a function of $A_V$ supports this interpretation. Both the $Z=3\,Z_\odot$ and $Z=1\,Z_\odot$ profiles show a sharp increase in the CO abundance at $A_V \gtrsim 1$, so the two curves line up in $A_V$ with an offset in the predicted $N_{\mathrm{CO,\,cloudy}}/N_{\mathrm{H_2,\,cloudy}}$. The remaining constant offset between the two profiles in $N_{\mathrm{CO,\,cloudy}}/N_{\mathrm{H_2,\,cloudy}}$ is likely due to the different carbon and oxygen abundances in the two sets of models, since both scale with metallicity and set the maximum CO abundance a cloud can reach at a given $\rm N_{H_2}$.

Our \LcoOneZero\ predictions therefore reproduce the observations when the shielding is high enough. At low shielding, however, our predictions in Figure~\ref{fig:lco_vs_nh2_mw_clouds} deviate systematically from the data. We attribute part of this deviation to CO-dark gas, which is molecular gas where H$_2$ is self-shielded but CO is photodissociated, so it cannot be traced with CO observations \citep{Wolfire_2010_coDarkGas, Feldmann2012_Xco_paperI, Madden_2020, bisbas_2015_co_dark_clouds}. In addition, Figure~9 of \cite{Miville-Deschenes_2017_MW_paper} shows that the atomic hydrogen column density dominates over the molecular one at $R_{\mathrm{gal}} > 6\,\mathrm{kpc}$. The discrepancy at low shielding may therefore also reflect the simple prescription we used to set $n_{\mathrm{H}}$ from the observed column densities, which likely underestimates the volume densities of the real clouds, therefore the shielding.

\section{Comparison of $N_{\rm CO}/N_{\rm H_2}$ versus $N_{\rm H_2}$ with Diffuse Milky Way Observations  \label{section:Nco_Nh2_comparison}}

The Milky Way cloud benchmark above shows that the CO abundance, and hence \LcoOneZero, rises sharply once dust shielding is sufficient ($\rm A_V\gtrsim1$). Here we test this using \fireTwo\ simulations. In this section we compare our predicted $\rm N(CO)/N(H_2)$ against ultraviolet absorption measurements along galactic translucent sight lines, compiled from \citet{Rachford_2002_Nco_Nh2}, \citet{Crenny_Federman_2004_Nco_Nh2}, \citet{Sheffer_2008_Nco_Nh2}, and \citet{Burgh_2010_Nco_Nh2}. Both $\rm N({CO})$ and $\rm N({H_2})$ are measured directly from the absorption profiles rather than inferred through a conversion factor. These individual surveys are described in Appendix~\ref{sec:MW_Nco_Nh2_data}.

\begin{figure}[h!]
    \centering
    \includegraphics[width=1\linewidth]{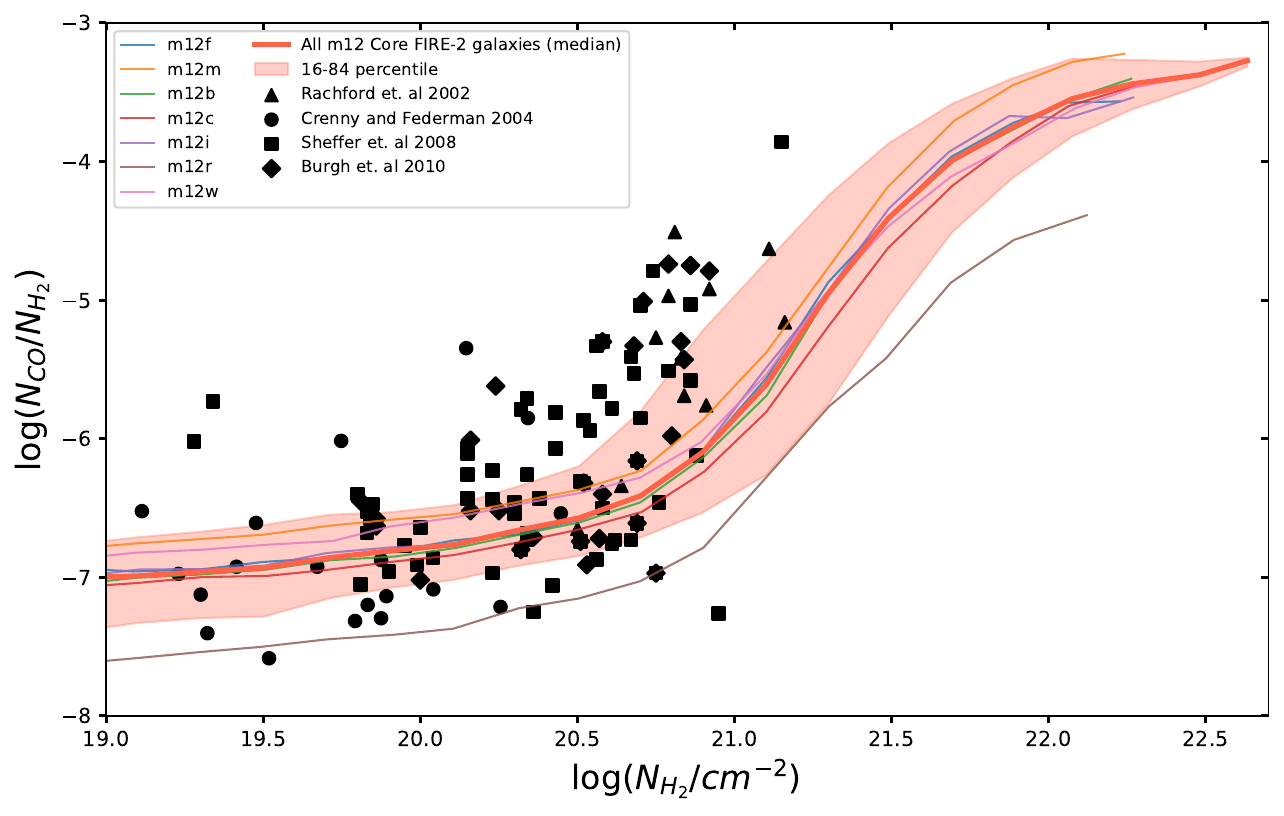}
    \caption{$\Sigma_{\rm CO}/\Sigma_{\rm H_2}$ as a function of $\Sigma_{\rm H_2}$ compared with observations of diffuse Milky Way clouds. The \fireTwo\ predictions are obtained from the MW-mass galaxies in \core\ \fireTwo\ simulation and a baryonic mass resolution of $7100~M_\odot$. Column densities are computed from a 100 pc pixelized galaxy map and binned in logarithmic $N_{\rm H_2}$ intervals using 51 bins over $5 < \log(N_{\rm H_2}/\mathrm{cm^{-2}}) < 23$. The red line shows the mean relation, and the shaded region shows the $1\sigma$ scatter. Black symbols show observational measurements from \citet{Rachford_2002_Nco_Nh2} (triangles), \citet{Crenny_Federman_2004_Nco_Nh2} (circles), \citet{Sheffer_2008_Nco_Nh2} (squares), and \citet{Burgh_2010_Nco_Nh2} (diamonds). \label{fig:Nco_Nh2_m12CoreOnly}}
\end{figure}

We first calculate $\rm H_2$ and $\rm CO$ mass for each gas cell in the simulation using Equation~\ref{eq:H2_mass_calculation_gas_cell} and Equation~\ref{eq:CO_mass_calculation_gas_cell}. We keep the gas cells in the inner galaxy and in the disc, defined as $\rm R_{gal} < 5\, kpc$, and height $\rm d_{\perp} < 500\,pc$ and view the galaxy face-on.  We cover the region $\rm -5 < x,y < 5$\,kpc with square pixels of side $\Delta = 100$\,pc. Each gas cell is assigned to the pixel that contains its position. The column density of each pixel is then: 
\begin{equation}
  N_{\rm H_2} = \frac{\sum_{j\in{\rm pix}} M_{{\rm H_2},j}}{m_{\rm H_2}\,\Delta^2}, \qquad
\end{equation}
\begin{equation}
  N_{\rm CO}  = \frac{\sum_{j\in{\rm pix}} M_{{\rm CO},j}}{m_{\rm CO}\,\Delta^2}
\end{equation}
with $m_{\rm H_2} = 2.016$\,u and $m_{\rm CO} = 28.01$\,u.

Figure~\ref{fig:Nco_Nh2_m12CoreOnly} shows our predicted CO to $\rm H_2$ column density ratio, $\rm N_{CO}/N_{H_2}$, plotted against $\rm N_{H_2}$, together with the observational data from MW sight lines. The \fireTwo\ predictions are for all of the \core\ \fireTwo\ galaxies at $z=0$ (m12b, m12c, m12f, m12i, m12m, m12r, and m12w) with mass resolution $7100\, \rm M_\odot$. We do a similar analysis for \firebox\ galaxies in Appendix~\ref{section:Nco_Nh2_comparison_firebox}. The thin colored lines show the median $\rm N_{CO}/N_{H_2}$ of each galaxy in bins of $\rm N_{H_2}$. The thick line shows the median of all the galaxies combined, and the shaded band marks the 16-84 percentile range of the combined sample of all galaxies. We note that the observations have much narrower beams. Increasing the pixel resolution in our \fireTwo\ galaxies that would match their beam sizes ($\sim 30'' \times 30''$ aperture) would require resolving the sub-resolution structure of our \fireTwo\ gas cells. Test calculations at somewhat higher resolution show that with the decreased pixel size the dispersion around the median line increases. This is expected as at large beam sizes the abundances will be smoothed.

Our predictions are consistent with the MW observations for $\rm \log(N_{H_2}/\mathrm{cm^{-2}}) \lesssim 20.5$.
The data show a sharp upturn in $\rm N_{\rm CO}$ at $\rm \log(N_{H_2}/\cm^{-2})\approx 20.5$, while our calculated $\rm N_{\rm CO}$ shows a slower upturn at a slightly higher column, $\rm \log(N_{H_2}/\cm^{-2})\approx20.8$. Above the break, the \core\ MW-like \fireTwo\ predictions fall below the observed values, by about an order of magnitude, although the two distributions overlap. It is nonetheless encouraging that the observed and predicted upturn fall within 0.3 dex (in $\rm N_{H_2}$) of one another.

\section{Comparison of $M_{H_2}$ and $X_{\rm CO}$ with Observations \label{sec:comparsion_mh2_mco_xco_with_obs}}

In this section we compare our predicted $\rm H_2$ masses, $M_{\rm H_2}$, and our CO-to-$\rm H_2$ conversion factor ($X_{\rm CO}$), with observations. We introduce each dataset in a sentence or two where it is first used. The sample selection, the observing setup, and for the galaxy-integrated surveys the assumptions used to convert the observed CO luminosities into the molecular gas masses we compare against, are presented in detail in Appendix~\ref{sec:observational_data}. The same datasets are used for the \LcoOneZero\ and \LcoTwoOne\ comparisons in \citetalias{Paper2}.

\subsection{Calculation of $\alpha_{\rm CO}$ and $X_{\rm CO}$}
\label{sec:Calculation of Xco}
We calculated  $\alpha_{\rm CO}$  \fireTwo\ galaxies using the definition in Equation~\ref{eq:alpha_co_definition}:
\begin{equation}
    \alpha_{\rm CO} \equiv \frac{\mu M_{\rm H_2}}{L'_{\rm CO(1-0)}},
\end{equation}
where $\mu=1.37$ is the mean molecular mass accounting for Helium \citep{1984ApJ...276..182S, 1991ARA&A..29..581Y}. 

We convert our $\alpha_{\rm CO}$ predictions to $X_{\rm CO}$ using the constant conversion factor  \cite{Narayanan_2012}: 
\begin{equation}
    \label{eq:conversion_Xco_alphaCO}
    X_{\rm CO} = 6.3 \times 10^{19} \times \alpha_{\rm CO},
\end{equation}

\subsection{Comparison with Observations at z=0}
\subsubsection{$M_{H_2}$ versus SFR}

At $z\approx0$ we compare against \cite{Saintonge_2017} (hereafter xCOLD GASS), a CO(1-0) survey of 532 stellar-mass-selected galaxies at $0.01<z<0.05$ spanning $9<\log(\Mstar/M_\odot)<11.3$. Their molecular gas masses are inferred from \LcoOneZero\ using the metallicity- and main-sequence-offset-dependent conversion factor of \citet{Accurso_2017}; see Appendix~\ref{sec:low_z_observational_data}.

\begin{figure}[h!]
    \centering
    \includegraphics[width=0.99\linewidth]{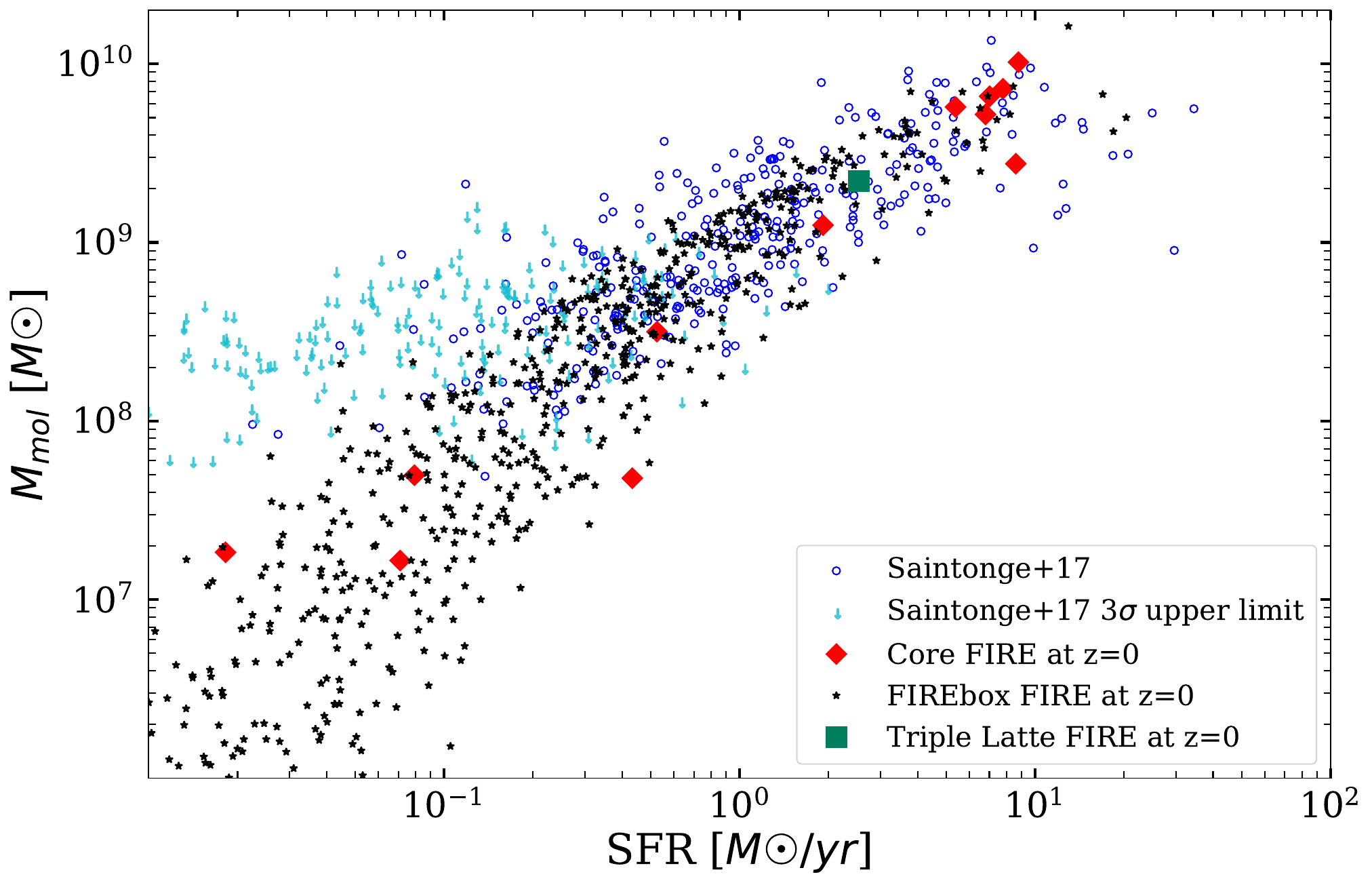}
    \caption{The molecular mass $\rm M_{mol}$ (including Helium contribution) as a function of SFR for post-processed \fireTwo\ galaxies and xCOLD GASS galaxies. We used the star formation rate averaged over 100 Myrs (\sfrHundredMyr) for our \fireTwo\ galaxies. \firebox, \core\ \fireTwo\, and \tripleLatte\ simulations are represented with  black stars, red diamonds, and a single green square, while xCOLD GASS detections are shown with open blue circles, upper limits with downward arrows. The xCOLD GASS masses $\rm M_{H_2}$ were calculated by those authors using the bimodal $\rm CO-to-H_2$ conversion factor ($\alpha_{\rm CO}$) of \cite{Accurso_2017}, which depends on the metallicity of the galaxy and the offset of the galaxy from star formation main sequence.}
    \label{fig:mh2_vs_xCOLDGASS}
\end{figure}

We use the reported molecular gas of xCOLD GASS galaxies to compare to our \fireTwo\ predictions in Figure \ref{fig:mh2_vs_xCOLDGASS}. Their reported molecular gas includes the helium contribution. Our predictions at $z=0$ lie within the scatter of the xCOLD GASS observational data for $SFR\gtrsim0.3\,M_\odot\,{\rm yr}^{-1}$. For that SFR range the xCOLD GASS galaxies exhibit a dispersion of $\sim0.5$ dex, while our \fireTwo\ predictions have a scatter of $\sim0.4$ dex, indicating comparable levels of variation for the data and in our predictions. This agreement is attained without tuning on our part but a prediction from the first principles using \skirt\ and \cloudy\ pipeline. At lower SFRs, the \fireTwo\ predictions are much lower than the few observational detections; most of the observations at low SFRs provide only upper limits. 

The xCOLD GASS sample is complete down to $\LcoOneZero\sim10^{7.5}\,\obsunits$. Galaxies with CO luminosities below this sensitivity limit enter the observational sample as upper limits rather than as measurements, and the true molecular gas masses of the non-detections may lie far below the quoted values. Using these data alone, we would therefore describe the low SFR regime as untested rather than discrepant, and settling it calls for more sensitive CO surveys rather than a better emission model. We return to this point in \citetalias{Paper2}, where we compare \LcoOneZero\ against SFR using data from more sensitive surveys, which do show a larger observed scatter extending to fainter luminosities.

Comparing the red diamonds in Figure~\ref{fig:mh2_vs_xCOLDGASS}, which depict the \core\ \fireTwo\ zooms, with the black stars, from the lower resolution \firebox\ simulations, suggests that the predicted \LcoOneZero\ are reasonably well converged. At all SFRs, the mean value of the molecular gas mass for the \core\ \fireTwo\ are within one standard deviation of the lower resolution \texttt{FIREbox} galaxies. The \core\ \fireTwo\ simulations show a smaller dispersion compared to \firebox\ galaxies, but this is due at least in part to the small number of \core\ \fireTwo\ simulations.

\subsubsection{$M_{H_2}$ at different redshifts}

\begin{figure}[t]
    \centering

    \begin{subfigure}[t]{0.48\textwidth}
        \centering
        \includegraphics[width=\linewidth]{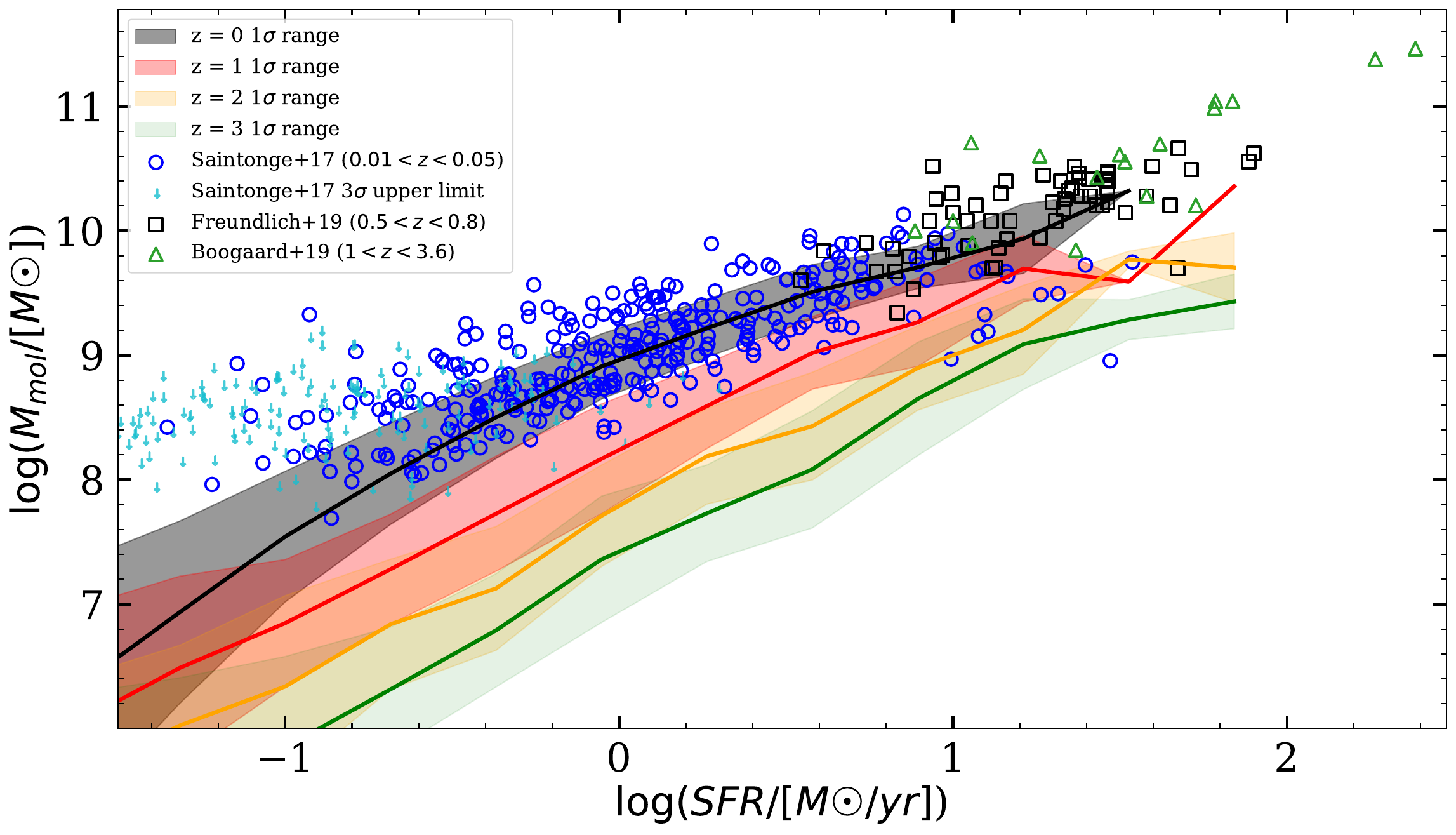}
    \end{subfigure}
    \hfill
    \begin{subfigure}[t]{0.48\textwidth}
        \centering
        \includegraphics[width=\linewidth]{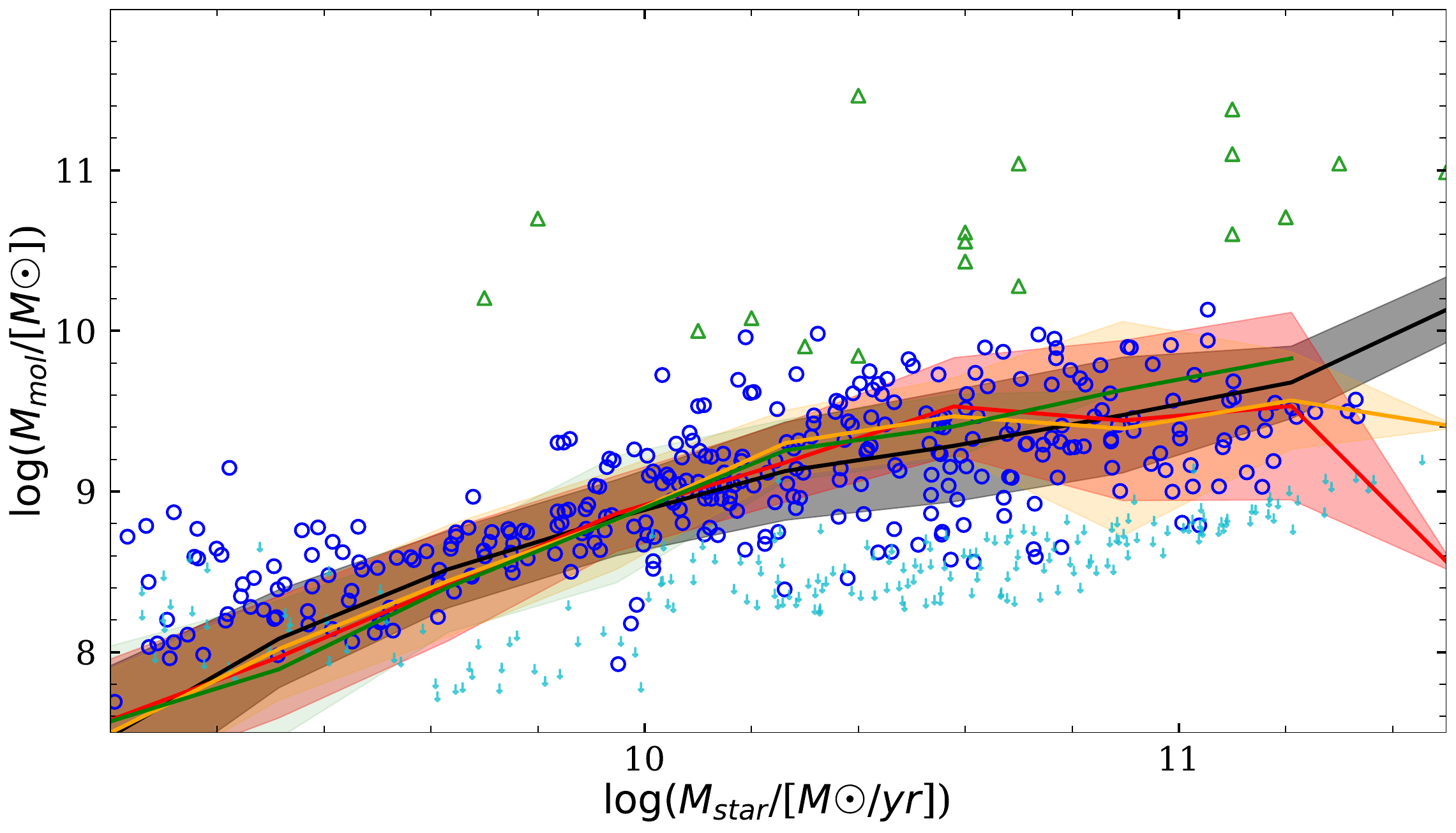}
    \end{subfigure}

    \caption{
      Molecular gas mass ($\rm M_{mol}$) at four different redshifts, plotted against SFR (top panel) and against \( M_{star} \) (bottom panel). For the \fireTwo\ galaxies we calculated SFR averaged over 100 Myrs (\( SFR_{10Myr} \)).   Redshift \( z=0 \) data from \citet{Saintonge_2017} is represented by blue circles, \( z=0.5-0.8 \) data from \citet{Freundlich_2019} by hollow black squares, \( z=1-3.6 \) data from \citet{boogaard_2019_alma_aspecs} by green triangles. Our simulation based predictions for all of our simulated \fireTwo\ galaxies are depicted by lines, with shading giving the 1$\sigma$ ranges. Redshift z=0 results are in black, z=1 in red, z=2 in orange, and redshift 3 in green.
      \label{fig:Mh2_vs_sfr_mstar_redshifts}
    }
    \label{fig:two_column_figure}
\end{figure}

The molecular gas masses of observed galaxies are not measured directly. Rather, they are inferred from CO luminosities by assuming a CO-to-H$2$ conversion factor, $\alpha_{\rm CO}$, see Equation~\ref{eq:alpha_co_definition}. A commonly adopted value is the Milky Way conversion factor, $\alpha_{\rm CO}=4.3~M_\odot\,(\obsunits)^{-1}$, or $3.6~M_\odot\,(\obsunits)^{-1}$ for $z\approx1.5$ star-forming galaxies \citep{Daddi2010_alphaCO} but $\alpha_{\rm CO}$ can vary strongly with the physical conditions of the gas, especially with metallicity, but also with radiation field and density \citep{Feldmann2012_Xco_paperI, Madden_2020, bisbas_2015_co_dark_clouds}. The observed molecular gas mass values should be interpreted with this in mind.

In the previous section we compared our predicted molecular gas masses against local observations from xCOLD GASS \citep{Saintonge_2017}. In this section we compare the molecular gas masses predicted for our \fireTwo\ galaxies against observational estimates from PHIBSS2 \citep{Freundlich_2019} at $0.5<z<0.8$, and ASPECS \citep{boogaard_2019_alma_aspecs} at $1<z<3.6$ as well as local xCOLD GASS \citep{Saintonge_2017}. PHIBSS2 is an IRAM Plateau de Bure CO(2-1) survey of massive star-forming galaxies, $10<\log(\Mstar/M_\odot)<11.8$, whose gas masses assume a fixed line ratio $r_{21}=0.77$ together with the metallicity-dependent conversion factor of \citet{Tacconi_2018}. Since both the excitation correction and $\alpha_{\rm CO}$ vary with galaxy properties, the higher-redshift gas masses carry the larger systematic uncertainty. We describe the two samples in detail in Appendix~\ref{sec:high_z_observational_data}. The comparison is shown as a function of star formation rate (top panel) and stellar mass (bottom panel) in Figure~\ref{fig:Mh2_vs_sfr_mstar_redshifts}.

The \fireTwo\ results shown in the top panel predict that molecular gas mass decreases with increasing redshift at fixed SFR. In contrast, the observations do not show much variation with redshift at fixed SFR. This lack of variation with redshift may partly arise from observational selection effects.

Current CO surveys are sensitivity-limited, so often report only upper limits for CO-dark galaxies. For example the xCOLD GASS \citep{Saintonge_2017}, one of the highest sensitivity observations used in this work, is only sensitive to sources with $\LcoOneZero > 10^{7.5} \, \obsunits$, and the corresponding limits for the higher-redshift samples are considerably brighter. For example, PHIBSS2 \citep{Freundlich_2019} detects no galaxy fainter than $\LcoTwoOne \approx 10^{8.9} \, \obsunits$ at $0.5<z<0.8$, while ASPECS \citep{boogaard_2019_alma_aspecs} reaches only  $\LcoOneZero \approx 10^{9.1}\,\obsunits$ at $1<z<1.7$ and $\approx10^{9.7}\,\obsunits$ at $2<z<3$. Galaxies with lower \LcoOneZero, and therefore lower inferred molecular gas masses, are more likely to fall below the detection limit. As a result, the observed samples may not represent the full galaxy population.

The bottom panel shows that the \fireTwo\ simulations predict molecular gas masses that are nearly independent of redshift at fixed \Mstar. In contrast, the observed redshift $\rm z=1-3.6$ galaxies (depicted by the green triangles) have molecular gas masses roughly an order of magnitude larger than lower-redshift galaxies at the same stellar mass. We caution, however, that the \cite{boogaard_2019_alma_aspecs} observations use high $J$ CO transitions. They convert the high-$J$ CO measurements to CO(1-0) using the excitation ladder from \cite{Daddi_2015_COexcitationLadder}. They then infer molecular gas masses using a constant conversion factor, $\alpha_{\rm CO}=3.6~M_\odot\,(\obsunits)^{-1}$. Since both the CO excitation correction and $\alpha_{\rm CO}$ can vary with galaxy properties, especially in low-metallicity systems, the inferred molecular gas masses may carry large systematic uncertainties.

\subsubsection{$X_{\rm CO}$ versus SFR}

We compared our $X_{\rm CO}$ predictions against xCOLD GASS sample again. We first determine the $\alpha_{\rm CO}$ of our \fireTwo\ predictions, then using Equation~\ref{eq:conversion_Xco_alphaCO} we calculate $X_{\rm CO}$. Then we compare against the
$X_{\rm CO}$ values those authors adopted by following \cite{Accurso_2017}. Appendix~\ref{sec:low_z_observational_data}
gives the details about xCOLD GASS data.

\begin{figure}[h!]
    \centering
    \includegraphics[width=0.99\linewidth]{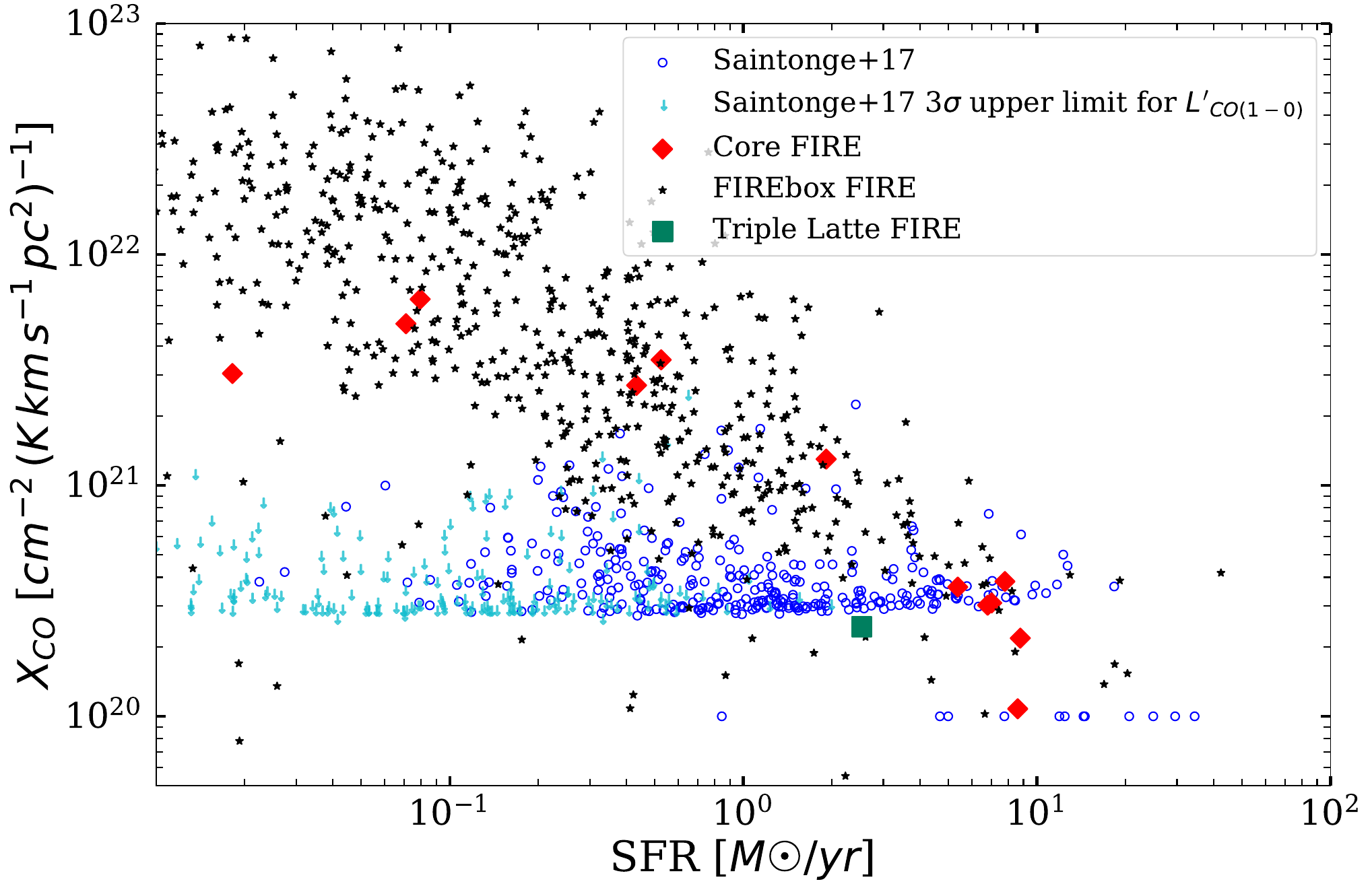}
    \caption{The $CO-to-H_2$ conversion factors  $X_{\rm CO}$ plotted as a function of SFR. Blue circles show the xCOLD GASS detections while arrows show $3\sigma$ upper limits \citep{Saintonge_2017}. The black stars, red diamonds, and the green square again represent \texttt{FIREbox}, \core\ \fireTwo\, and \tripleLatte\ resimulation results. The $X_{\rm CO}$ values found from the \fireTwo\ simulations vary by three orders of magnitude for both low (\texttt{FIREbox}) and medium resolution (\core\ \fireTwo) simulations. In contrast, the $X_{\rm CO}$ values reported by xCOLD GASS scatter by about one order of magnitude at a given SFR. However, the plot clearly shows that for many objects, $X_{\rm CO}$ was effectively held fixed at $4\times10^{20}\,{\rm cm}^{-2}(\obsunits)^{-1}$ in obtaining the xCOLD GASS results.}
    \label{fig:xco_vs_xCOLDGASS}
\end{figure}

Our calculations predict substantial variations in $X_{\rm CO}$ at fixed SFR, as shown in Figure~\ref{fig:xco_vs_xCOLDGASS}; the stars, diamonds, squares, circles and arrows denote the same sources as in Figure~\ref{fig:mh2_vs_xCOLDGASS}. The \fireTwo\ predicted $X_{\rm CO}$ can reach values two orders of magnitude higher than the value $\approx 4\times 10^{20}\,{\rm cm}^{-2}\,({\rm K-km}\, {\rm s}^{-1})^{-1}$ seen in nearby MW-like galaxies \citep{Narayanan_2012, Bolatto_2013}.

This result is not surprising, since many of the low SFR galaxies in our simulations have low metallicity and it is well known that low metallicity galaxies have high $X_{\rm CO}$ (or equivalently $\alpha_{\rm CO}$), e.g., \citet{Leroy_2011_constantXco, Bolatto_2013}. Similarly, theoretical and simulation-based studies consistently predict that $X_{\rm CO}$ increases substantially as metallicity decreases \citep{Wolfire_2010_coDarkGas, Feldmann2012_Xco_paperI, Madden_2020, bisbas_2015_co_dark_clouds}. This increase reflects the growing fraction of CO-dark molecular gas: H$_2$ can remain abundant in regions where CO is efficiently photodissociated and therefore no longer traces the full molecular gas reservoir.

\section{Caveats}
\label{sec:discussion_caveats}

Several assumptions may limit the accuracy of our results; we list some of them here.

One important simplification is the geometry assumed in our \cloudy\ runs. We model individual simulated gas cells as one-dimensional, constant-density, plane-parallel slabs illuminated from one side, with a stopping length matched to the smoothing length of the gas cell. Real clouds are three dimensional and clumpy, and show strong surface density variations within cloud boundaries. In fact the boundaries in images of GMCs are not observationally (or physically) well-defined, but they are definitely not spherical, e.g., \citep{Dame2001_COdata_for_MivilleDeschenes}. Our choice to use a slab geometry for \cloudy, then map the result onto gas cells (see Section~\ref{sec:cloudy_lineLuminosityPerMass}) avoids the spherical volume weighting commonly commonly in the literature \citep{Liang_2023, simonLunjun_jasonGuachao_extendedCII}; the latter could underweight the luminosity produces by the cloud's shielded interior where the bulk of the CO lives. Relaxing this assumption would require much higher resolution in the simulations.

The grid for the \cloudy\ look-up table is finite. We tabulate emission per unit gas mass at 27,324 points for each redshift, amounting to total of 109,296 \cloudy\ simulation for four different redshifts. We then interpolate onto a distributions of gas properties. Roughly $0.6\%$ of these \cloudy\ runs fail to converge (see Appendix~\ref{sec:problematic_cloudy_runs}). We conducted a self-consistency check on the interpolation itself with the Milky Way (MW) clouds in Section~\ref{sec:cloudy_runs_for_mw_molecular_clouds}: the interpolated total \LcoOneZero\ differs by $\sim2\%$ from what we obtain by running \cloudy\ on each cloud individually, which we consider to be tolerable.

An another simplification is related with the continuum spectral energy density (SED) used in our \cloudy\ runs. Although \skirt\ returns the full local radiation field for every gas cell, we only pass its FUV normalization $G_{\rm FUV}$ (see Equation~\ref{eq:radiation_field_calculation_fuv_habing}) to \cloudy. We then use the predefined Milky Way spectral shape (the ``table ism'' option) in our runs, scaled up or down in strength alone. In reality, the shape of the SED will vary from one location to another within a galaxy. A gas cell sitting close to a young, massive stellar population faces considerably harder field than one in the outskirts, where the light has be reprocessed and reddened. The hardness of the radiation field also depends on the age and metallicity of the stars, so that we would expect systematically harder spectra in the younger and more metal-poor galaxies, especially at high-$z$. This will be particularly important for recombination lines such as H$\alpha$ emission, but the effects may also be important for CO emission.  Photons between $11.1$ and $13.6$ eV can photodissociate CO, so the rate at which CO is destroyed, and hence \LcoOneZero, will depend on the shape of the SED, not just the overall flux.

Using the correct SEDs from \skirt\ in  \cloudy\ runs is possible in principle, but it would require greatly expanding the precomputed lookup table as one would have to specify the spectral shape which could not be reduced to a single interpolation variable in the way that $G_{\rm FUV}$ can. We note that the choice of a fixed SED is one of the less well controlled assumptions in the pipeline. 

Another simplifying assumption is our adoption of a fixed dust-to-metal ratio of $0.4$ and a single grain mixture \citep{Weingartner_2001} for all galaxies at all redshifts. In addition, we remove the effects of dust in all gas with $T$ above $10^{6}\,{\rm K}$.  Any deviation from these assumptions could change the dust extinction and alter the local radiation field, so it would change our predictions directly. We also assumed a MW cosmic ray background in all simulated galaxies which could penetrate deep inside the molecular clouds and dissociate CO molecules.

The Milky Way cloud test in Section~\ref{sec:cloudy_runs_for_mw_molecular_clouds} shows how sensitive our CO predictions are to the way metallicity is assigned. Changing the assumed metallicity of the \citet{Miville-Deschenes_2017_MW_paper} clouds from $Z=1\,Z_\odot$ to $Z=3\,Z_\odot$, a factor of three, changes the predicted total \LcoOneZero\ by a factor of $\approx10$ (Table~\ref{tab:mw_cloud_lco_comparison}). For the simulated galaxies, we scale the abundances of all elements in \cloudy\ using only the total metal mass fraction of each gas cell, and we do not use the individual carbon and oxygen abundances that \fireTwo\ tracks. This could matter a great deal for \LcoOneZero\ predictions, whose formation depends on the C and O abundances directly rather than on the total metal content. Gas with the same total metallicity but a different C/H or O/H ratio may therefore produce a different CO abundance and luminosity than we predict. Such differences may be systematic with the star formation history, assumed initial mass function and the redshift of the galaxy. We have not tested this effect. Doing so would require extending the \cloudy\ table with separate carbon and oxygen abundances, and we leave it for future work.

Our treatment of line radiative transfer is local for every gas cell. \cloudy\ accounts for the optical depth of the CO lines within each slab, including the line broadening from our `turbulent' velocities (see Equation~\ref{eq:turbulence}). However, we then sum the luminosities of each gas cell over the galaxy without accounting for line transfer between gas cells. Our calculation, therefore, does not include absorption of CO photons along the line of sight or the dependence of the observed emission on viewing angle. A full three-dimensional line radiative transfer calculation (e.g., with \texttt{RADMC-3D}) for a representative subsample of our galaxies would test this caveat directly, yet it is beyond the scope of this work.

Nevertheless, the MW cloud observations \citep{Miville-Deschenes_2017_MW_paper} suggests that the missing inter-cloud line transfer is a small effect for CO(1-0) emission. For self-absorption along the line of sight two clouds must overlap both on the sky and in velocity. \citet{Miville-Deschenes_2017_MW_paper}, the catalog we used to benchmark our \cloudy\ predictions in Section~\ref{sec:cloudy_runs_for_mw_molecular_clouds}, report the number of clouds along the line of sight across the Galactic plane (their Figure~5). They found that this distribution peaks at 1, while there is a tail reaching to six or even more clouds per line of sight. But even these extreme sight lines do not lead to significant self-absorption because they are well separated in velocity space as shown by their Figure~3. Because CO emission from distinct clouds along a line of sight is generally separated in velocity, summing the luminosities of individual elements is a reasonable approximation for MW like galaxies.

Finally, all of the galaxies used here come from \fireTwo\ simulations, so our results carry the biases that \fireTwo\ star formation and feedback prescriptions may introduce, and we have no way testing this from within our own sample. The resolution of the baryonic particles at least does not appear to be limiting: the mean molecular mass ($\rm M_{mol}$) in the \core\ \fireTwo\ and \tripleLatte\ \fireTwo\ galaxies agrees with the lower-resolution \firebox\ galaxies within one standard deviation at all SFRs.

\section{Conclusion \label{sec:conclusion}}

We have presented a pipeline for computing CO and nebular lines from \fireTwo\ cosmological simulations and tested the molecular gas content predicted by the pipeline using observations. The pipeline employs the three-dimensional Monte Carlo dust radiative transfer code \skirt, which gives the radiation field reaching each gas cell, followed by the photoionization code \cloudy. The latter returns line emission and species abundance, given a density, metallicity, turbulence, Habing radiation field strength and gas cell size. Rather than running \cloudy\ on individual gas cells, we tabulate the emission per unit mass on a grid of 109,296 \cloudy\ models and interpolate onto the simulated gas, obtaining CO(1-0), CO(2-1), $\mathrm{H\alpha}$, $\mathrm{H\beta}$, $\mathrm{[OII]}$, $\mathrm{[OIII]}$ and $\mathrm{[NII]}$ luminosities together with $\rm H_2$ and CO masses for every galaxy in our sample.

Benchmarking against the \citet{Miville-Deschenes_2017_MW_paper} catalog of galactic molecular clouds in Section~\ref{sec:cloudy_runs_for_mw_molecular_clouds}, we recover the total MW CO luminosity to within a factor of $\sim 3$ (the predicted total is about a third of the observed value) using the observed Galactic metallicity gradient \citep{rudolph_2006_MW_metallicity_profile}. We show that the onset of CO luminosity is when the individual clouds reach $\rm A_v\approx1$, where the CO molecules can be shielded from the surrounding radiation field efficiently.

Next we turn to the $\rm H_2$ and $\rm CO$ abundances. We predicted $\rm N_{CO}/N_{H_2}$ versus $\rm N_{H_2}$ our \fireTwo\ galaxies and compared them with Galactic sight lines in Section~\ref{section:Nco_Nh2_comparison}. We find that the predictions match observations below $\rm \log(N_{H_2}/cm^{-2})\approx20.5$, and recover the fairly sharp upturn at $\rm N_{H_2}\approx3-5\times10^{20}\,cm^{-2}$ for \core\ MW-mass \fireTwo\ galaxies. 

Finally in Section~\ref{sec:comparsion_mh2_mco_xco_with_obs} we compared the molecular gas mass ($\rm M_{mol}$) predictions against local xCOLD GASS \citep{Saintonge_2017} observations and high-redshift \texttt{PHIBBS2} \citep{Freundlich_2019} and \texttt{ASPECS} \citep{boogaard_2019_alma_aspecs} observations. At $z=0$ our molecular gas masses agree with xCOLD GASS within the observed scatter for $\mathrm{SFR}\gtrsim0.3\,M_\odot\,\yr^{-1}$. Below this value, the galaxies fall well below the detections. However, since the survey is complete only above $\LcoOneZero\sim10^{7.5}\,\obsunits$ we regard this regime as untested by this data set. At higher redshifts, our predictions show an evolution in $\rm M_{mol}$ that is not apparent in the observations. This difference may be related to systematic uncertainties in the inferred CO luminosities from assumed spectral line energy distribution and $\rm CO$-to-$\rm H_2$ conversion factors, although selection effects may also contribute. In \citetalias{Paper2}, we compare to other observations suggesting that selection effect may contribute. 

The main caveats in our approach are the following. First, we treat each gas cell as a one-dimensional, constant-density, plane-parallel slab, and we do not transport line photons between cells. Second, we adopt a fixed dust-to-metal ratio and a single grain mixture at all redshifts. Third, we pass only the FUV normalization of the \skirt\ radiation field to \cloudy, which assumes a fixed Milky Way spectral shape. Fourth, we scale all elemental abundances using only the total metal mass fraction of each gas cell, relative to an adopted $Z_\odot=0.0134$, rather than using the carbon and oxygen abundances tracked by \fireTwo. This last assumption matters because our predictions are very sensitive to metallicity. Finally, all of our galaxies come from a single galaxy formation model. \citetalias{Paper2} and \citetalias{Paper3} apply this pipeline to predict CO luminosities and scaling relations across cosmic time and to build the response functions needed for CO line intensity mapping forecasts.

\vspace{5mm}

\begin{acknowledgments}

This research was supported in Canada by NSERC and CIFAR. AW received support via NSF CAREER award AST-2045928. The \skirt\ and \cloudy\ calculations used to post-process the \fireTwo\ simulations were carried out on the Niagara supercomputer and its successor, Trillium, at the SciNet HPC Consortium, on the Fir supercomputer at the Cedar Supercomputing Centre at Simon Fraser University, operated through the Digital Research Alliance of Canada, and on the Sunnyvale computing cluster at CITA. We parallelized the runs across many nodes using GNU Parallel \citep{GNUparallel} while running on Trillium. SciNet receives funding from the Canada Foundation for Innovation under the auspices of the Digital Research Alliance of Canada, the Government of Ontario, the Ontario Research Fund--Research Excellence, and the University of Toronto. We are grateful to John Dubinski, MJ Huang, Jay Li and Daniel Xiao for their tireless work maintaining these resources. We also thank the Digital Research Alliance of Canada team, and the staff of the Cedar Supercomputing Centre at Simon Fraser University, for their support and for maintaining the Fir and Trillium supercomputers. We used snapshots of \fireTwo\ simulations that were run on the following resources: XSEDE, supported by NSF grant ACI-1548562; Blue Waters, supported by the NSF; Frontera allocations AST21010 and AST20016, supported by the NSF and the Texas Advanced Computing Center; and Pleiades, via the NASA HEC program through the NAS Division at Ames Research Center.

During the preparation of this work, the first author (D.T.) used the large language model (LLM) assistants Claude (Anthropic), ChatGPT (OpenAI), and Gemini (Google). These tools were used to help draft and restructure portions of the manuscript, to correct grammar and improve clarity of the text, and to assist in writing and debugging the Python code used for the analysis and figures. All text, code, and analysis produced with the help of these tools were reviewed, tested and verified by the authors. No AI tools were used to generate the simulation data. The scientific results, interpretations, and conclusions are those of the authors, who take full responsibility for the content of this paper. 
\end{acknowledgments}

\begin{contribution}
DT led the project, performed all of the coding, analysis, and wrote the manuscript. NM and RB supervised DT. NM provided edits to an early draft, which DT then revised. DC contributed substantially to the coding and analysis. NC produced the first three panels of Figure~\ref{fig:peakpatch_firebox_zoom_stacked_6panel}. All authors provided comments on the manuscript.
\end{contribution}


\software{astropy \citep{2013A&A...558A..33A,2018AJ....156..123A},
SKIRT \citep{Baes_2011, Baes_2015, Camps_2015, Camps_and_baes_2020_skirt9},
          Cloudy \citep{2013RMxAA..49..137F, Ferland_2017, cloudy_version_c25_release},
          SciPy \citep{scipy_2020},
          Meshoid (\url{https://github.com/mikegrudic/meshoid}),
          GNU Parallel \citep{GNUparallel},
          GIZMO \citep{Hopkins_2015},
          ROCKSTAR \citep{Behroozi_2013},
          AHF \citep{Gill_2004, Knollmann_2009},
          MUSIC \citep{Hahn_2011},
          STARBURST99 \citep{Leitherer_1999}}
          
\appendix

\twocolumngrid

\section{Problematic \cloudy\ runs}
\label{sec:problematic_cloudy_runs}

\begin{figure}[h!]
    \centering
    \includegraphics[width=0.99\linewidth]{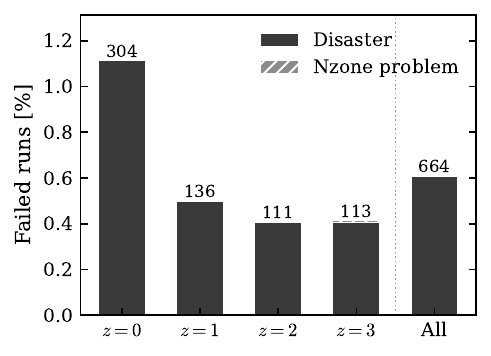}
    \caption{Fraction of failed \cloudy\ runs at each redshift ($z = 0, 1, 2, 3$) and for all redshifts combined (rightmost bar). Each redshift uses the same grid of $27{,}324$ parameter combinations after the column-density cut $n_{\rm H} R < 10^{5.6}\,{\rm cm^{-3}\,pc}$. A run counts as failed if it ended with a DISASTER error (dark grey) or stopped at the default maximum number of zones (hatched grey). The number above each bar is the number of failed runs. Overall, $664$ of $109{,}296$ runs ($0.61$ per cent) failed. Only two runs, both at $z = 3$, hit the zone limit, so that category is too small to see.}
    \label{fig:cloudy_failure_rate_by_redshift}
\end{figure}

\begin{figure}[h!]
    \centering
    \includegraphics[width=0.99\linewidth]{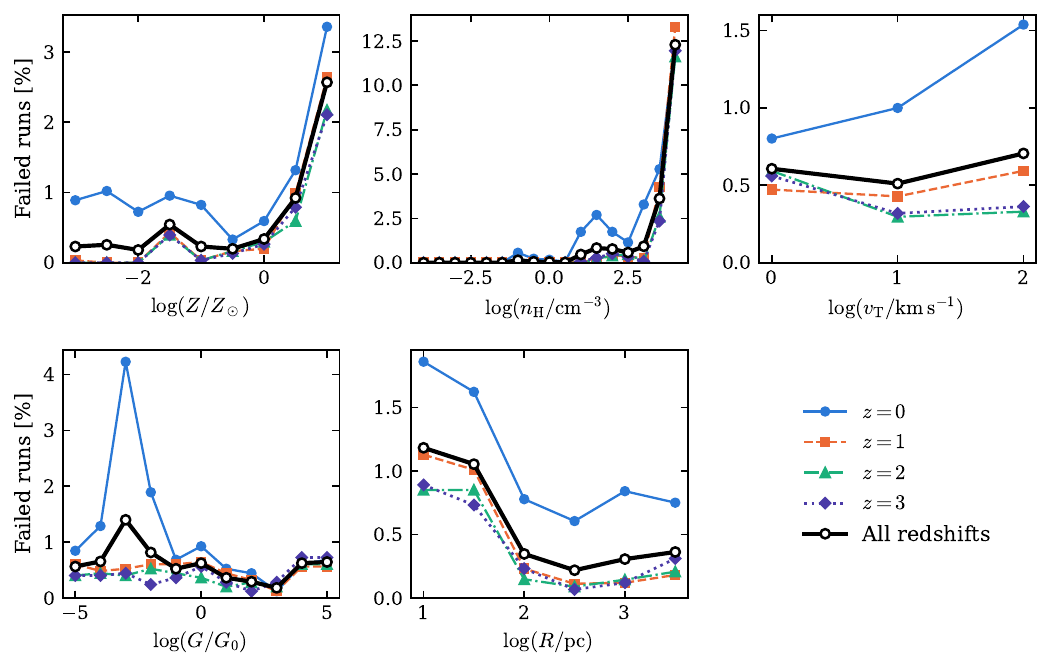}
    \caption{Failure rate of the \cloudy\ runs as a function of each grid parameter: metallicity ($Z$), hydrogen density ($n_{\rm H}$), turbulent velocity ($v_{\rm T}$), ISRF strength ($G$) and stopping length ($R$). At each parameter value, the failure rate is the number of failed runs divided by the total number of runs at that value, summed over the other four parameters. Coloured lines show the individual redshifts, $z = 0$ (blue circles), $z = 1$ (orange squares), $z = 2$ (green triangles) and $z = 3$ (purple diamonds). The thick black line with open circles shows all redshifts combined. Note that the vertical axis range differs between panels. Failures are concentrated at the highest densities ($\log (n_{\rm H}/\rm cm^{-3}) \geq 3.5$), super-solar metallicities ($\log (Z/Z_{\odot}) \geq 0$) and small $R$ ($\log (R/\rm {pc}) \leq 2$).}
    \label{fig:cloudy_failure_rate_vs_parameters}
\end{figure}

\begin{figure}[h!]
    \centering
    \includegraphics[width=0.99\linewidth]{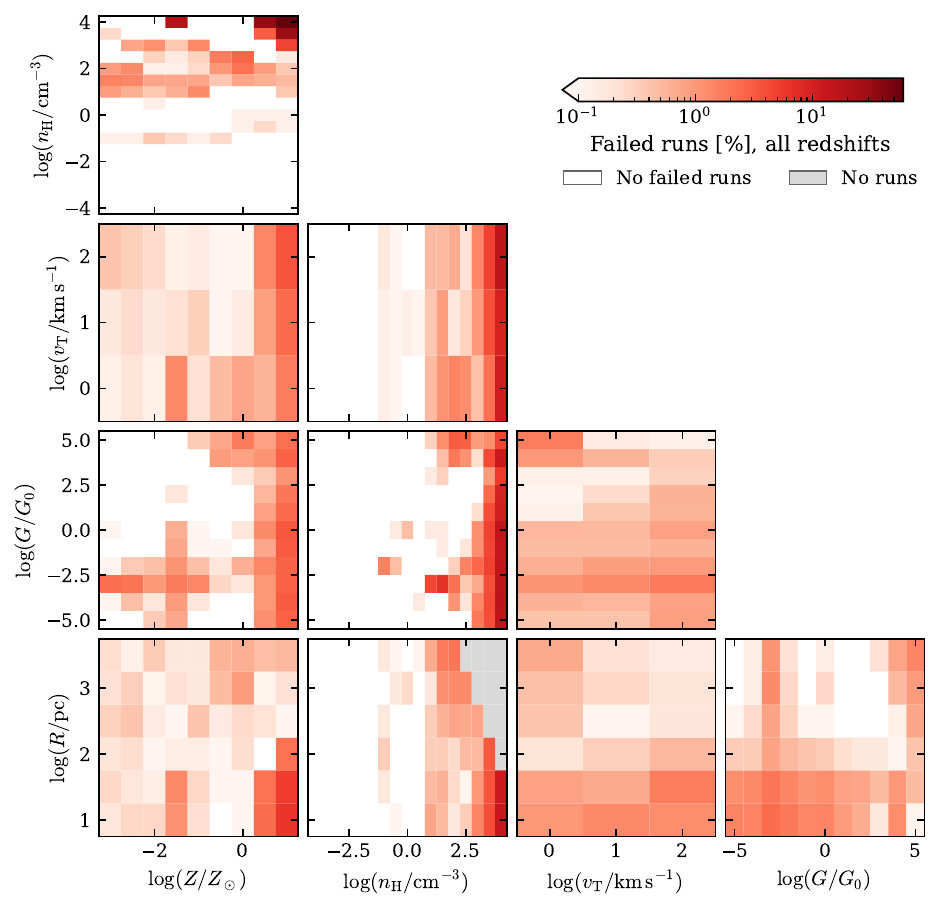}
    \caption{Failure rate of the \cloudy\ runs in every two-parameter plane of the grid, combining all redshifts. Each cell gives the fraction of failed runs at that pair of values, summed over the remaining three parameters. The colour scale is logarithmic, and cells with a failure rate below 0.1 per cent use the lightest colour. White cells contain no failed runs. Grey cells contain no runs because they are removed by the column-density cut $n_{\rm H} R < 10^{5.6}\,{\rm cm^{-3}\,pc}$. The failures concentrate at high density combined with super-solar metallicity.}
    \label{fig:cloudy_failure_rate_parameter_pairs_all_redshifts}
\end{figure}

To build the line-flux tables that we interpolate over to assign line fluxes to every gas cell in our galaxies, we ran $27{,}324$ \cloudy\ simulations at each of four redshifts, $z = 0, 1, 2$ and $3$, for a total of $109{,}296$ runs. A small fraction of these simulations did not complete successfully. The two errors we encountered were the \texttt{DISASTER} error and the \texttt{NZONE} problem.

A \texttt{DISASTER} error occurs when \cloudy\ meets a problem that prevents it from continuing reliably, and the code terminates. One example is the gas kinetic temperature falling below the lowest temperature the code allows. Another is a strong population inversion (maser) in a line, which drives that line's optical depth to large negative values.

\cloudy\ divides the illuminated slab into thin zones and iteratively solves for the molecular, atomic and ionic abundances in each zone. An \texttt{NZONE} problem occurs when a simulation needs more zones than the limit set by the user. The default limit is 1400 zones. Raising it increases the memory the simulation requires.

The number of zones a simulation needs grows with its column density. We first ran every simulation with the default limit of 1400 zones, then reran the high-column-density simulations that hit an \texttt{NZONE} problem with a larger limit. After these reruns, only two simulations still stopped at the zone limit. Both are at $z = 3$ and lie at the highest column density in our grid, $n_{\rm H} R = 10^{5.5}\,{\rm cm^{-3}\,pc}$. Very few gas cells have such properties, so rerunning these two with an even larger zone limit would not change our results, and we did not do so.

In total, $664$ of the $109{,}296$ simulations ($0.61$ per cent) failed: $662$ with a \texttt{DISASTER} error and two with an \texttt{NZONE} problem. The failure rate is highest at $z = 0$ ($1.11$ per cent) and falls to $0.4$--$0.5$ per cent at $z = 1$--$3$ (Fig.~\ref{fig:cloudy_failure_rate_by_redshift}). Figs.~\ref{fig:cloudy_failure_rate_vs_parameters} and \ref{fig:cloudy_failure_rate_parameter_pairs_all_redshifts} show how the failures are distributed across the simulation parameters. Most of them occur at the highest densities ($\log n_{\rm H} \geq 3.5$), especially at super-solar metallicity.

\section{Observational Data \label{sec:observational_data}}

In Section~\ref{sec:cloudy_runs_for_mw_molecular_clouds}, Section~\ref{section:Nco_Nh2_comparison}, Section~\ref{sec:comparsion_mh2_mco_xco_with_obs} and Appendix~\ref{section:Nco_Nh2_comparison_firebox}, we compare our simulation results with several observational studies. Specifically, we compare our predictions for the molecular gas mass as a function of galactic star formation rate (SFR), stellar mass, and gas-phase metallicity and CO column ($\rm N_{CO}$) as a function of $\rm N_{H_2}$. In \citetalias{Paper2}, we compare the predicted CO luminosities of our simulated \fireTwo\ galaxies with the same observational datasets. In this section we describe the observational data sets we will use.

\subsection{The Milky Way molecular cloud catalog of
\citet{Miville-Deschenes_2017_MW_paper} \label{sec:appendix_mw_cloud_catalog}}

The benchmark in Section~\ref{sec:cloudy_runs_for_mw_molecular_clouds} uses the Galactic molecular cloud catalog of \citet{Miville-Deschenes_2017_MW_paper}. The catalog is built from the \citet{Dame2001_COdata_for_MivilleDeschenes} composite $^{12}$CO(1-0) survey, obtained with the two CfA $1.2\,$meter telescopes, which covers the full Galactic plane within $|b|<5^{\circ}$ at an angular resolution of $\approx8\farcm5$ and a velocity channel width of $1.3\,\kms$. 

Clouds are identified in two steps. Each CO spectrum is first decomposed into a sum of at most twelve Gaussian components, yielding $\approx5.4\times10^{5}$ components over the survey. In the second step, the Gaussian components are grouped into clouds by a hierarchical clustering with a descending threshold, applied to their integrated emission $W_{\rm CO}(l,b,v)$ in position-position-velocity space. The resulting catalog contains 8107 clouds and accounts for $98\%$ of the CO emission in the surveyed strip. Because the clustering is performed on the Gaussian components rather than on the brightness temperature directly, the catalog includes faint clouds that a brightness threshold would miss. The method does not solve the velocity-crowding problem, e.g. structures that are unrelated in three dimensions but overlap in position-position-velocity space can be assigned to a single cloud.

Distances are kinematic, computed using the \citet{BrandBlitz1993_velocityFields_MivilleDeschenesData} rotation curve with $R_0=8.5\,\kpc$ and $V_0=220\,\kms$. In the inner Galaxy, where each velocity admits a near and a far solution, the ambiguity is resolved using the observed correlation between the velocity dispersion and $\Sigma R$ (the catalog selects the distance that brings the cloud closest to $\sigma_v\propto(\Sigma R)^{0.43}$). The catalog reports both solutions together with a flag indicating the more likely one. In Section~\ref{sec:cloudy_runs_for_mw_molecular_clouds} we adopt the solution favored by this flag for every cloud. Distance errors propagate into the physical size, area, and mass of each cloud ($R\propto D$, $S\propto D^2$, $M\propto D^2$), and are the dominant uncertainty on those quantities.
 
For each cloud the catalog reports the integrated CO emission $W_{\rm CO}$ summed over the $N_{\rm pix}$ pixels the cloud covers, the mean $\rm H_2$ column density, the surface density $\Sigma$, the velocity dispersion $\sigma_v$, the angular and physical radii, the physical area $S$, the Galactocentric radius, and the mass. We use the $\rm H_2$ column density, velocity dispersion, and size as inputs to the \cloudy\ runs described in Section~\ref{sec:cloudy_runs_for_mw_molecular_clouds}, while we assume three different metallicity profiles and constant MW radiation field ($\rm G_{FUV}=1$). 

The clouds span a wide range of properties. The most probable radius is $\approx30\,\pc$, surface densities run from $2$ to $300\,M_\odot\,\pc^{-2}$ with a median of $16.5\,M_\odot\,\pc^{-2}$, the median mass is $3.8\times10^{4}\,M_\odot$, and the median velocity dispersion is $3.6\,\kms$. The total $\rm H_2$ mass in the catalog is $1.2\times10^{9}\,M_\odot$. These values describe the full catalog; in Section~\ref{sec:cloudy_runs_for_mw_molecular_clouds} we use only the clouds within $15\,\kpc$ of the Galactic center.

\subsection{Milky Way diffuse molecular clouds for CO column density observations \label{sec:MW_Nco_Nh2_data}}

We compare our predictions for the CO abundance relative to molecular hydrogen with observations of diffuse molecular clouds in the Milky Way. These observations provide a complementary test of the molecular chemistry of our predictions at the cloud scale, where both CO and H$_2$ column densities can be measured directly through ultraviolet (UV) absorption spectroscopy.

We use the CO and H$_2$ column-density measurements compiled by \citet{Rachford_2002_Nco_Nh2}, \citet{Crenny_Federman_2004_Nco_Nh2}, \citet{Sheffer_2008_Nco_Nh2}, and \citet{Burgh_2010_Nco_Nh2} from galactic sight lines. \citet{Rachford_2002_Nco_Nh2} derived H$_2$ column densities from FUSE far-ultraviolet absorption measurements toward 23 Galactic sources selected to probe clouds with $A_V \gtrsim 1$, the type of cloud  expected to contain $\gtrsim99\%$ of the total H$_2$ in the Milky Way. \citet{Crenny_Federman_2004_Nco_Nh2} reanalyzed archival Copernicus observations of CO absorption toward Galactic stars, focusing on sight lines containing modest amounts of molecular material and relatively low CO column densities. These measurements therefore provide a useful constraint on the CO abundance in the low-column-density, diffuse molecular regime.

The more extensive survey of \citet{Sheffer_2008_Nco_Nh2} combines CO measurements from HST with H$_2$ measurements from FUSE and includes 114 Galactic sight lines with UV data. Their results demonstrate that the relation between $\rm N(CO)$ and $\rm N(H_2)$ is not described by a single power law. Instead, the relation steepens at approximately $\rm \log(N(H_2)/\mathrm{cm}^{-2})\approx20.4$, corresponding to $\rm \log(N(CO)/cm^{-2})\approx14.1$. This change is associated with the transition from lower-density diffuse gas to denser molecular material and with the increasing importance of CO shielding. \citet{Burgh_2010_Nco_Nh2} further investigated 34 Galactic sight lines using high-resolution Hubble Space Telescope (HST)/Space Telescope Imaging Spectrograph (STIS) measurements of neutral carbon together with CO and H$_2$ column densities. They found that the ratio $\rm N(CO)/N(H_2)$ increases with molecular fraction, with values below $\sim10^{-6}$ characteristic of predominantly diffuse molecular material and larger values indicating progressively more shielded, translucent material.

We compare our predicted $\rm N(CO)/N(H_2)$ ratios with these Milky Way measurements to test whether our models reproduce the transition from CO-poor diffuse molecular gas to more CO-rich, shielded molecular material. This comparison is particularly useful for assessing the dependence of CO abundance on molecular column density and shielding, which is not directly constrained by galaxy-integrated CO observations.

\subsection{Low-redshift surveys \label{sec:low_z_observational_data}}

At low redshift, we compare our predicted molecular gas masses with the xCOLD GASS survey \citep{Saintonge_2017}. xCOLD GASS is one of the most comprehensive surveys of molecular gas in nearby galaxies, providing CO(1--0) observations for 532 stellar-mass-selected galaxies over the redshift range $0.01 < z < 0.05$. The sample spans a broad range of galaxy properties, with stellar masses of $9 < \log(M_\star/M_\odot) < 11.3$ and star formation rates ranging from approximately $0$ to $35\,M_\odot\,\mathrm{yr}^{-1}$. The authors state that their reported molecular gas masses are inferred from the observed CO(1-0) luminosities using a metallicity dependent conversion factor ($\alpha_{\rm CO}$, \cite{Accurso_2017}). Their inferred molecular gas mass contains Helium contribution. We use the molecular gas masses, SFRs, and stellar masses reported by \citet{Saintonge_2017} to compare with our model predictions at $z\approx0$.

\subsection{Intermediate- and high-redshift surveys \label{sec:high_z_observational_data}}

To test our predictions at higher redshifts, we use the PHIBSS2 survey \citep{Freundlich_2019} and the ASPECS survey \citep{boogaard_2019_alma_aspecs}.

The PHIBSS2 survey provides CO(2-1) observations for galaxies at $0.5 < z < 0.8$, observed with the IRAM Plateau de Bure Interferometer. The sample includes galaxies spanning a broad range of stellar masses, $\rm 10 < \log(M_\star/M_\odot) < 11.8$, and star formation rates of approximately $3.5$-$\rm 500\,M_\odot\,\mathrm{yr}^{-1}$. Molecular gas masses are inferred from the observed CO luminosities by assuming constant CO(2-1)/CO(1-0) ratio ($r21=0.77$) and using the metallicity dependent conversion factor as modeled in \cite{Tacconi_2018}. Their reported molecular gas masses include helium contribution which is about $36\%$.

At higher redshifts, we use the ASPECS survey \citep{boogaard_2019_alma_aspecs}, which provides CO detections for galaxies in the Hubble Ultra Deep Field over $1 \lesssim z \lesssim 3.6$. The sample contains galaxies detected in several CO transitions, including CO(2-1) emitters at $1 < z < 2$, CO(3-2) emitters at $2 < z < 3$, and a CO(5-4) emitter at $z\approx3.6$. The detected galaxies span SFRs of approximately $8$--$\rm 318\,M_\odot\,\mathrm{yr}^{-1}$ and stellar masses of $\rm9.7 < \log(M_\star/M_\odot) < 11.5$. They derived the molecular gas masses of the observed galaxies by assuming excitation ratios using \cite{Daddi_2015_COexcitationLadder} and MW mass to luminosity ratio ($\alpha_{CO}=3.6\,M_\odot(\obsunits)^{-1}$). Their molecular gas measurements include the Helium contribution.

\citet{boogaard_2019_alma_aspecs} report the depth of the survey as a $5\sigma$ limiting molecular gas mass rather than as a limiting CO luminosity: $M_{\rm H_2} \geq 10^{9.7}\,M_\odot$ at $1.0<z<1.7$ for a line FWHM of $100\,{\rm km\,s^{-1}}$, and $M_{\rm H_2} \geq 10^{10.3}\,M_\odot$ at $2.0<z<3.1$ for $300\,{\rm km\,s^{-1}}$. 


\section{Comparison of $N_{\rm CO}/N_{\rm H_2}$ versus $N_{\rm H_2}$ with Diffuse Milky Way Observations For \firebox\ Galaxies \label{section:Nco_Nh2_comparison_firebox}}

\begin{figure}[h]
    \centering
    \includegraphics[width=1\linewidth]{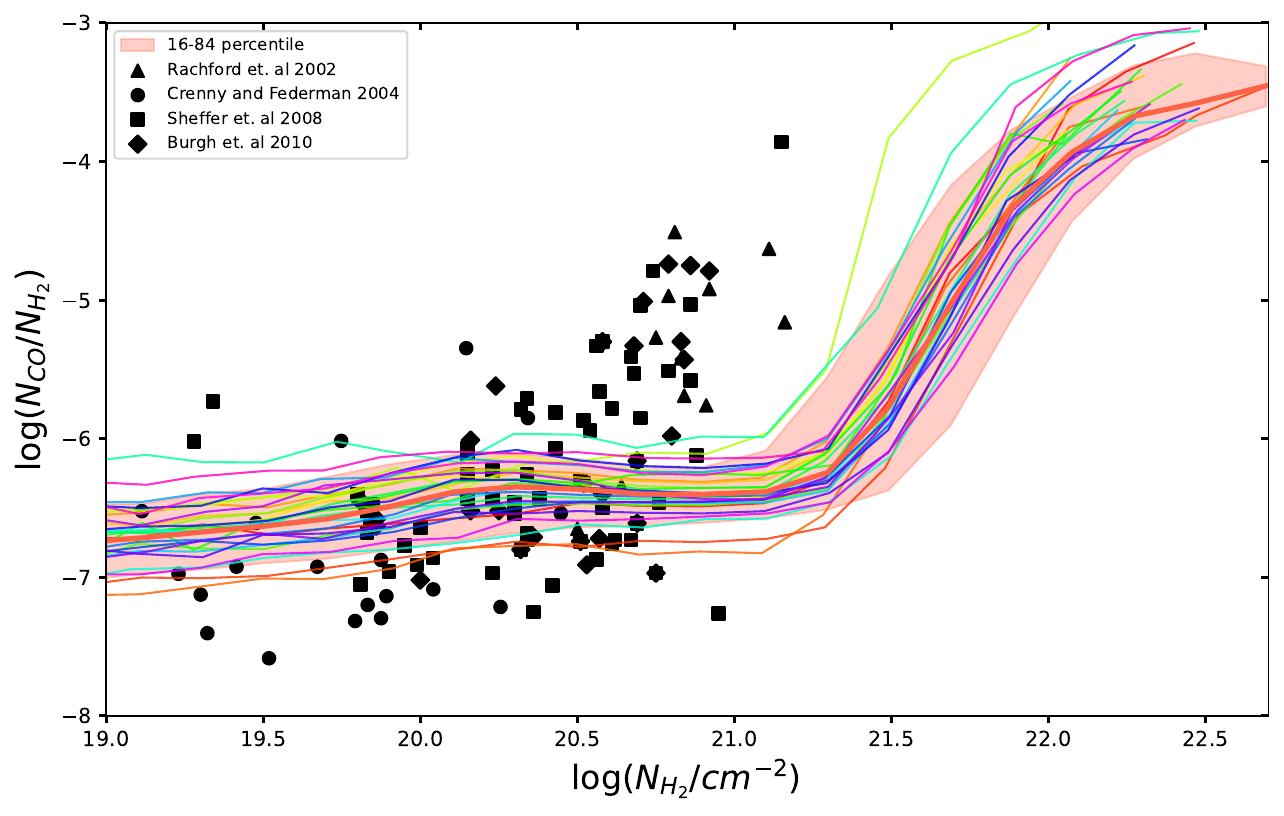}
    \caption{$\rm N_{\rm CO}/N_{\rm H_2}$ as a function of $\rm N_{\rm H_2}$ for Milky Way mass galaxies in the \firebox\ cosmological volume at $z=0$, compared with observations of diffuse Milky Way clouds. The predictions are for the 29 \firebox\ galaxies with stellar masses of $\rm 4\times10^{10} < M_\star/M_\odot < 8\times10^{10}$. Column densities are computed from a 100 pc pixelized face-on map of each galaxy (galactocentric radius $r<5$ kpc and $d_\perp<500$ pc) and binned in 50 logarithmic $\rm N_{\rm H_2}$ intervals of 0.2 dex over $\rm 15 < \log(N_{\rm H_2}/\mathrm{cm^{-2}}) < 25$. The thin colored lines show the median relation of each individual galaxy. The thick red line shows the median relation of all the galaxies combined, and the shaded region shows the 16--84 percentile range. Black symbols show observational measurements from \citet{Rachford_2002_Nco_Nh2} (triangles), \citet{Crenny_Federman_2004_Nco_Nh2} (circles), \citet{Sheffer_2008_Nco_Nh2} (squares), and \citet{Burgh_2010_Nco_Nh2} (diamonds). \label{fig:Nco_Nh2_fireboxOnly}}
\end{figure}

\begin{figure}[h]
    \centering
    \includegraphics[width=1\linewidth]{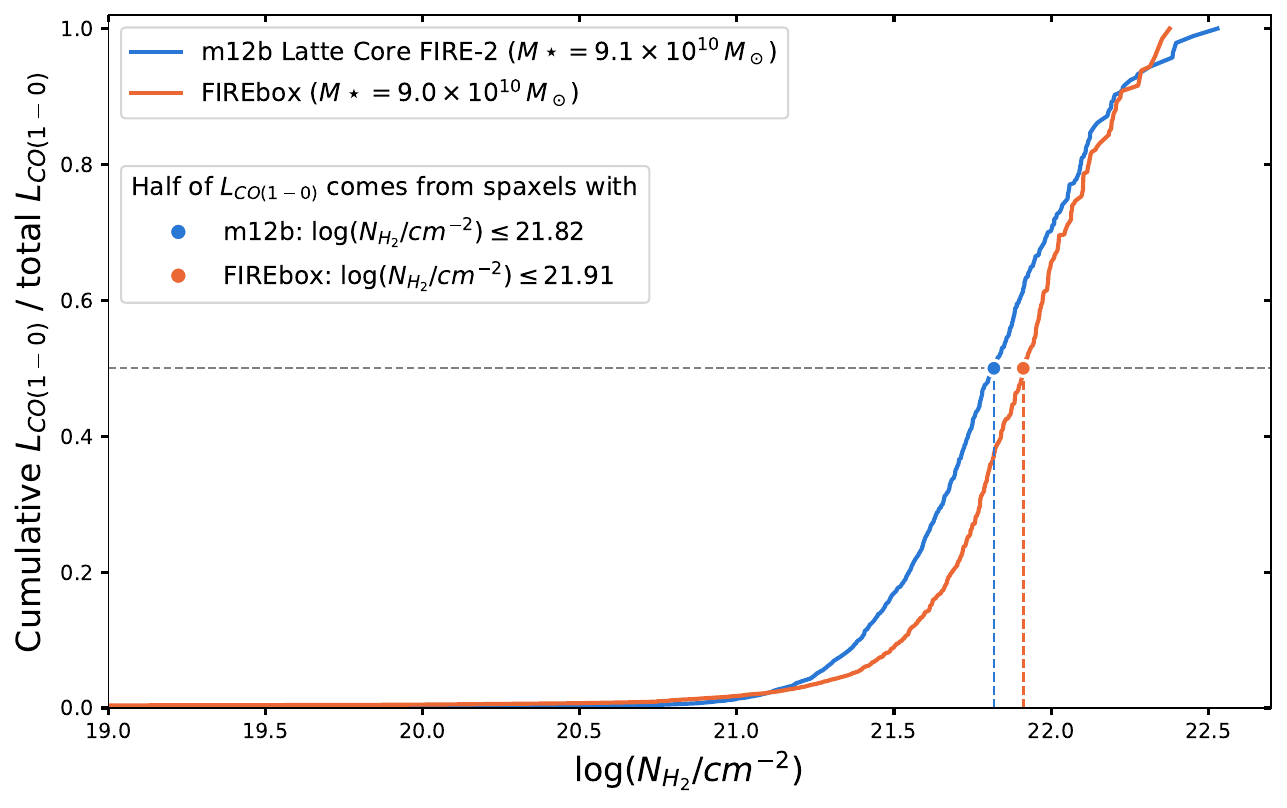}
    \caption{Cumulative \LcoOneZero\ as a function of H$_2$ column density, normalized to each galaxy's total CO luminosity. Gas particles in the disk ($r < 5$~kpc, $d_\perp < 500$~pc) are projected face-on onto 100~pc spaxels. The spaxels are sorted from low to high $N_{\mathrm{H_2}}$, and the curve shows the fraction of the total \LcoOneZero\ contributed by spaxels at or below a given $N_{\mathrm{H_2}}$. The blue curve is the m12b \core\ \fireTwo\ galaxy ($M_\star = 9.1\times10^{10}\,M_\odot$) and the orange curve is the \firebox\ galaxy with the closest stellar mass ($M_\star = 9.0\times10^{10}\,M_\odot$). The dashed lines and markers show where each galaxy reaches half of its total CO luminosity. Half of the CO luminosity comes from spaxels with $\log(N_{\mathrm{H_2}}/\mathrm{cm}^{-2}) \lesssim 21.8$ in m12b and $\lesssim 21.9$ in \firebox, a difference of only $\sim$0.1~dex. Most of the CO luminosity in both galaxies comes from spaxels with $\log(N_{\mathrm{H_2}}/\mathrm{cm}^{-2}) \gtrsim 21.8$. \label{fig:cumulative_Lco10_vs_NH2}}
\end{figure}

In Section~\ref{section:Nco_Nh2_comparison} we show the predicted CO-to-H$_2$ column density ratio, $\rm N_{CO}/N_{H_2}$, as a function of $\rm N_{H_2}$ for the MW-mass \core\ \fireTwo\ galaxies, which have a mass resolution of $7100\,\rm M_\odot$. Figure~\ref{fig:Nco_Nh2_fireboxOnly} shows the same comparison for \firebox\ galaxies, which have a mass resolution of $6.26\times10^{4}\,\rm M_\odot$. For \firebox\ we use only galaxies with stellar masses $\rm 4\times10^{10} < M_\star/M_\odot < 8\times10^{10}$. As before, the column densities are computed by projecting each galaxy face-on onto 100~pc spaxels. The \firebox\ galaxies show a sharp upturn in $\rm N_{\rm CO}/N_{\rm H_2}$ at $\rm \log(N_{\rm H_2}/{cm^{-2}})\approx21.5$, whereas the \core\ \fireTwo\ galaxies show it at $\rm \log(N_{H_2}/{cm^{-2}})\approx20.8$ (see Figure~\ref{fig:Nco_Nh2_m12CoreOnly}). Consequently, at $\rm N_{ H_2}\lesssim10^{22}\,{\rm cm^{-2}}$ the \firebox\ galaxies systematically predict lower $\rm N_{\rm CO}$ than the higher-resolution \core\ \fireTwo\ galaxies at the same $\rm N_{\rm H_2}$. One would therefore naively expect the \firebox\ galaxies to produce systematically lower \LcoOneZero. However, galaxies from the two suites with similar stellar mass and star formation rate produce similar \LcoOneZero\ as we show in \citetalias{Paper2}. This raises the question of why the two suites produce similar \LcoOneZero\ despite these differences.

To address this, Figure~\ref{fig:cumulative_Lco10_vs_NH2} shows the cumulative \LcoOneZero, normalized to the total, as a function of $\rm N_{\rm H_2}$ for one \core\ \fireTwo\ galaxy (m12b, $M_\star = 9.1\times10^{10}\,\rm M_\odot$) and for the \firebox\ galaxy with the closest stellar mass ($M_\star = 9.0\times10^{10}\,\rm M_\odot$). In both galaxies the \LcoOneZero\ is dominated by high column density spaxels. Half of it is produced in spaxels with $\rm \log(N_{H_2}/cm^{-2})\gtrsim21.8$, and the two cumulative curves closely follow each other above $\rm \log(N_{H_2}/cm^{-2})\approx21$. At these column densities the two suites predict $\rm N_{\rm CO}/N_{\rm H_2}$ values that agree within a factor of $\approx3$. The differences in $\rm N_{\rm CO}/N_{\rm H_2}$ between the suites occur at lower column densities, where spaxels contribute only a small fraction of the total \LcoOneZero. Because the CO luminosity is set by spaxels at column densities where the two suites predict similar values of $\rm N_{\rm CO}$ and the total \LcoOneZero\ is insensitive to the differences at lower $N_{\rm H_2}$, and the two suites therefore produce similar \LcoOneZero.

\clearpage
\newpage

\bibliography{bibliography}{}

@ARTICLE{2011MNRAS.412..337G,
       author = {{Glover}, S.~C.~O. and {Mac Low}, M.-M.},
        title = "{On the relationship between molecular hydrogen and carbon monoxide abundances in molecular clouds}",
      journal = {\mnras},
         year = 2011,
        month = mar,
       volume = {412},
       number = {1},
        pages = {337-350},
          doi = {10.1111/j.1365-2966.2010.17907.x},
archivePrefix = {arXiv},
       eprint = {1003.1340},
 primaryClass = {astro-ph.GA},
       adsurl = {https://ui.adsabs.harvard.edu/abs/2011MNRAS.412..337G}
}

@ARTICLE{1984ApJ...276..182S,
       author = {{Sanders}, D.~B. and {Solomon}, P.~M. and {Scoville}, N.~Z.},
        title = "{Giant molecular clouds in the Galaxy. I - The axisymmetric distribution of H2}",
      journal = {\apj},
         year = 1984,
        month = jan,
       volume = {276},
        pages = {182-203},
          doi = {10.1086/161602},
       adsurl = {https://ui.adsabs.harvard.edu/abs/1984ApJ...276..182S}
}

@ARTICLE{2019ApJ...872..126M,
       author = {{Moradinezhad Dizgah}, Azadeh and {Keating}, Garrett K.},
        title = "{Line Intensity Mapping with [C II] and CO(1-0) as Probes of Primordial Non-Gaussianity}",
      journal = {\apj},
         year = 2019,
        month = feb,
       volume = {872},
       number = {2},
          eid = {126},
        pages = {126},
          doi = {10.3847/1538-4357/aafd36},
archivePrefix = {arXiv},
       eprint = {1810.02850},
 primaryClass = {astro-ph.CO},
       adsurl = {https://ui.adsabs.harvard.edu/abs/2019ApJ...872..126M}
}

@ARTICLE{2013A&A...558A..33A,
       author = {{Astropy Collaboration} and {Robitaille}, Thomas P. and {Tollerud}, Erik J. and {Greenfield}, Perry and {Droettboom}, Michael and {Bray}, Erik and {Aldcroft}, Tom and {Davis}, Matt and {Ginsburg}, Adam and {Price-Whelan}, Adrian M. and {Kerzendorf}, Wolfgang E. and {Conley}, Alexander and {Crighton}, Neil and {Barbary}, Kyle and {Muna}, Demitri and {Ferguson}, Henry and {Grollier}, Fr{\'e}d{\'e}ric and {Parikh}, Madhura M. and {Nair}, Prasanth H. and {Unther}, Hans M. and {Deil}, Christoph and {Woillez}, Julien and {Conseil}, Simon and {Kramer}, Roban and {Turner}, James E.~H. and {Singer}, Leo and {Fox}, Ryan and {Weaver}, Benjamin A. and {Zabalza}, Victor and {Edwards}, Zachary I. and {Azalee Bostroem}, K. and {Burke}, D.~J. and {Casey}, Andrew R. and {Crawford}, Steven M. and {Dencheva}, Nadia and {Ely}, Justin and {Jenness}, Tim and {Labrie}, Kathleen and {Lim}, Pey Lian and {Pierfederici}, Francesco and {Pontzen}, Andrew and {Ptak}, Andy and {Refsdal}, Brian and {Servillat}, Mathieu and {Streicher}, Ole},
        title = "{Astropy: A community Python package for astronomy}",
      journal = {\aap},
         year = 2013,
        month = oct,
       volume = {558},
          eid = {A33},
        pages = {A33},
          doi = {10.1051/0004-6361/201322068},
archivePrefix = {arXiv},
       eprint = {1307.6212},
 primaryClass = {astro-ph.IM},
       adsurl = {https://ui.adsabs.harvard.edu/abs/2013A&A...558A..33A}
}

@ARTICLE{2018AJ....156..123A,
       author = {{Astropy Collaboration} and {Price-Whelan}, A.~M. and {Sip{\H{o}}cz}, B.~M. and {G{\"u}nther}, H.~M. and {Lim}, P.~L. and {Crawford}, S.~M. and {Conseil}, S. and {Shupe}, D.~L. and {Craig}, M.~W. and {Dencheva}, N. and {Ginsburg}, A. and {VanderPlas}, J.~T. and {Bradley}, L.~D. and {P{\'e}rez-Su{\'a}rez}, D. and {de Val-Borro}, M. and {Aldcroft}, T.~L. and {Cruz}, K.~L. and {Robitaille}, T.~P. and {Tollerud}, E.~J. and {Ardelean}, C. and {Babej}, T. and {Bach}, Y.~P. and {Bachetti}, M. and {Bakanov}, A.~V. and {Bamford}, S.~P. and {Barentsen}, G. and {Barmby}, P. and {Baumbach}, A. and {Berry}, K.~L. and {Biscani}, F. and {Boquien}, M. and {Bostroem}, K.~A. and {Bouma}, L.~G. and {Brammer}, G.~B. and {Bray}, E.~M. and {Breytenbach}, H. and {Buddelmeijer}, H. and {Burke}, D.~J. and {Calderone}, G. and {Cano Rodr{\'\i}guez}, J.~L. and {Cara}, M. and {Cardoso}, J.~V.~M. and {Cheedella}, S. and {Copin}, Y. and {Corrales}, L. and {Crichton}, D. and {D'Avella}, D. and {Deil}, C. and {Depagne}, {\'E}. and {Dietrich}, J.~P. and {Donath}, A. and {Droettboom}, M. and {Earl}, N. and {Erben}, T. and {Fabbro}, S. and {Ferreira}, L.~A. and {Finethy}, T. and {Fox}, R.~T. and {Garrison}, L.~H. and {Gibbons}, S.~L.~J. and {Goldstein}, D.~A. and {Gommers}, R. and {Greco}, J.~P. and {Greenfield}, P. and {Groener}, A.~M. and {Grollier}, F. and {Hagen}, A. and {Hirst}, P. and {Homeier}, D. and {Horton}, A.~J. and {Hosseinzadeh}, G. and {Hu}, L. and {Hunkeler}, J.~S. and {Ivezi{\'c}}, {\v{Z}}. and {Jain}, A. and {Jenness}, T. and {Kanarek}, G. and {Kendrew}, S. and {Kern}, N.~S. and {Kerzendorf}, W.~E. and {Khvalko}, A. and {King}, J. and {Kirkby}, D. and {Kulkarni}, A.~M. and {Kumar}, A. and {Lee}, A. and {Lenz}, D. and {Littlefair}, S.~P. and {Ma}, Z. and {Macleod}, D.~M. and {Mastropietro}, M. and {McCully}, C. and {Montagnac}, S. and {Morris}, B.~M. and {Mueller}, M. and {Mumford}, S.~J. and {Muna}, D. and {Murphy}, N.~A. and {Nelson}, S. and {Nguyen}, G.~H. and {Ninan}, J.~P. and {N{\"o}the}, M. and {Ogaz}, S. and {Oh}, S. and {Parejko}, J.~K. and {Parley}, N. and {Pascual}, S. and {Patil}, R. and {Patil}, A.~A. and {Plunkett}, A.~L. and {Prochaska}, J.~X. and {Rastogi}, T. and {Reddy Janga}, V. and {Sabater}, J. and {Sakurikar}, P. and {Seifert}, M. and {Sherbert}, L.~E. and {Sherwood-Taylor}, H. and {Shih}, A.~Y. and {Sick}, J. and {Silbiger}, M.~T. and {Singanamalla}, S. and {Singer}, L.~P. and {Sladen}, P.~H. and {Sooley}, K.~A. and {Sornarajah}, S. and {Streicher}, O. and {Teuben}, P. and {Thomas}, S.~W. and {Tremblay}, G.~R. and {Turner}, J.~E.~H. and {Terr{\'o}n}, V. and {van Kerkwijk}, M.~H. and {de la Vega}, A. and {Watkins}, L.~L. and {Weaver}, B.~A. and {Whitmore}, J.~B. and {Woillez}, J. and {Zabalza}, V. and {Astropy Contributors}},
        title = "{The Astropy Project: Building an Open-science Project and Status of the v2.0 Core Package}",
      journal = {\aj},
         year = 2018,
        month = sep,
       volume = {156},
       number = {3},
          eid = {123},
        pages = {123},
          doi = {10.3847/1538-3881/aabc4f},
archivePrefix = {arXiv},
       eprint = {1801.02634},
 primaryClass = {astro-ph.IM},
       adsurl = {https://ui.adsabs.harvard.edu/abs/2018AJ....156..123A}
}

@ARTICLE{2013RMxAA..49..137F,
       author = {{Ferland}, G.~J. and {Porter}, R.~L. and {van Hoof}, P.~A.~M. and {Williams}, R.~J.~R. and {Abel}, N.~P. and {Lykins}, M.~L. and {Shaw}, G. and {Henney}, W.~J. and {Stancil}, P.~C.},
        title = "{The 2013 Release of Cloudy}",
      journal = {\rmxaa},
         year = 2013,
        month = apr,
       volume = {49},
        pages = {137-163},
          doi = {10.48550/arXiv.1302.4485},
archivePrefix = {arXiv},
       eprint = {1302.4485},
 primaryClass = {astro-ph.GA},
       adsurl = {https://ui.adsabs.harvard.edu/abs/2013RMxAA..49..137F}
}

@ARTICLE{1991ARA&A..29..581Y,
       author = {{Young}, J.~S. and {Scoville}, N.~Z.},
        title = "{Molecular gas in galaxies.}",
      journal = {\araa},
         year = 1991,
        month = jan,
       volume = {29},
        pages = {581-625},
          doi = {10.1146/annurev.aa.29.090191.003053},
       adsurl = {https://ui.adsabs.harvard.edu/abs/1991ARA&A..29..581Y}
}

@ARTICLE{Leroy_2011_constantXco,
       author = {{Leroy}, Adam K. and {Bolatto}, Alberto and {Gordon}, Karl and {Sandstrom}, Karin and {Gratier}, Pierre and {Rosolowsky}, Erik and {Engelbracht}, Charles W. and {Mizuno}, Norikazu and {Corbelli}, Edvige and {Fukui}, Yasuo and {Kawamura}, Akiko},
        title = "{The CO-to-H$_{2}$ Conversion Factor from Infrared Dust Emission across the Local Group}",
      journal = {\apj},
         year = 2011,
        month = aug,
       volume = {737},
       number = {1},
          eid = {12},
        pages = {12},
          doi = {10.1088/0004-637X/737/1/12},
archivePrefix = {arXiv},
       eprint = {1102.4618},
 primaryClass = {astro-ph.CO},
       adsurl = {https://ui.adsabs.harvard.edu/abs/2011ApJ...737...12L}
}

@ARTICLE{Solomon_1997_ULIRGs_Xco,
       author = {{Solomon}, P.~M. and {Downes}, D. and {Radford}, S.~J.~E. and {Barrett}, J.~W.},
        title = "{The Molecular Interstellar Medium in Ultraluminous Infrared Galaxies}",
      journal = {\apj},
         year = 1997,
        month = mar,
       volume = {478},
       number = {1},
        pages = {144-161},
          doi = {10.1086/303765},
archivePrefix = {arXiv},
       eprint = {astro-ph/9610166},
 primaryClass = {astro-ph},
       adsurl = {https://ui.adsabs.harvard.edu/abs/1997ApJ...478..144S}
}

@ARTICLE{Tacconi_2020_annualReviews,
       author = {{Tacconi}, Linda J. and {Genzel}, Reinhard and {Sternberg}, Amiel},
        title = "{The Evolution of the Star-Forming Interstellar Medium Across Cosmic Time}",
      journal = {\araa},
         year = 2020,
        month = aug,
       volume = {58},
        pages = {157-203},
          doi = {10.1146/annurev-astro-082812-141034},
archivePrefix = {arXiv},
       eprint = {2003.06245},
 primaryClass = {astro-ph.GA},
       adsurl = {https://ui.adsabs.harvard.edu/abs/2020ARA&A..58..157T}
}

@ARTICLE{Wolfire_2010_coDarkGas,
       author = {{Wolfire}, Mark G. and {Hollenbach}, David and {McKee}, Christopher F.},
        title = "{The Dark Molecular Gas}",
      journal = {\apj},
         year = 2010,
        month = jun,
       volume = {716},
       number = {2},
        pages = {1191-1207},
          doi = {10.1088/0004-637X/716/2/1191},
archivePrefix = {arXiv},
       eprint = {1004.5401},
 primaryClass = {astro-ph.GA},
       adsurl = {https://ui.adsabs.harvard.edu/abs/2010ApJ...716.1191W}
}

@ARTICLE{Lidz_2011_coModelling,
       author = {{Lidz}, Adam and {Furlanetto}, Steven R. and {Oh}, S. Peng and {Aguirre}, James and {Chang}, Tzu-Ching and {Dor{\'e}}, Olivier and {Pritchard}, Jonathan R.},
        title = "{Intensity Mapping with Carbon Monoxide Emission Lines and the Redshifted 21 cm Line}",
      journal = {\apj},
         year = 2011,
        month = nov,
       volume = {741},
       number = {2},
          eid = {70},
        pages = {70},
          doi = {10.1088/0004-637X/741/2/70},
archivePrefix = {arXiv},
       eprint = {1104.4800},
 primaryClass = {astro-ph.CO},
       adsurl = {https://ui.adsabs.harvard.edu/abs/2011ApJ...741...70L}
}

@ARTICLE{Daddi2010_alphaCO,
       author = {{Daddi}, E. and {Bournaud}, F. and {Walter}, F. and {Dannerbauer}, H. and {Carilli}, C.~L. and {Dickinson}, M. and {Elbaz}, D. and {Morrison}, G.~E. and {Riechers}, D. and {Onodera}, M. and {Salmi}, F. and {Krips}, M. and {Stern}, D.},
        title = "{Very High Gas Fractions and Extended Gas Reservoirs in z = 1.5 Disk Galaxies}",
      journal = {\apj},
         year = 2010,
        month = apr,
       volume = {713},
       number = {1},
        pages = {686-707},
          doi = {10.1088/0004-637X/713/1/686},
archivePrefix = {arXiv},
       eprint = {0911.2776},
 primaryClass = {astro-ph.CO},
       adsurl = {https://ui.adsabs.harvard.edu/abs/2010ApJ...713..686D}
}

@ARTICLE{Li_2016_coModelling,
       author = {{Li}, Tony Y. and {Wechsler}, Risa H. and {Devaraj}, Kiruthika and {Church}, Sarah E.},
        title = "{Connecting CO Intensity Mapping to Molecular Gas and Star Formation in the Epoch of Galaxy Assembly}",
      journal = {\apj},
         year = 2016,
        month = feb,
       volume = {817},
       number = {2},
          eid = {169},
        pages = {169},
          doi = {10.3847/0004-637X/817/2/169},
archivePrefix = {arXiv},
       eprint = {1503.08833},
 primaryClass = {astro-ph.CO},
       adsurl = {https://ui.adsabs.harvard.edu/abs/2016ApJ...817..169L}
}

@ARTICLE{Keating_2020_coModelling,
       author = {{Keating}, Garrett K. and {Marrone}, Daniel P. and {Bower}, Geoffrey C. and {Keenan}, Ryan P.},
        title = "{An Intensity Mapping Detection of Aggregate CO Line Emission at 3 mm}",
      journal = {\apj},
         year = 2020,
        month = oct,
       volume = {901},
       number = {2},
          eid = {141},
        pages = {141},
          doi = {10.3847/1538-4357/abb08e},
archivePrefix = {arXiv},
       eprint = {2008.08087},
 primaryClass = {astro-ph.GA},
       adsurl = {https://ui.adsabs.harvard.edu/abs/2020ApJ...901..141K}
}

@ARTICLE{Gong_2011_coModelling,
       author = {{Gong}, Yan and {Cooray}, Asantha and {Silva}, Marta B. and {Santos}, Mario G. and {Lubin}, Phillip},
        title = "{Probing Reionization with Intensity Mapping of Molecular and Fine-structure Lines}",
      journal = {\apjl},
         year = 2011,
        month = feb,
       volume = {728},
       number = {2},
          eid = {L46},
        pages = {L46},
          doi = {10.1088/2041-8205/728/2/L46},
archivePrefix = {arXiv},
       eprint = {1101.2892},
 primaryClass = {astro-ph.CO},
       adsurl = {https://ui.adsabs.harvard.edu/abs/2011ApJ...728L..46G}
}

@ARTICLE{masRibas2023_LIMfast1,
       author = {{Mas-Ribas}, Llu{\'\i}s and {Sun}, Guochao and {Chang}, Tzu-Ching and {Gonzalez}, Michael O. and {Mebane}, Richard H.},
        title = "{LIMFAST. I. A Seminumerical Tool for Line Intensity Mapping}",
      journal = {\apj},
         year = 2023,
        month = jun,
       volume = {950},
       number = {1},
          eid = {39},
        pages = {39},
          doi = {10.3847/1538-4357/acc9b2},
archivePrefix = {arXiv},
       eprint = {2206.14185},
 primaryClass = {astro-ph.CO},
       adsurl = {https://ui.adsabs.harvard.edu/abs/2023ApJ...950...39M}
}

@ARTICLE{jasonSun2023_LIMfast2_coModelling,
       author = {{Sun}, Guochao and {Mas-Ribas}, Llu{\'\i}s and {Chang}, Tzu-Ching and {Furlanetto}, Steven R. and {Mebane}, Richard H. and {Gonzalez}, Michael O. and {Parsons}, Jasmine and {Trapp}, A.~C.},
        title = "{LIMFAST. II. Line Intensity Mapping as a Probe of High-redshift Galaxy Formation}",
      journal = {\apj},
         year = 2023,
        month = jun,
       volume = {950},
       number = {1},
          eid = {40},
        pages = {40},
          doi = {10.3847/1538-4357/acc9b3},
archivePrefix = {arXiv},
       eprint = {2206.14186},
 primaryClass = {astro-ph.CO},
       adsurl = {https://ui.adsabs.harvard.edu/abs/2023ApJ...950...40S}
}

@ARTICLE{Yang_2022_coModelling,
       author = {{Yang}, Shengqi and {Popping}, Gerg{\"o} and {Somerville}, Rachel S. and {Pullen}, Anthony R. and {Breysse}, Patrick C. and {Maniyar}, Abhishek S.},
        title = "{An Empirical Representation of a Physical Model for the ISM [C II], CO, and [C I] Emission at Redshift 1 {\ensuremath{\leq}} z {\ensuremath{\leq}} 9}",
      journal = {\apj},
         year = 2022,
        month = apr,
       volume = {929},
       number = {2},
          eid = {140},
        pages = {140},
          doi = {10.3847/1538-4357/ac5d57},
archivePrefix = {arXiv},
       eprint = {2108.07716},
 primaryClass = {astro-ph.GA},
       adsurl = {https://ui.adsabs.harvard.edu/abs/2022ApJ...929..140Y}
}

@ARTICLE{pullen_2013_coModelling,
       author = {{Pullen}, Anthony R. and {Chang}, Tzu-Ching and {Dor{\'e}}, Olivier and {Lidz}, Adam},
        title = "{Cross-correlations as a Cosmological Carbon Monoxide Detector}",
      journal = {\apj},
         year = 2013,
        month = may,
       volume = {768},
       number = {1},
          eid = {15},
        pages = {15},
          doi = {10.1088/0004-637X/768/1/15},
archivePrefix = {arXiv},
       eprint = {1211.1397},
 primaryClass = {astro-ph.CO},
       adsurl = {https://ui.adsabs.harvard.edu/abs/2013ApJ...768...15P}
}

@ARTICLE{Sun_2019_coModelling,
       author = {{Sun}, Guochao and {Hensley}, Brandon S. and {Chang}, Tzu-Ching and {Dor{\'e}}, Olivier and {Serra}, Paolo},
        title = "{A Self-consistent Framework for Multiline Modeling in Line Intensity Mapping Experiments}",
      journal = {\apj},
         year = 2019,
        month = dec,
       volume = {887},
       number = {2},
          eid = {142},
        pages = {142},
          doi = {10.3847/1538-4357/ab55df},
archivePrefix = {arXiv},
       eprint = {1907.02999},
 primaryClass = {astro-ph.GA},
       adsurl = {https://ui.adsabs.harvard.edu/abs/2019ApJ...887..142S}
}

@ARTICLE{Valentino_2018_H0_tension_wz,
       author = {{Di Valentino}, Eleonora and {Linder}, Eric V. and {Melchiorri}, Alessandro},
        title = "{Vacuum phase transition solves the H$_{0}$ tension}",
      journal = {\prd},
         year = 2018,
        month = feb,
       volume = {97},
       number = {4},
          eid = {043528},
        pages = {043528},
          doi = {10.1103/PhysRevD.97.043528},
archivePrefix = {arXiv},
       eprint = {1710.02153},
 primaryClass = {astro-ph.CO},
       adsurl = {https://ui.adsabs.harvard.edu/abs/2018PhRvD..97d3528D}
}

@ARTICLE{Dinda_2018_LIM_and_cosmology_wz,
       author = {{Dinda}, Bikash R. and {Sen}, Anjan A and {Choudhury}, Tirthankar Roy},
        title = "{Dark energy constraints from the 21\raisebox{-0.5ex}\textasciitildecm intensity mapping surveys with SKA1}",
      journal = {arXiv e-prints},
         year = 2018,
        month = apr,
          eid = {arXiv:1804.11137},
        pages = {arXiv:1804.11137},
          doi = {10.48550/arXiv.1804.11137},
archivePrefix = {arXiv},
       eprint = {1804.11137},
 primaryClass = {astro-ph.CO},
       adsurl = {https://ui.adsabs.harvard.edu/abs/2018arXiv180411137D}
}

@ARTICLE{Zhao_2017_H0_tension_wz,
       author = {{Zhao}, Gong-Bo and {Raveri}, Marco and {Pogosian}, Levon and {Wang}, Yuting and {Crittenden}, Robert G. and {Handley}, Will J. and {Percival}, Will J. and {Beutler}, Florian and {Brinkmann}, Jonathan and {Chuang}, Chia-Hsun and {Cuesta}, Antonio J. and {Eisenstein}, Daniel J. and {Kitaura}, Francisco-Shu and {Koyama}, Kazuya and {L'Huillier}, Benjamin and {Nichol}, Robert C. and {Pieri}, Matthew M. and {Rodriguez-Torres}, Sergio and {Ross}, Ashley J. and {Rossi}, Graziano and {S{\'a}nchez}, Ariel G. and {Shafieloo}, Arman and {Tinker}, Jeremy L. and {Tojeiro}, Rita and {Vazquez}, Jose A. and {Zhang}, Hanyu},
        title = "{Dynamical dark energy in light of the latest observations}",
      journal = {Nature Astronomy},
         year = 2017,
        month = aug,
       volume = {1},
        pages = {627-632},
          doi = {10.1038/s41550-017-0216-z},
archivePrefix = {arXiv},
       eprint = {1701.08165},
 primaryClass = {astro-ph.CO},
       adsurl = {https://ui.adsabs.harvard.edu/abs/2017NatAs...1..627Z}
}

@ARTICLE{Poulin_2023_H0_tension,
       author = {{Poulin}, Vivian and {Smith}, Tristan L. and {Karwal}, Tanvi},
        title = "{The Ups and Downs of Early Dark Energy solutions to the Hubble tension: A review of models, hints and constraints circa 2023}",
      journal = {Physics of the Dark Universe},
         year = 2023,
        month = dec,
       volume = {42},
          eid = {101348},
        pages = {101348},
          doi = {10.1016/j.dark.2023.101348},
archivePrefix = {arXiv},
       eprint = {2302.09032},
 primaryClass = {astro-ph.CO},
       adsurl = {https://ui.adsabs.harvard.edu/abs/2023PDU....4201348P}
}

@ARTICLE{Riess_2016_H0_tension,
       author = {{Riess}, Adam G. and {Macri}, Lucas M. and {Hoffmann}, Samantha L. and {Scolnic}, Dan and {Casertano}, Stefano and {Filippenko}, Alexei V. and {Tucker}, Brad E. and {Reid}, Mark J. and {Jones}, David O. and {Silverman}, Jeffrey M. and {Chornock}, Ryan and {Challis}, Peter and {Yuan}, Wenlong and {Brown}, Peter J. and {Foley}, Ryan J.},
        title = "{A 2.4\% Determination of the Local Value of the Hubble Constant}",
      journal = {\apj},
         year = 2016,
        month = jul,
       volume = {826},
       number = {1},
          eid = {56},
        pages = {56},
          doi = {10.3847/0004-637X/826/1/56},
archivePrefix = {arXiv},
       eprint = {1604.01424},
 primaryClass = {astro-ph.CO},
       adsurl = {https://ui.adsabs.harvard.edu/abs/2016ApJ...826...56R}
}

@ARTICLE{Krolewski2024_desi_fnl,
       author = {{Krolewski}, Alex and {Percival}, Will J. and {Ferraro}, Simone and {Chaussidon}, Edmond and {Rezaie}, Mehdi and {Aguilar}, Jessica Nicole and {Ahlen}, Steven and {Brooks}, David and {Dawson}, Kyle and {de la Macorra}, Axel and {Doel}, Peter and {Fanning}, Kevin and {Font-Ribera}, Andreu and {a Gontcho}, Satya Gontcho and {Guy}, Julien and {Honscheid}, Klaus and {Kehoe}, Robert and {Kisner}, Theodore and {Kremin}, Anthony and {Landriau}, Martin and {Levi}, Michael E. and {Martini}, Paul and {Meisner}, Aaron M. and {Miquel}, Ramon and {Nie}, Jundan and {Poppett}, Claire and {Ross}, Ashley J. and {Rossi}, Graziano and {Schubnell}, Michael and {Seo}, Hee-Jong and {Tarl{\'e}}, Gregory and {Vargas-Maga{\~n}a}, Mariana and {Weaver}, Benjamin Alan and {Y{\`e}che}, Christophe and {Zhou}, Rongpu and {Zhou}, Zhimin},
        title = "{Constraining primordial non-Gaussianity from DESI quasar targets and Planck CMB lensing}",
      journal = {\jcap},
         year = 2024,
        month = mar,
       volume = {2024},
       number = {3},
          eid = {021},
        pages = {021},
          doi = {10.1088/1475-7516/2024/03/021},
archivePrefix = {arXiv},
       eprint = {2305.07650},
 primaryClass = {astro-ph.CO},
       adsurl = {https://ui.adsabs.harvard.edu/abs/2024JCAP...03..021K}
}

@ARTICLE{Chiarenza2026_planckAndDESIcross_fnl,
       author = {{Chiarenza}, S. and {Krolewski}, A. and {Bonici}, M. and {Chaussidon}, E. and {de Belsunce}, R. and {Percival}, W.~J. and {Aguilar}, J. and {Ahlen}, S. and {Baleato Lizancos}, A. and {Bianchi}, D. and {Brooks}, D. and {Claybaugh}, T. and {Cuceu}, A. and {Dawson}, K.~S. and {de la Macorra}, A. and {Doel}, P. and {Ferraro}, S. and {Font-Ribera}, A. and {Forero-Romero}, J.~E. and {Gazta{\~n}aga}, E. and {Gontcho}, S. Gontcho A. and {Gutierrez}, G. and {Herrera-Alcantar}, H.~K. and {Honscheid}, K. and {Huterer}, D. and {Ishak}, M. and {Joyce}, R. and {Kirkby}, D. and {Kremin}, A. and {Lahav}, O. and {Lamman}, C. and {Landriau}, M. and {Le Guillou}, L. and {Levi}, M.~E. and {Manera}, M. and {Martini}, P. and {Meisner}, A. and {Miquel}, R. and {Nadathur}, S. and {Newman}, J.~A. and {Niz}, G. and {Palanque-Delabrouille}, N. and {Poppett}, C. and {Prada}, F. and {P{\'e}rez-R{\`a}fols}, I. and {Rossi}, G. and {Sanchez}, E. and {Schlegel}, D. and {Schubnell}, M. and {Seo}, H. and {Silber}, J. and {Sprayberry}, D. and {Tarl{\'e}}, G. and {Weaver}, B.~A. and {Y{\`e}che}, C. and {Zhou}, R. and {Zou}, H.},
        title = "{Constraining primordial non-Gaussianity from DESI DR1 quasars and Planck PR4 CMB lensing}",
      journal = {\jcap},
         year = 2026,
        month = may,
       volume = {2026},
       number = {5},
          eid = {098},
        pages = {098},
          doi = {10.1088/1475-7516/2026/05/098},
archivePrefix = {arXiv},
       eprint = {2512.17865},
 primaryClass = {astro-ph.CO},
       adsurl = {https://ui.adsabs.harvard.edu/abs/2026JCAP...05..098C}
}

@ARTICLE{planck_2018,
       author = {{Planck Collaboration} and {Aghanim}, N. and {Akrami}, Y. and {Ashdown}, M. and {Aumont}, J. and {Baccigalupi}, C. and {Ballardini}, M. and {Banday}, A.~J. and {Barreiro}, R.~B. and {Bartolo}, N. and {Basak}, S. and {Battye}, R. and {Benabed}, K. and {Bernard}, J.-P. and {Bersanelli}, M. and {Bielewicz}, P. and {Bock}, J.~J. and {Bond}, J.~R. and {Borrill}, J. and {Bouchet}, F.~R. and {Boulanger}, F. and {Bucher}, M. and {Burigana}, C. and {Butler}, R.~C. and {Calabrese}, E. and {Cardoso}, J.-F. and {Carron}, J. and {Challinor}, A. and {Chiang}, H.~C. and {Chluba}, J. and {Colombo}, L.~P.~L. and {Combet}, C. and {Contreras}, D. and {Crill}, B.~P. and {Cuttaia}, F. and {de Bernardis}, P. and {de Zotti}, G. and {Delabrouille}, J. and {Delouis}, J.-M. and {Di Valentino}, E. and {Diego}, J.~M. and {Dor{\'e}}, O. and {Douspis}, M. and {Ducout}, A. and {Dupac}, X. and {Dusini}, S. and {Efstathiou}, G. and {Elsner}, F. and {En{\ss}lin}, T.~A. and {Eriksen}, H.~K. and {Fantaye}, Y. and {Farhang}, M. and {Fergusson}, J. and {Fernandez-Cobos}, R. and {Finelli}, F. and {Forastieri}, F. and {Frailis}, M. and {Fraisse}, A.~A. and {Franceschi}, E. and {Frolov}, A. and {Galeotta}, S. and {Galli}, S. and {Ganga}, K. and {G{\'e}nova-Santos}, R.~T. and {Gerbino}, M. and {Ghosh}, T. and {Gonz{\'a}lez-Nuevo}, J. and {G{\'o}rski}, K.~M. and {Gratton}, S. and {Gruppuso}, A. and {Gudmundsson}, J.~E. and {Hamann}, J. and {Handley}, W. and {Hansen}, F.~K. and {Herranz}, D. and {Hildebrandt}, S.~R. and {Hivon}, E. and {Huang}, Z. and {Jaffe}, A.~H. and {Jones}, W.~C. and {Karakci}, A. and {Keih{\"a}nen}, E. and {Keskitalo}, R. and {Kiiveri}, K. and {Kim}, J. and {Kisner}, T.~S. and {Knox}, L. and {Krachmalnicoff}, N. and {Kunz}, M. and {Kurki-Suonio}, H. and {Lagache}, G. and {Lamarre}, J.-M. and {Lasenby}, A. and {Lattanzi}, M. and {Lawrence}, C.~R. and {Le Jeune}, M. and {Lemos}, P. and {Lesgourgues}, J. and {Levrier}, F. and {Lewis}, A. and {Liguori}, M. and {Lilje}, P.~B. and {Lilley}, M. and {Lindholm}, V. and {L{\'o}pez-Caniego}, M. and {Lubin}, P.~M. and {Ma}, Y.-Z. and {Mac{\'\i}as-P{\'e}rez}, J.~F. and {Maggio}, G. and {Maino}, D. and {Mandolesi}, N. and {Mangilli}, A. and {Marcos-Caballero}, A. and {Maris}, M. and {Martin}, P.~G. and {Martinelli}, M. and {Mart{\'\i}nez-Gonz{\'a}lez}, E. and {Matarrese}, S. and {Mauri}, N. and {McEwen}, J.~D. and {Meinhold}, P.~R. and {Melchiorri}, A. and {Mennella}, A. and {Migliaccio}, M. and {Millea}, M. and {Mitra}, S. and {Miville-Desch{\^e}nes}, M.-A. and {Molinari}, D. and {Montier}, L. and {Morgante}, G. and {Moss}, A. and {Natoli}, P. and {N{\o}rgaard-Nielsen}, H.~U. and {Pagano}, L. and {Paoletti}, D. and {Partridge}, B. and {Patanchon}, G. and {Peiris}, H.~V. and {Perrotta}, F. and {Pettorino}, V. and {Piacentini}, F. and {Polastri}, L. and {Polenta}, G. and {Puget}, J.-L. and {Rachen}, J.~P. and {Reinecke}, M. and {Remazeilles}, M. and {Renzi}, A. and {Rocha}, G. and {Rosset}, C. and {Roudier}, G. and {Rubi{\~n}o-Mart{\'\i}n}, J.~A. and {Ruiz-Granados}, B. and {Salvati}, L. and {Sandri}, M. and {Savelainen}, M. and {Scott}, D. and {Shellard}, E.~P.~S. and {Sirignano}, C. and {Sirri}, G. and {Spencer}, L.~D. and {Sunyaev}, R. and {Suur-Uski}, A.-S. and {Tauber}, J.~A. and {Tavagnacco}, D. and {Tenti}, M. and {Toffolatti}, L. and {Tomasi}, M. and {Trombetti}, T. and {Valenziano}, L. and {Valiviita}, J. and {Van Tent}, B. and {Vibert}, L. and {Vielva}, P. and {Villa}, F. and {Vittorio}, N. and {Wandelt}, B.~D. and {Wehus}, I.~K. and {White}, M. and {White}, S.~D.~M. and {Zacchei}, A. and {Zonca}, A.},
        title = "{Planck 2018 results. VI. Cosmological parameters}",
      journal = {\aap},
         year = 2020,
        month = sep,
       volume = {641},
          eid = {A6},
        pages = {A6},
          doi = {10.1051/0004-6361/201833910},
archivePrefix = {arXiv},
       eprint = {1807.06209},
 primaryClass = {astro-ph.CO},
       adsurl = {https://ui.adsabs.harvard.edu/abs/2020A&A...641A...6P}
}

@ARTICLE{Planck_2014,
       author = {{Planck Collaboration} and {Ade}, P.~A.~R. and {Aghanim}, N. and {Armitage-Caplan}, C. and {Arnaud}, M. and {Ashdown}, M. and {Atrio-Barandela}, F. and {Aumont}, J. and {Baccigalupi}, C. and {Banday}, A.~J. and {Barreiro}, R.~B. and {Bartlett}, J.~G. and {Battaner}, E. and {Benabed}, K. and {Beno{\^\i}t}, A. and {Benoit-L{\'e}vy}, A. and {Bernard}, J. -P. and {Bersanelli}, M. and {Bethermin}, M. and {Bielewicz}, P. and {Blagrave}, K. and {Bobin}, J. and {Bock}, J.~J. and {Bonaldi}, A. and {Bond}, J.~R. and {Borrill}, J. and {Bouchet}, F.~R. and {Boulanger}, F. and {Bridges}, M. and {Bucher}, M. and {Burigana}, C. and {Butler}, R.~C. and {Cardoso}, J. -F. and {Catalano}, A. and {Challinor}, A. and {Chamballu}, A. and {Chen}, X. and {Chiang}, H.~C. and {Chiang}, L. -Y. and {Christensen}, P.~R. and {Church}, S. and {Clements}, D.~L. and {Colombi}, S. and {Colombo}, L.~P.~L. and {Couchot}, F. and {Coulais}, A. and {Crill}, B.~P. and {Curto}, A. and {Cuttaia}, F. and {Danese}, L. and {Davies}, R.~D. and {Davis}, R.~J. and {de Bernardis}, P. and {de Rosa}, A. and {de Zotti}, G. and {Delabrouille}, J. and {Delouis}, J. -M. and {D{\'e}sert}, F. -X. and {Dickinson}, C. and {Diego}, J.~M. and {Dole}, H. and {Donzelli}, S. and {Dor{\'e}}, O. and {Douspis}, M. and {Dupac}, X. and {Efstathiou}, G. and {En{\ss}lin}, T.~A. and {Eriksen}, H.~K. and {Finelli}, F. and {Forni}, O. and {Frailis}, M. and {Franceschi}, E. and {Galeotta}, S. and {Ganga}, K. and {Ghosh}, T. and {Giard}, M. and {Giraud-H{\'e}raud}, Y. and {Gonz{\'a}lez-Nuevo}, J. and {G{\'o}rski}, K.~M. and {Gratton}, S. and {Gregorio}, A. and {Gruppuso}, A. and {Hansen}, F.~K. and {Hanson}, D. and {Harrison}, D. and {Helou}, G. and {Henrot-Versill{\'e}}, S. and {Hern{\'a}ndez-Monteagudo}, C. and {Herranz}, D. and {Hildebrandt}, S.~R. and {Hivon}, E. and {Hobson}, M. and {Holmes}, W.~A. and {Hornstrup}, A. and {Hovest}, W. and {Huffenberger}, K.~M. and {Jaffe}, A.~H. and {Jaffe}, T.~R. and {Jones}, W.~C. and {Juvela}, M. and {Kalberla}, P. and {Keih{\"a}nen}, E. and {Kerp}, J. and {Keskitalo}, R. and {Kisner}, T.~S. and {Kneissl}, R. and {Knoche}, J. and {Knox}, L. and {Kunz}, M. and {Kurki-Suonio}, H. and {Lacasa}, F. and {Lagache}, G. and {L{\"a}hteenm{\"a}ki}, A. and {Lamarre}, J. -M. and {Langer}, M. and {Lasenby}, A. and {Laureijs}, R.~J. and {Lawrence}, C.~R. and {Leonardi}, R. and {Le{\'o}n-Tavares}, J. and {Lesgourgues}, J. and {Liguori}, M. and {Lilje}, P.~B. and {Linden-V{\o}rnle}, M. and {L{\'o}pez-Caniego}, M. and {Lubin}, P.~M. and {Mac{\'\i}as-P{\'e}rez}, J.~F. and {Maffei}, B. and {Maino}, D. and {Mandolesi}, N. and {Maris}, M. and {Marshall}, D.~J. and {Martin}, P.~G. and {Mart{\'\i}nez-Gonz{\'a}lez}, E. and {Masi}, S. and {Massardi}, M. and {Matarrese}, S. and {Matthai}, F. and {Mazzotta}, P. and {Melchiorri}, A. and {Mendes}, L. and {Mennella}, A. and {Migliaccio}, M. and {Mitra}, S. and {Miville-Desch{\^e}nes}, M. -A. and {Moneti}, A. and {Montier}, L. and {Morgante}, G. and {Mortlock}, D. and {Munshi}, D. and {Murphy}, J.~A. and {Naselsky}, P. and {Nati}, F. and {Natoli}, P. and {Netterfield}, C.~B. and {N{\o}rgaard-Nielsen}, H.~U. and {Noviello}, F. and {Novikov}, D. and {Novikov}, I. and {Osborne}, S. and {Oxborrow}, C.~A. and {Paci}, F. and {Pagano}, L. and {Pajot}, F. and {Paladini}, R. and {Paoletti}, D. and {Partridge}, B. and {Pasian}, F. and {Patanchon}, G. and {Perdereau}, O. and {Perotto}, L. and {Perrotta}, F. and {Piacentini}, F. and {Piat}, M. and {Pierpaoli}, E. and {Pietrobon}, D. and {Plaszczynski}, S. and {Pointecouteau}, E. and {Polenta}, G. and {Ponthieu}, N. and {Popa}, L. and {Poutanen}, T. and {Pratt}, G.~W. and {Pr{\'e}zeau}, G. and {Prunet}, S. and {Puget}, J. -L. and {Rachen}, J.~P. and {Reach}, W.~T. and {Rebolo}, R. and {Reinecke}, M. and {Remazeilles}, M. and {Renault}, C. and {Ricciardi}, S. and {Riller}, T. and {Ristorcelli}, I. and {Rocha}, G. and {Rosset}, C. and {Roudier}, G. and {Rowan-Robinson}, M. and {Rubi{\~n}o-Mart{\'\i}n}, J.~A.},
        title = "{Planck 2013 results. XXX. Cosmic infrared background measurements and implications for star formation}",
      journal = {\aap},
         year = 2014,
        month = nov,
       volume = {571},
          eid = {A30},
        pages = {A30},
          doi = {10.1051/0004-6361/201322093},
archivePrefix = {arXiv},
       eprint = {1309.0382},
 primaryClass = {astro-ph.CO},
       adsurl = {https://ui.adsabs.harvard.edu/abs/2014A&A...571A..30P}
}

@ARTICLE{Hopkins_2015,
       author = {{Hopkins}, Philip F.},
        title = "{A new class of accurate, mesh-free hydrodynamic simulation methods}",
      journal = {\mnras},
         year = 2015,
        month = jun,
       volume = {450},
       number = {1},
        pages = {53-110},
          doi = {10.1093/mnras/stv195},
archivePrefix = {arXiv},
       eprint = {1409.7395},
 primaryClass = {astro-ph.CO},
       adsurl = {https://ui.adsabs.harvard.edu/abs/2015MNRAS.450...53H}
}

@ARTICLE{Hopkins2020_Lebron,
       author = {{Hopkins}, Philip F. and {Grudi{\'c}}, Michael Y. and {Wetzel}, Andrew and {Kere{\v{s}}}, Du{\v{s}}an and {Faucher-Gigu{\`e}re}, Claude-Andr{\'e} and {Ma}, Xiangcheng and {Murray}, Norman and {Butcher}, Nathan},
        title = "{Radiative stellar feedback in galaxy formation: Methods and physics}",
      journal = {\mnras},
         year = 2020,
        month = jan,
       volume = {491},
       number = {3},
        pages = {3702-3729},
          doi = {10.1093/mnras/stz3129},
archivePrefix = {arXiv},
       eprint = {1811.12462},
 primaryClass = {astro-ph.GA},
       adsurl = {https://ui.adsabs.harvard.edu/abs/2020MNRAS.491.3702H}
}

@ARTICLE{Kim2014_agoraCosmology_FIRE,
       author = {{Kim}, Ji-hoon and {Abel}, Tom and {Agertz}, Oscar and {Bryan}, Greg L. and {Ceverino}, Daniel and {Christensen}, Charlotte and {Conroy}, Charlie and {Dekel}, Avishai and {Gnedin}, Nickolay Y. and {Goldbaum}, Nathan J. and {Guedes}, Javiera and {Hahn}, Oliver and {Hobbs}, Alexander and {Hopkins}, Philip F. and {Hummels}, Cameron B. and {Iannuzzi}, Francesca and {Keres}, Dusan and {Klypin}, Anatoly and {Kravtsov}, Andrey V. and {Krumholz}, Mark R. and {Kuhlen}, Michael and {Leitner}, Samuel N. and {Madau}, Piero and {Mayer}, Lucio and {Moody}, Christopher E. and {Nagamine}, Kentaro and {Norman}, Michael L. and {Onorbe}, Jose and {O'Shea}, Brian W. and {Pillepich}, Annalisa and {Primack}, Joel R. and {Quinn}, Thomas and {Read}, Justin I. and {Robertson}, Brant E. and {Rocha}, Miguel and {Rudd}, Douglas H. and {Shen}, Sijing and {Smith}, Britton D. and {Szalay}, Alexander S. and {Teyssier}, Romain and {Thompson}, Robert and {Todoroki}, Keita and {Turk}, Matthew J. and {Wadsley}, James W. and {Wise}, John H. and {Zolotov}, Adi and {AGORA Collaboration29},the},
        title = "{The AGORA High-resolution Galaxy Simulations Comparison Project}",
      journal = {\apjs},
         year = 2014,
        month = jan,
       volume = {210},
       number = {1},
          eid = {14},
        pages = {14},
          doi = {10.1088/0067-0049/210/1/14},
archivePrefix = {arXiv},
       eprint = {1308.2669},
 primaryClass = {astro-ph.GA},
       adsurl = {https://ui.adsabs.harvard.edu/abs/2014ApJS..210...14K}
}

@ARTICLE{Wetzel_2023,
       author = {{Wetzel}, Andrew and {Hayward}, Christopher C. and {Sanderson}, Robyn E. and {Ma}, Xiangcheng and {Angl{\'e}s-Alc{\'a}zar}, Daniel and {Feldmann}, Robert and {Chan}, T.~K. and {El-Badry}, Kareem and {Wheeler}, Coral and {Garrison-Kimmel}, Shea and {Nikakhtar}, Farnik and {Panithanpaisal}, Nondh and {Arora}, Arpit and {Gurvich}, Alexander B. and {Samuel}, Jenna and {Sameie}, Omid and {Pandya}, Viraj and {Hafen}, Zachary and {Hummels}, Cameron and {Loebman}, Sarah and {Boylan-Kolchin}, Michael and {Bullock}, James S. and {Faucher-Gigu{\`e}re}, Claude-Andr{\'e} and {Kere{\v{s}}}, Du{\v{s}}an and {Quataert}, Eliot and {Hopkins}, Philip F.},
        title = "{Public Data Release of the FIRE-2 Cosmological Zoom-in Simulations of Galaxy Formation}",
      journal = {\apjs},
         year = 2023,
        month = apr,
       volume = {265},
       number = {2},
          eid = {44},
        pages = {44},
          doi = {10.3847/1538-4365/acb99a},
archivePrefix = {arXiv},
       eprint = {2202.06969},
 primaryClass = {astro-ph.GA},
       adsurl = {https://ui.adsabs.harvard.edu/abs/2023ApJS..265...44W}
}

@ARTICLE{Wetzel2025_secondDataRelease_FIRE2,
       author = {{Wetzel}, Andrew and {Samuel}, Jenna and {Gandhi}, Pratik J. and {Ponnada}, Sam B. and {Su}, Kung-Yi and {Arora}, Arpit and {Angles-Alcazar}, Daniel and {Hayward}, Christopher C. and {Sanderson}, Robyn E. and {Feldmann}, Robert and {Cochrane}, Rachel and {Nikakhtar}, Farnik and {Panithanpaisal}, Nondh and {Benavides}, Jose A. and {Pandya}, Viraj and {Grudic}, Mike and {Hummels}, Cameron and {Gurvich}, Alexander B. and {Hafen}, Zachary and {Ma}, Xiangcheng and {Garrison-Kimmel}, Shea and {Sameie}, Omid and {Chan}, T. K and {El-Badry}, Kareem and {Necib}, Lina and {Loebman}, Sarah and {Wellons}, Sarah and {Robles}, Victor H. and {Wheeler}, Coral and {Moreno}, Jorge and {Stern}, Jonathan and {Boylan-Kolchin}, Michael and {Bullock}, James S. and {Faucher-Giguere}, Claude-Andre and {Keres}, Dusan and {Quataert}, Eliot and {Hopkins}, Philip F.},
        title = "{Second public data release of the FIRE-2 cosmological zoom-in simulations of galaxy formation}",
      journal = {arXiv e-prints},
         year = 2025,
        month = aug,
          eid = {arXiv:2508.06608},
        pages = {arXiv:2508.06608},
          doi = {10.48550/arXiv.2508.06608},
archivePrefix = {arXiv},
       eprint = {2508.06608},
 primaryClass = {astro-ph.GA},
       adsurl = {https://ui.adsabs.harvard.edu/abs/2025arXiv250806608W}
}

@ARTICLE{Hopkins_2018_fire2,
       author = {{Hopkins}, Philip F. and {Wetzel}, Andrew and {Kere{\v{s}}}, Du{\v{s}}an and {Faucher-Gigu{\`e}re}, Claude-Andr{\'e} and {Quataert}, Eliot and {Boylan-Kolchin}, Michael and {Murray}, Norman and {Hayward}, Christopher C. and {Garrison-Kimmel}, Shea and {Hummels}, Cameron and {Feldmann}, Robert and {Torrey}, Paul and {Ma}, Xiangcheng and {Angl{\'e}s-Alc{\'a}zar}, Daniel and {Su}, Kung-Yi and {Orr}, Matthew and {Schmitz}, Denise and {Escala}, Ivanna and {Sanderson}, Robyn and {Grudi{\'c}}, Michael Y. and {Hafen}, Zachary and {Kim}, Ji-Hoon and {Fitts}, Alex and {Bullock}, James S. and {Wheeler}, Coral and {Chan}, T.~K. and {Elbert}, Oliver D. and {Narayanan}, Desika},
        title = "{FIRE-2 simulations: physics versus numerics in galaxy formation}",
      journal = {\mnras},
         year = 2018,
        month = oct,
       volume = {480},
       number = {1},
        pages = {800-863},
          doi = {10.1093/mnras/sty1690},
archivePrefix = {arXiv},
       eprint = {1702.06148},
 primaryClass = {astro-ph.GA},
       adsurl = {https://ui.adsabs.harvard.edu/abs/2018MNRAS.480..800H}
}

@ARTICLE{ElBadry2018_m11halosPlanck,
       author = {{El-Badry}, Kareem and {Quataert}, Eliot and {Wetzel}, Andrew and {Hopkins}, Philip F. and {Weisz}, Daniel R. and {Chan}, T.~K. and {Fitts}, Alex and {Boylan-Kolchin}, Michael and {Kere{\v{s}}}, Du{\v{s}}an and {Faucher-Gigu{\`e}re}, Claude-Andr{\'e} and {Garrison-Kimmel}, Shea},
        title = "{Gas kinematics, morphology and angular momentum in the FIRE simulations}",
      journal = {\mnras},
         year = 2018,
        month = jan,
       volume = {473},
       number = {2},
        pages = {1930-1955},
          doi = {10.1093/mnras/stx2482},
archivePrefix = {arXiv},
       eprint = {1705.10321},
 primaryClass = {astro-ph.GA},
       adsurl = {https://ui.adsabs.harvard.edu/abs/2018MNRAS.473.1930E}
}

@ARTICLE{Hopkins_2014,
       author = {{Hopkins}, Philip F. and {Kere{\v{s}}}, Du{\v{s}}an and {O{\~n}orbe}, Jos{\'e} and {Faucher-Gigu{\`e}re}, Claude-Andr{\'e} and {Quataert}, Eliot and {Murray}, Norman and {Bullock}, James S.},
        title = "{Galaxies on FIRE (Feedback In Realistic Environments): stellar feedback explains cosmologically inefficient star formation}",
      journal = {\mnras},
         year = 2014,
        month = nov,
       volume = {445},
       number = {1},
        pages = {581-603},
          doi = {10.1093/mnras/stu1738},
archivePrefix = {arXiv},
       eprint = {1311.2073},
 primaryClass = {astro-ph.CO},
       adsurl = {https://ui.adsabs.harvard.edu/abs/2014MNRAS.445..581H}
}

@ARTICLE{Feldmann_2017_massiveFIRE,
       author = {{Feldmann}, Robert and {Quataert}, Eliot and {Hopkins}, Philip F. and {Faucher-Gigu{\`e}re}, Claude-Andr{\'e} and {Kere{\v{s}}}, Du{\v{s}}an},
        title = "{Colours, star formation rates and environments of star-forming and quiescent galaxies at the cosmic noon}",
      journal = {\mnras},
         year = 2017,
        month = sep,
       volume = {470},
       number = {1},
        pages = {1050-1072},
          doi = {10.1093/mnras/stx1120},
archivePrefix = {arXiv},
       eprint = {1610.02411},
 primaryClass = {astro-ph.GA},
       adsurl = {https://ui.adsabs.harvard.edu/abs/2017MNRAS.470.1050F}
}

@ARTICLE{Springel2005Gadget2,
       author = {{Springel}, Volker},
        title = "{The cosmological simulation code GADGET-2}",
      journal = {\mnras},
         year = 2005,
        month = dec,
       volume = {364},
       number = {4},
        pages = {1105-1134},
          doi = {10.1111/j.1365-2966.2005.09655.x},
archivePrefix = {arXiv},
       eprint = {astro-ph/0505010},
 primaryClass = {astro-ph},
       adsurl = {https://ui.adsabs.harvard.edu/abs/2005MNRAS.364.1105S}
}

@ARTICLE{Feldmann_2023,
       author = {{Feldmann}, Robert and {Quataert}, Eliot and {Faucher-Gigu{\`e}re}, Claude-Andr{\'e} and {Hopkins}, Philip F. and {{\c{C}}atmabacak}, Onur and {Kere{\v{s}}}, Du{\v{s}}an and {Bassini}, Luigi and {Bernardini}, Mauro and {Bullock}, James S. and {Cenci}, Elia and {Gensior}, Jindra and {Liang}, Lichen and {Moreno}, Jorge and {Wetzel}, Andrew},
        title = "{FIREbox: simulating galaxies at high dynamic range in a cosmological volume}",
      journal = {\mnras},
         year = 2023,
        month = jul,
       volume = {522},
       number = {3},
        pages = {3831-3860},
          doi = {10.1093/mnras/stad1205},
archivePrefix = {arXiv},
       eprint = {2205.15325},
 primaryClass = {astro-ph.GA},
       adsurl = {https://ui.adsabs.harvard.edu/abs/2023MNRAS.522.3831F}
}

@ARTICLE{AnglesAlcazar2017_massiveFIRE_FIRE2,
       author = {{Angl{\'e}s-Alc{\'a}zar}, Daniel and {Faucher-Gigu{\`e}re}, Claude-Andr{\'e} and {Quataert}, Eliot and {Hopkins}, Philip F. and {Feldmann}, Robert and {Torrey}, Paul and {Wetzel}, Andrew and {Kere{\v{s}}}, Du{\v{s}}an},
        title = "{Black holes on FIRE: stellar feedback limits early feeding of galactic nuclei}",
      journal = {\mnras},
         year = 2017,
        month = nov,
       volume = {472},
       number = {1},
        pages = {L109-L114},
          doi = {10.1093/mnrasl/slx161},
archivePrefix = {arXiv},
       eprint = {1707.03832},
 primaryClass = {astro-ph.GA},
       adsurl = {https://ui.adsabs.harvard.edu/abs/2017MNRAS.472L.109A}
}

@ARTICLE{Feldmann2016_massiveFIRE_FIRE1,
       author = {{Feldmann}, Robert and {Hopkins}, Philip F. and {Quataert}, Eliot and {Faucher-Gigu{\`e}re}, Claude-Andr{\'e} and {Kere{\v{s}}}, Du{\v{s}}an},
        title = "{The formation of massive, quiescent galaxies at cosmic noon}",
      journal = {\mnras},
         year = 2016,
        month = may,
       volume = {458},
       number = {1},
        pages = {L14-L18},
          doi = {10.1093/mnrasl/slw014},
archivePrefix = {arXiv},
       eprint = {1601.04704},
 primaryClass = {astro-ph.GA},
       adsurl = {https://ui.adsabs.harvard.edu/abs/2016MNRAS.458L..14F}
}

@ARTICLE{Wetzel_2016,
       author = {{Wetzel}, Andrew R. and {Hopkins}, Philip F. and {Kim}, Ji-hoon and {Faucher-Gigu{\`e}re}, Claude-Andr{\'e} and {Kere{\v{s}}}, Du{\v{s}}an and {Quataert}, Eliot},
        title = "{Reconciling Dwarf Galaxies with {\ensuremath{\Lambda}}CDM Cosmology: Simulating a Realistic Population of Satellites around a Milky Way-mass Galaxy}",
      journal = {\apjl},
         year = 2016,
        month = aug,
       volume = {827},
       number = {2},
          eid = {L23},
        pages = {L23},
          doi = {10.3847/2041-8205/827/2/L23},
archivePrefix = {arXiv},
       eprint = {1602.05957},
 primaryClass = {astro-ph.GA},
       adsurl = {https://ui.adsabs.harvard.edu/abs/2016ApJ...827L..23W}
}

@UNPUBLISHED{Wetzel_tripleLatte,
       author = {{Wetzel}, Andrew R. and others},
        title = "{}",
         year = {in prep.},
         note = {}
}

@UNPUBLISHED{Paper2,
       author = {{Tolgay}, Do{\u{g}}a and others},
        title = "{}",
         year = {in prep.},
         note = {Modeling CO Line Intensity with FIRE simulations: II. Comparison With Observations, CO Luminosity Functions and Cosmic Evolution}
}

@UNPUBLISHED{Paper3,
       author = {{Tolgay}, Do{\u{g}}a and others},
        title = "{}",
         year = {in prep.},
         note = {Modeling CO Line Intensity with FIRE simulations: III. Random Forest Tree Feature Importance, Response Functions and Implications for CO LIM models}
}

@ARTICLE{Samuel2020_m12w_m12r,
       author = {{Samuel}, Jenna and {Wetzel}, Andrew and {Tollerud}, Erik and {Garrison-Kimmel}, Shea and {Loebman}, Sarah and {El-Badry}, Kareem and {Hopkins}, Philip F. and {Boylan-Kolchin}, Michael and {Faucher-Gigu{\`e}re}, Claude-Andr{\'e} and {Bullock}, James S. and {Benincasa}, Samantha and {Bailin}, Jeremy},
        title = "{A profile in FIRE: resolving the radial distributions of satellite galaxies in the Local Group with simulations}",
      journal = {\mnras},
         year = 2020,
        month = jan,
       volume = {491},
       number = {1},
        pages = {1471-1490},
          doi = {10.1093/mnras/stz3054},
archivePrefix = {arXiv},
       eprint = {1904.11508},
 primaryClass = {astro-ph.GA},
       adsurl = {https://ui.adsabs.harvard.edu/abs/2020MNRAS.491.1471S}
}

@ARTICLE{Garrison-Kimmel2019_m12b_m12c,
       author = {{Garrison-Kimmel}, Shea and {Hopkins}, Philip F. and {Wetzel}, Andrew and {Bullock}, James S. and {Boylan-Kolchin}, Michael and {Kere{\v{s}}}, Du{\v{s}}an and {Faucher-Gigu{\`e}re}, Claude-Andr{\'e} and {El-Badry}, Kareem and {Lamberts}, Astrid and {Quataert}, Eliot and {Sanderson}, Robyn},
        title = "{The Local Group on FIRE: dwarf galaxy populations across a suite of hydrodynamic simulations}",
      journal = {\mnras},
         year = 2019,
        month = jul,
       volume = {487},
       number = {1},
        pages = {1380-1399},
          doi = {10.1093/mnras/stz1317},
archivePrefix = {arXiv},
       eprint = {1806.04143},
 primaryClass = {astro-ph.GA},
       adsurl = {https://ui.adsabs.harvard.edu/abs/2019MNRAS.487.1380G}
}

@ARTICLE{Garrison-Kimmel2017_m12f,
       author = {{Garrison-Kimmel}, Shea and {Wetzel}, Andrew and {Bullock}, James S. and {Hopkins}, Philip F. and {Boylan-Kolchin}, Michael and {Faucher-Gigu{\`e}re}, Claude-Andr{\'e} and {Kere{\v{s}}}, Du{\v{s}}an and {Quataert}, Eliot and {Sanderson}, Robyn E. and {Graus}, Andrew S. and {Kelley}, Tyler},
        title = "{Not so lumpy after all: modelling the depletion of dark matter subhaloes by Milky Way-like galaxies}",
      journal = {\mnras},
         year = 2017,
        month = oct,
       volume = {471},
       number = {2},
        pages = {1709-1727},
          doi = {10.1093/mnras/stx1710},
archivePrefix = {arXiv},
       eprint = {1701.03792},
 primaryClass = {astro-ph.GA},
       adsurl = {https://ui.adsabs.harvard.edu/abs/2017MNRAS.471.1709G}
}

@ARTICLE{Planck_2016,
       author = {{Planck Collaboration} and {Ade}, P.~A.~R. and {Aghanim}, N. and {Arnaud}, M. and {Ashdown}, M. and {Aumont}, J. and {Baccigalupi}, C. and {Banday}, A.~J. and {Barreiro}, R.~B. and {Bartlett}, J.~G. and {Bartolo}, N. and {Battaner}, E. and {Battye}, R. and {Benabed}, K. and {Beno{\^\i}t}, A. and {Benoit-L{\'e}vy}, A. and {Bernard}, J. -P. and {Bersanelli}, M. and {Bielewicz}, P. and {Bock}, J.~J. and {Bonaldi}, A. and {Bonavera}, L. and {Bond}, J.~R. and {Borrill}, J. and {Bouchet}, F.~R. and {Boulanger}, F. and {Bucher}, M. and {Burigana}, C. and {Butler}, R.~C. and {Calabrese}, E. and {Cardoso}, J. -F. and {Catalano}, A. and {Challinor}, A. and {Chamballu}, A. and {Chary}, R. -R. and {Chiang}, H.~C. and {Chluba}, J. and {Christensen}, P.~R. and {Church}, S. and {Clements}, D.~L. and {Colombi}, S. and {Colombo}, L.~P.~L. and {Combet}, C. and {Coulais}, A. and {Crill}, B.~P. and {Curto}, A. and {Cuttaia}, F. and {Danese}, L. and {Davies}, R.~D. and {Davis}, R.~J. and {de Bernardis}, P. and {de Rosa}, A. and {de Zotti}, G. and {Delabrouille}, J. and {D{\'e}sert}, F. -X. and {Di Valentino}, E. and {Dickinson}, C. and {Diego}, J.~M. and {Dolag}, K. and {Dole}, H. and {Donzelli}, S. and {Dor{\'e}}, O. and {Douspis}, M. and {Ducout}, A. and {Dunkley}, J. and {Dupac}, X. and {Efstathiou}, G. and {Elsner}, F. and {En{\ss}lin}, T.~A. and {Eriksen}, H.~K. and {Farhang}, M. and {Fergusson}, J. and {Finelli}, F. and {Forni}, O. and {Frailis}, M. and {Fraisse}, A.~A. and {Franceschi}, E. and {Frejsel}, A. and {Galeotta}, S. and {Galli}, S. and {Ganga}, K. and {Gauthier}, C. and {Gerbino}, M. and {Ghosh}, T. and {Giard}, M. and {Giraud-H{\'e}raud}, Y. and {Giusarma}, E. and {Gjerl{\o}w}, E. and {Gonz{\'a}lez-Nuevo}, J. and {G{\'o}rski}, K.~M. and {Gratton}, S. and {Gregorio}, A. and {Gruppuso}, A. and {Gudmundsson}, J.~E. and {Hamann}, J. and {Hansen}, F.~K. and {Hanson}, D. and {Harrison}, D.~L. and {Helou}, G. and {Henrot-Versill{\'e}}, S. and {Hern{\'a}ndez-Monteagudo}, C. and {Herranz}, D. and {Hildebrandt}, S.~R. and {Hivon}, E. and {Hobson}, M. and {Holmes}, W.~A. and {Hornstrup}, A. and {Hovest}, W. and {Huang}, Z. and {Huffenberger}, K.~M. and {Hurier}, G. and {Jaffe}, A.~H. and {Jaffe}, T.~R. and {Jones}, W.~C. and {Juvela}, M. and {Keih{\"a}nen}, E. and {Keskitalo}, R. and {Kisner}, T.~S. and {Kneissl}, R. and {Knoche}, J. and {Knox}, L. and {Kunz}, M. and {Kurki-Suonio}, H. and {Lagache}, G. and {L{\"a}hteenm{\"a}ki}, A. and {Lamarre}, J. -M. and {Lasenby}, A. and {Lattanzi}, M. and {Lawrence}, C.~R. and {Leahy}, J.~P. and {Leonardi}, R. and {Lesgourgues}, J. and {Levrier}, F. and {Lewis}, A. and {Liguori}, M. and {Lilje}, P.~B. and {Linden-V{\o}rnle}, M. and {L{\'o}pez-Caniego}, M. and {Lubin}, P.~M. and {Mac{\'\i}as-P{\'e}rez}, J.~F. and {Maggio}, G. and {Maino}, D. and {Mandolesi}, N. and {Mangilli}, A. and {Marchini}, A. and {Maris}, M. and {Martin}, P.~G. and {Martinelli}, M. and {Mart{\'\i}nez-Gonz{\'a}lez}, E. and {Masi}, S. and {Matarrese}, S. and {McGehee}, P. and {Meinhold}, P.~R. and {Melchiorri}, A. and {Melin}, J. -B. and {Mendes}, L. and {Mennella}, A. and {Migliaccio}, M. and {Millea}, M. and {Mitra}, S. and {Miville-Desch{\^e}nes}, M. -A. and {Moneti}, A. and {Montier}, L. and {Morgante}, G. and {Mortlock}, D. and {Moss}, A. and {Munshi}, D. and {Murphy}, J.~A. and {Naselsky}, P. and {Nati}, F. and {Natoli}, P. and {Netterfield}, C.~B. and {N{\o}rgaard-Nielsen}, H.~U. and {Noviello}, F. and {Novikov}, D. and {Novikov}, I. and {Oxborrow}, C.~A. and {Paci}, F. and {Pagano}, L. and {Pajot}, F. and {Paladini}, R. and {Paoletti}, D. and {Partridge}, B. and {Pasian}, F. and {Patanchon}, G. and {Pearson}, T.~J. and {Perdereau}, O. and {Perotto}, L. and {Perrotta}, F. and {Pettorino}, V. and {Piacentini}, F. and {Piat}, M. and {Pierpaoli}, E. and {Pietrobon}, D. and {Plaszczynski}, S. and {Pointecouteau}, E. and {Polenta}, G. and {Popa}, L. and {Pratt}, G.~W. and {Pr{\'e}zeau}, G. and {Prunet}, S. and {Puget}, J. -L. and {Rachen}, J.~P. and {Reach}, W.~T. and {Rebolo}, R. and {Reinecke}, M. and {Remazeilles}, M. and {Renault}, C. and {Renzi}, A. and {Ristorcelli}, I. and {Rocha}, G. and {Rosset}, C. and {Rossetti}, M. and {Roudier}, G. and {Rouill{\'e} d'Orfeuil}, B. and {Rowan-Robinson}, M. and {Rubi{\~n}o-Mart{\'\i}n}, J.~A. and {Rusholme}, B. and {Said}, N. and {Salvatelli}, V. and {Salvati}, L. and {Sandri}, M. and {Santos}, D. and {Savelainen}, M. and {Savini}, G. and {Scott}, D. and {Seiffert}, M.~D. and {Serra}, P. and {Shellard}, E.~P.~S. and {Spencer}, L.~D. and {Spinelli}, M. and {Stolyarov}, V. and {Stompor}, R. and {Sudiwala}, R. and {Sunyaev}, R. and {Sutton}, D. and {Suur-Uski}, A. -S. and {Sygnet}, J. -F. and {Tauber}, J.~A. and {Terenzi}, L. and {Toffolatti}, L. and {Tomasi}, M. and {Tristram}, M. and {Trombetti}, T. and {Tucci}, M. and {Tuovinen}, J. and {T{\"u}rler}, M. and {Umana}, G. and {Valenziano}, L. and {Valiviita}, J. and {Van Tent}, F. and {Vielva}, P. and {Villa}, F. and {Wade}, L.~A. and {Wandelt}, B.~D. and {Wehus}, I.~K. and {White}, M. and {White}, S.~D.~M. and {Wilkinson}, A. and {Yvon}, D. and {Zacchei}, A. and {Zonca}, A.},
        title = "{Planck 2015 results. XIII. Cosmological parameters}",
      journal = {\aap},
         year = 2016,
        month = sep,
       volume = {594},
          eid = {A13},
        pages = {A13},
          doi = {10.1051/0004-6361/201525830},
archivePrefix = {arXiv},
       eprint = {1502.01589},
 primaryClass = {astro-ph.CO},
       adsurl = {https://ui.adsabs.harvard.edu/abs/2016A&A...594A..13P}
}

@ARTICLE{Kroupa_2001,
       author = {{Kroupa}, Pavel},
        title = "{On the variation of the initial mass function}",
      journal = {\mnras},
         year = 2001,
        month = apr,
       volume = {322},
       number = {2},
        pages = {231-246},
          doi = {10.1046/j.1365-8711.2001.04022.x},
archivePrefix = {arXiv},
       eprint = {astro-ph/0009005},
 primaryClass = {astro-ph},
       adsurl = {https://ui.adsabs.harvard.edu/abs/2001MNRAS.322..231K}
}

@ARTICLE{Leitherer_1999,
       author = {{Leitherer}, Claus and {Schaerer}, Daniel and {Goldader}, Jeffrey D. and {Delgado}, Rosa M. Gonz{\'a}lez and {Robert}, Carmelle and {Kune}, Denis Foo and {de Mello}, Du{\'\i}lia F. and {Devost}, Daniel and {Heckman}, Timothy M.},
        title = "{Starburst99: Synthesis Models for Galaxies with Active Star Formation}",
      journal = {\apjs},
         year = 1999,
        month = jul,
       volume = {123},
       number = {1},
        pages = {3-40},
          doi = {10.1086/313233},
archivePrefix = {arXiv},
       eprint = {astro-ph/9902334},
 primaryClass = {astro-ph},
       adsurl = {https://ui.adsabs.harvard.edu/abs/1999ApJS..123....3L}
}

@ARTICLE{Hahn_2011,
       author = {{Hahn}, Oliver and {Abel}, Tom},
        title = "{Multi-scale initial conditions for cosmological simulations}",
      journal = {\mnras},
         year = 2011,
        month = aug,
       volume = {415},
       number = {3},
        pages = {2101-2121},
          doi = {10.1111/j.1365-2966.2011.18820.x},
archivePrefix = {arXiv},
       eprint = {1103.6031},
 primaryClass = {astro-ph.CO},
       adsurl = {https://ui.adsabs.harvard.edu/abs/2011MNRAS.415.2101H}
}

@ARTICLE{Behroozi_2013,
       author = {{Behroozi}, Peter S. and {Wechsler}, Risa H. and {Wu}, Hao-Yi},
        title = "{The ROCKSTAR Phase-space Temporal Halo Finder and the Velocity Offsets of Cluster Cores}",
      journal = {\apj},
         year = 2013,
        month = jan,
       volume = {762},
       number = {2},
          eid = {109},
        pages = {109},
          doi = {10.1088/0004-637X/762/2/109},
archivePrefix = {arXiv},
       eprint = {1110.4372},
 primaryClass = {astro-ph.CO},
       adsurl = {https://ui.adsabs.harvard.edu/abs/2013ApJ...762..109B}
}

@ARTICLE{Gill_2004,
       author = {{Gill}, Stuart P.~D. and {Knebe}, Alexander and {Gibson}, Brad K.},
        title = "{The evolution of substructure - I. A new identification method}",
      journal = {\mnras},
         year = 2004,
        month = jun,
       volume = {351},
       number = {2},
        pages = {399-409},
          doi = {10.1111/j.1365-2966.2004.07786.x},
archivePrefix = {arXiv},
       eprint = {astro-ph/0404258},
 primaryClass = {astro-ph},
       adsurl = {https://ui.adsabs.harvard.edu/abs/2004MNRAS.351..399G}
}

@ARTICLE{Knollmann_2009,
       author = {{Knollmann}, Steffen R. and {Knebe}, Alexander},
        title = "{AHF: Amiga's Halo Finder}",
      journal = {\apjs},
         year = 2009,
        month = jun,
       volume = {182},
       number = {2},
        pages = {608-624},
          doi = {10.1088/0067-0049/182/2/608},
archivePrefix = {arXiv},
       eprint = {0904.3662},
 primaryClass = {astro-ph.CO},
       adsurl = {https://ui.adsabs.harvard.edu/abs/2009ApJS..182..608K}
}

@ARTICLE{cloudy_version_c25_release,
       author = {{Gunasekera}, C.~M. and {van Hoof}, P.~A.~M. and {Dehghanian}, M. and {Chakraborty}, P. and {Shaw}, G. and {Bianchi}, S. and {Chatzikos}, M. and {Tsujimoto}, M. and {Ferland}, G.~J.},
        title = "{The 2025 release of Cloudy}",
      journal = {\rmxaa},
         year = 2025,
        month = nov,
       volume = {61},
        pages = {120-133},
          doi = {10.22201/ia.01851101p.2025.61.03.01},
archivePrefix = {arXiv},
       eprint = {2508.01102},
 primaryClass = {astro-ph.GA},
       adsurl = {https://ui.adsabs.harvard.edu/abs/2025RMxAA..61c.120G}
}

@ARTICLE{Ferland_2017,
       author = {{Ferland}, G.~J. and {Chatzikos}, M. and {Guzm{\'a}n}, F. and {Lykins}, M.~L. and {van Hoof}, P.~A.~M. and {Williams}, R.~J.~R. and {Abel}, N.~P. and {Badnell}, N.~R. and {Keenan}, F.~P. and {Porter}, R.~L. and {Stancil}, P.~C.},
        title = "{The 2017 Release Cloudy}",
      journal = {\rmxaa},
         year = 2017,
        month = oct,
       volume = {53},
        pages = {385-438},
          doi = {10.48550/arXiv.1705.10877},
archivePrefix = {arXiv},
       eprint = {1705.10877},
 primaryClass = {astro-ph.GA},
       adsurl = {https://ui.adsabs.harvard.edu/abs/2017RMxAA..53..385F}
}

@software{GNUparallel,
      author       = {Tange, Ole},
      title        = {GNU Parallel 20230522 ('Charles')},
      month        = May,
      year         = 2023,
      note         = {{GNU Parallel is a general parallelizer to run
                       multiple serial command line programs in parallel
                       without changing them.}},
      publisher    = {Zenodo},
      doi          = {10.5281/zenodo.7958356},
      url          = {https://doi.org/10.5281/zenodo.7958356}
}

@ARTICLE{Cowie_songaila_1986_ismAbundance,
       author = {{Cowie}, Lennox L. and {Songaila}, Antoinette},
        title = "{High-resolution optical and ultraviolet absorption-line studies of interstellar gas.}",
      journal = {\araa},
         year = 1986,
        month = jan,
       volume = {24},
        pages = {499-535},
          doi = {10.1146/annurev.aa.24.090186.002435},
       adsurl = {https://ui.adsabs.harvard.edu/abs/1986ARA&A..24..499C}
}

@ARTICLE{Savage_sembach_1996_ismAbundance,
       author = {{Savage}, Blair D. and {Sembach}, Kenneth R.},
        title = "{Interstellar Abundances from Absorption-Line Observations with the Hubble Space Telescope}",
      journal = {\araa},
         year = 1996,
        month = jan,
       volume = {34},
        pages = {279-330},
          doi = {10.1146/annurev.astro.34.1.279},
       adsurl = {https://ui.adsabs.harvard.edu/abs/1996ARA&A..34..279S}
}

@ARTICLE{Abel_2005,
       author = {{Abel}, N.~P. and {Ferland}, G.~J. and {Shaw}, G. and {van Hoof}, P.~A.~M.},
        title = "{The H II Region/PDR Connection: Self-consistent Calculations of Physical Conditions in Star-forming Regions}",
      journal = {\apjs},
         year = 2005,
        month = nov,
       volume = {161},
       number = {1},
        pages = {65-95},
          doi = {10.1086/432913},
archivePrefix = {arXiv},
       eprint = {astro-ph/0506514},
 primaryClass = {astro-ph},
       adsurl = {https://ui.adsabs.harvard.edu/abs/2005ApJS..161...65A}
}

@ARTICLE{Schoier_2005,
       author = {{Sch{\"o}ier}, F.~L. and {van der Tak}, F.~F.~S. and {van Dishoeck}, E.~F. and {Black}, J.~H.},
        title = "{An atomic and molecular database for analysis of submillimetre line observations}",
      journal = {\aap},
         year = 2005,
        month = mar,
       volume = {432},
       number = {1},
        pages = {369-379},
          doi = {10.1051/0004-6361:20041729},
archivePrefix = {arXiv},
       eprint = {astro-ph/0411110},
 primaryClass = {astro-ph},
       adsurl = {https://ui.adsabs.harvard.edu/abs/2005A&A...432..369S}
}

@ARTICLE{Spitzer_1968,
       author = {{Spitzer}, Lyman, Jr. and {Tomasko}, Martin G.},
        title = "{Heating of H i Regions by Energetic Particles}",
      journal = {\apj},
         year = 1968,
        month = jun,
       volume = {152},
        pages = {971},
          doi = {10.1086/149610},
       adsurl = {https://ui.adsabs.harvard.edu/abs/1968ApJ...152..971S}
}

@ARTICLE{Ferland_1984,
       author = {{Ferland}, G.~J. and {Mushotzky}, R.~F.},
        title = "{Cosmic rays and the emission-line regions of active galactic nuclei.}",
      journal = {\apj},
         year = 1984,
        month = nov,
       volume = {286},
        pages = {42-52},
          doi = {10.1086/162574},
       adsurl = {https://ui.adsabs.harvard.edu/abs/1984ApJ...286...42F}
}

@ARTICLE{Ferland_2009,
       author = {{Ferland}, G.~J. and {Fabian}, A.~C. and {Hatch}, N.~A. and {Johnstone}, R.~M. and {Porter}, R.~L. and {van Hoof}, P.~A.~M. and {Williams}, R.~J.~R.},
        title = "{Collisional heating as the origin of filament emission in galaxy clusters}",
      journal = {\mnras},
         year = 2009,
        month = feb,
       volume = {392},
       number = {4},
        pages = {1475-1502},
          doi = {10.1111/j.1365-2966.2008.14153.x},
archivePrefix = {arXiv},
       eprint = {0810.5372},
 primaryClass = {astro-ph},
       adsurl = {https://ui.adsabs.harvard.edu/abs/2009MNRAS.392.1475F}
}

@ARTICLE{scipy_2020,
  author  = {Virtanen, Pauli and Gommers, Ralf and Oliphant, Travis E. and
            Haberland, Matt and Reddy, Tyler and Cournapeau, David and
            Burovski, Evgeni and Peterson, Pearu and Weckesser, Warren and
            Bright, Jonathan and {van der Walt}, St{\'e}fan J. and
            Brett, Matthew and Wilson, Joshua and Millman, K. Jarrod and
            Mayorov, Nikolay and Nelson, Andrew R. J. and Jones, Eric and
            Kern, Robert and Larson, Eric and Carey, C J and
            Polat, {\.I}lhan and Feng, Yu and Moore, Eric W. and
            {VanderPlas}, Jake and Laxalde, Denis and Perktold, Josef and
            Cimrman, Robert and Henriksen, Ian and Quintero, E. A. and
            Harris, Charles R. and Archibald, Anne M. and
            Ribeiro, Ant{\^o}nio H. and Pedregosa, Fabian and
            {van Mulbregt}, Paul and {SciPy 1.0 Contributors}},
  title   = {{{SciPy} 1.0: Fundamental Algorithms for Scientific
            Computing in Python}},
  journal = {Nature Methods},
  year    = {2020},
  volume  = {17},
  pages   = {261--272},
  adsurl  = {https://rdcu.be/b08Wh},
  doi     = {10.1038/s41592-019-0686-2},
}

@ARTICLE{simonLunjun_jasonGuachao_extendedCII,
       author = {{Liu}, Lun-Jun and {Sun}, Guochao and {Faisst}, Andreas L. and {Faucher-Gigu{\`e}re}, Claude-Andr{\'e} and {Lidz}, Adam},
        title = "{Widespread Extended [C II] Emission in High-redshift Galaxies: Insights from the FIRE-2 Cosmological Zoom-in Simulations}",
      journal = {\apj},
         year = 2026,
        month = aug,
       volume = {1007},
       number = {2},
          eid = {123},
        pages = {123},
          doi = {10.3847/1538-4357/ae8190},
archivePrefix = {arXiv},
       eprint = {2601.00959},
 primaryClass = {astro-ph.GA},
       adsurl = {https://ui.adsabs.harvard.edu/abs/2026ApJ..1007..123L}
}

@ARTICLE{Baes_2011,
       author = {{Baes}, Maarten and {Verstappen}, Joris and {De Looze}, Ilse and {Fritz}, Jacopo and {Saftly}, Waad and {Vidal P{\'e}rez}, Edgardo and {Stalevski}, Marko and {Valcke}, Sander},
        title = "{Efficient Three-dimensional NLTE Dust Radiative Transfer with SKIRT}",
      journal = {\apjs},
         year = 2011,
        month = oct,
       volume = {196},
       number = {2},
          eid = {22},
        pages = {22},
          doi = {10.1088/0067-0049/196/2/22},
archivePrefix = {arXiv},
       eprint = {1108.5056},
 primaryClass = {astro-ph.CO},
       adsurl = {https://ui.adsabs.harvard.edu/abs/2011ApJS..196...22B}
}

@ARTICLE{Baes_2015,
       author = {{Baes}, M. and {Camps}, P.},
        title = "{SKIRT: The design of a suite of input models for Monte Carlo radiative transfer simulations}",
      journal = {Astronomy and Computing},
         year = 2015,
        month = sep,
       volume = {12},
        pages = {33-44},
          doi = {10.1016/j.ascom.2015.05.006},
archivePrefix = {arXiv},
       eprint = {1505.07708},
 primaryClass = {astro-ph.IM},
       adsurl = {https://ui.adsabs.harvard.edu/abs/2015A&C....12...33B}
}

@ARTICLE{Camps_2015,
       author = {{Camps}, P. and {Baes}, M.},
        title = "{SKIRT: An advanced dust radiative transfer code with a user-friendly architecture}",
      journal = {Astronomy and Computing},
         year = 2015,
        month = mar,
       volume = {9},
        pages = {20-33},
          doi = {10.1016/j.ascom.2014.10.004},
archivePrefix = {arXiv},
       eprint = {1410.1629},
 primaryClass = {astro-ph.IM},
       adsurl = {https://ui.adsabs.harvard.edu/abs/2015A&C.....9...20C}
}

@ARTICLE{Camps_and_baes_2020_skirt9,
       author = {{Camps}, P. and {Baes}, M.},
        title = "{SKIRT 9: Redesigning an advanced dust radiative transfer code to allow kinematics, line transfer and polarization by aligned dust grains}",
      journal = {Astronomy and Computing},
         year = 2020,
        month = apr,
       volume = {31},
          eid = {100381},
        pages = {100381},
          doi = {10.1016/j.ascom.2020.100381},
archivePrefix = {arXiv},
       eprint = {2003.00721},
 primaryClass = {astro-ph.GA},
       adsurl = {https://ui.adsabs.harvard.edu/abs/2020A&C....3100381C}
}

@ARTICLE{Weingartner_2001,
       author = {{Weingartner}, Joseph C. and {Draine}, B.~T.},
        title = "{Photoelectric Emission from Interstellar Dust: Grain Charging and Gas Heating}",
      journal = {\apjs},
         year = 2001,
        month = jun,
       volume = {134},
       number = {2},
        pages = {263-281},
          doi = {10.1086/320852},
archivePrefix = {arXiv},
       eprint = {astro-ph/9907251},
 primaryClass = {astro-ph},
       adsurl = {https://ui.adsabs.harvard.edu/abs/2001ApJS..134..263W}
}

@ARTICLE{Draine_1978_ISRF_MW_value_habingUnits,
       author = {{Draine}, B.~T.},
        title = "{Photoelectric heating of interstellar gas.}",
      journal = {\apjs},
         year = 1978,
        month = apr,
       volume = {36},
        pages = {595-619},
          doi = {10.1086/190513},
       adsurl = {https://ui.adsabs.harvard.edu/abs/1978ApJS...36..595D}
}

@ARTICLE{Tielens_1994,
       author = {{Tielens}, A.~G.~G.~M. and {McKee}, C.~F. and {Seab}, C.~G. and {Hollenbach}, D.~J.},
        title = "{The Physics of Grain-Grain Collisions and Gas-Grain Sputtering in Interstellar Shocks}",
      journal = {\apj},
         year = 1994,
        month = aug,
       volume = {431},
        pages = {321},
          doi = {10.1086/174488},
       adsurl = {https://ui.adsabs.harvard.edu/abs/1994ApJ...431..321T}
}

@ARTICLE{Miville-Deschenes_2017_MW_paper,
       author = {{Miville-Desch{\^e}nes}, Marc-Antoine and {Murray}, Norman and {Lee}, Eve J.},
        title = "{Physical Properties of Molecular Clouds for the Entire Milky Way Disk}",
      journal = {\apj},
         year = 2017,
        month = jan,
       volume = {834},
       number = {1},
          eid = {57},
        pages = {57},
          doi = {10.3847/1538-4357/834/1/57},
archivePrefix = {arXiv},
       eprint = {1610.05918},
 primaryClass = {astro-ph.GA},
       adsurl = {https://ui.adsabs.harvard.edu/abs/2017ApJ...834...57M}
}

@ARTICLE{Dame2001_COdata_for_MivilleDeschenes,
       author = {{Dame}, T.~M. and {Hartmann}, Dap and {Thaddeus}, P.},
        title = "{The Milky Way in Molecular Clouds: A New Complete CO Survey}",
      journal = {\apj},
         year = 2001,
        month = feb,
       volume = {547},
       number = {2},
        pages = {792-813},
          doi = {10.1086/318388},
archivePrefix = {arXiv},
       eprint = {astro-ph/0009217},
 primaryClass = {astro-ph},
       adsurl = {https://ui.adsabs.harvard.edu/abs/2001ApJ...547..792D}
}

@ARTICLE{Benincasa2020_GMCs_FIRE2,
       author = {{Benincasa}, Samantha M. and {Loebman}, Sarah R. and {Wetzel}, Andrew and {Hopkins}, Philip F. and {Murray}, Norman and {Bellardini}, Matthew A. and {Faucher-Gigu{\`e}re}, Claude-Andr{\'e} and {Guszejnov}, D{\'a}vid and {Orr}, Matthew},
        title = "{Live fast, die young: GMC lifetimes in the FIRE cosmological simulations of Milky Way mass galaxies}",
      journal = {\mnras},
         year = 2020,
        month = sep,
       volume = {497},
       number = {3},
        pages = {3993-3999},
          doi = {10.1093/mnras/staa2116},
archivePrefix = {arXiv},
       eprint = {1911.05251},
 primaryClass = {astro-ph.GA},
       adsurl = {https://ui.adsabs.harvard.edu/abs/2020MNRAS.497.3993B}
}

@ARTICLE{Guszejnov2020_GMCs_FIRE,
       author = {{Guszejnov}, D{\'a}vid and {Grudi{\'c}}, Michael Y. and {Offner}, Stella S.~R. and {Boylan-Kolchin}, Michael and {Faucher-Gigu{\`e}re}, Claude-Andr{\'e} and {Wetzel}, Andrew and {Benincasa}, Samantha M. and {Loebman}, Sarah},
        title = "{Evolution of giant molecular clouds across cosmic time}",
      journal = {\mnras},
         year = 2020,
        month = feb,
       volume = {492},
       number = {1},
        pages = {488-502},
          doi = {10.1093/mnras/stz3527},
archivePrefix = {arXiv},
       eprint = {1910.01163},
 primaryClass = {astro-ph.GA},
       adsurl = {https://ui.adsabs.harvard.edu/abs/2020MNRAS.492..488G}
}

@ARTICLE{Colombo2014_M51_GMC_properties,
       author = {{Colombo}, Dario and {Hughes}, Annie and {Schinnerer}, Eva and {Meidt}, Sharon E. and {Leroy}, Adam K. and {Pety}, J{\'e}r{\^o}me and {Dobbs}, Clare L. and {Garc{\'\i}a-Burillo}, Santiago and {Dumas}, Ga{\"e}lle and {Thompson}, Todd A. and {Schuster}, Karl F. and {Kramer}, Carsten},
        title = "{The PdBI Arcsecond Whirlpool Survey (PAWS): Environmental Dependence of Giant Molecular Cloud Properties in M51}",
      journal = {\apj},
         year = 2014,
        month = mar,
       volume = {784},
       number = {1},
          eid = {3},
        pages = {3},
          doi = {10.1088/0004-637X/784/1/3},
archivePrefix = {arXiv},
       eprint = {1401.1505},
 primaryClass = {astro-ph.GA},
       adsurl = {https://ui.adsabs.harvard.edu/abs/2014ApJ...784....3C}
}

@ARTICLE{BrandBlitz1993_velocityFields_MivilleDeschenesData,
       author = {{Brand}, J. and {Blitz}, L.},
        title = "{The velocity field of the outer galaxy.}",
      journal = {\aap},
         year = 1993,
        month = aug,
       volume = {275},
        pages = {67-90},
       adsurl = {https://ui.adsabs.harvard.edu/abs/1993A&A...275...67B}
}

@ARTICLE{rudolph_2006_MW_metallicity_profile,
       author = {{Rudolph}, Alexander L. and {Fich}, Michel and {Bell}, Gwendolyn R. and {Norsen}, Travis and {Simpson}, Janet P. and {Haas}, Michael R. and {Erickson}, Edwin F.},
        title = "{Abundance Gradients in the Galaxy}",
      journal = {\apjs},
         year = 2006,
        month = feb,
       volume = {162},
       number = {2},
        pages = {346-374},
          doi = {10.1086/498869},
       adsurl = {https://ui.adsabs.harvard.edu/abs/2006ApJS..162..346R}
}

@ARTICLE{asplund_2009_Mw_metallicity,
       author = {{Asplund}, Martin and {Grevesse}, Nicolas and {Sauval}, A. Jacques and {Scott}, Pat},
        title = "{The Chemical Composition of the Sun}",
      journal = {\araa},
         year = 2009,
        month = sep,
       volume = {47},
       number = {1},
        pages = {481-522},
          doi = {10.1146/annurev.astro.46.060407.145222},
archivePrefix = {arXiv},
       eprint = {0909.0948},
 primaryClass = {astro-ph.SR},
       adsurl = {https://ui.adsabs.harvard.edu/abs/2009ARA&A..47..481A}
}

@article{Yue_cii_LIM,
    author = {Yue, B. and Ferrara, A. and Pallottini, A. and Gallerani, S. and Vallini, L.},
    title = {Intensity mapping of [C ii] emission from early galaxies},
    journal = {Monthly Notices of the Royal Astronomical Society},
    volume = {450},
    number = {4},
    pages = {3829-3839},
    year = {2015},
    month = {07},
    issn = {0035-8711},
    doi = {10.1093/mnras/stv933},
    url = {https://doi.org/10.1093/mnras/stv933},
    eprint = {https://academic.oup.com/mnras/article-pdf/450/4/3829/5838660/stv933.pdf},
}

@ARTICLE{ccat_2023,
       author = {{CCAT-Prime Collaboration} and {Aravena}, Manuel and {Austermann}, Jason E. and {Basu}, Kaustuv and {Battaglia}, Nicholas and {Beringue}, Benjamin and {Bertoldi}, Frank and {Bigiel}, Frank and {Bond}, J. Richard and {Breysse}, Patrick C. and {Broughton}, Colton and {Bustos}, Ricardo and {Chapman}, Scott C. and {Charmetant}, Maude and {Choi}, Steve K. and {Chung}, Dongwoo T. and {Clark}, Susan E. and {Cothard}, Nicholas F. and {Crites}, Abigail T. and {Dev}, Ankur and {Douglas}, Kaela and {Duell}, Cody J. and {D{\"u}nner}, Rolando and {Ebina}, Haruki and {Erler}, Jens and {Fich}, Michel and {Fissel}, Laura M. and {Foreman}, Simon and {Freundt}, R.~G. and {Gallardo}, Patricio A. and {Gao}, Jiansong and {Garc{\'\i}a}, Pablo and {Giovanelli}, Riccardo and {Golec}, Joseph E. and {Groppi}, Christopher E. and {Haynes}, Martha P. and {Henke}, Douglas and {Hensley}, Brandon and {Herter}, Terry and {Higgins}, Ronan and {Hlo{\v{z}}ek}, Ren{\'e}e and {Huber}, Anthony and {Huber}, Zachary and {Hubmayr}, Johannes and {Jackson}, Rebecca and {Johnstone}, Douglas and {Karoumpis}, Christos and {Keating}, Laura C. and {Komatsu}, Eiichiro and {Li}, Yaqiong and {Magnelli}, Benjamin and {Matthews}, Brenda C. and {Mauskopf}, Philip D. and {McMahon}, Jeffrey J. and {Meerburg}, P. Daniel and {Meyers}, Joel and {Muralidhara}, Vyoma and {Murray}, Norman W. and {Niemack}, Michael D. and {Nikola}, Thomas and {Okada}, Yoko and {Puddu}, Roberto and {Riechers}, Dominik A. and {Rosolowsky}, Erik and {Rossi}, Kayla and {Rotermund}, Kaja and {Roy}, Anirban and {Sadavoy}, Sarah I. and {Schaaf}, Reinhold and {Schilke}, Peter and {Scott}, Douglas and {Simon}, Robert and {Sinclair}, Adrian K. and {Sivakoff}, Gregory R. and {Stacey}, Gordon J. and {Stutz}, Amelia M. and {Stutzki}, Juergen and {Tahani}, Mehrnoosh and {Thanjavur}, Karun and {Timmermann}, Ralf A. and {Ullom}, Joel N. and {van Engelen}, Alexander and {Vavagiakis}, Eve M. and {Vissers}, Michael R. and {Wheeler}, Jordan D. and {White}, Simon D.~M. and {Zhu}, Yijie and {Zou}, Bugao},
        title = "{CCAT-prime Collaboration: Science Goals and Forecasts with Prime-Cam on the Fred Young Submillimeter Telescope}",
      journal = {\apjs},
         year = 2023,
        month = jan,
       volume = {264},
       number = {1},
          eid = {7},
        pages = {7},
          doi = {10.3847/1538-4365/ac9838},
archivePrefix = {arXiv},
       eprint = {2107.10364},
 primaryClass = {astro-ph.CO},
       adsurl = {https://ui.adsabs.harvard.edu/abs/2023ApJS..264....7C}
}

@ARTICLE{Silva_2018_ha_lim,
       author = {{Silva}, B. Marta and {Zaroubi}, Saleem and {Kooistra}, Robin and {Cooray}, Asantha},
        title = "{Tomographic intensity mapping versus galaxy surveys: observing the Universe in H {\ensuremath{\alpha}} emission with new generation instruments}",
      journal = {\mnras},
         year = 2018,
        month = apr,
       volume = {475},
       number = {2},
        pages = {1587-1608},
          doi = {10.1093/mnras/stx3265},
archivePrefix = {arXiv},
       eprint = {1711.09902},
 primaryClass = {astro-ph.GA},
       adsurl = {https://ui.adsabs.harvard.edu/abs/2018MNRAS.475.1587S}
}

@ARTICLE{spherex_2026,
       author = {{Bock}, James J. and {Aboobaker}, Asad M. and {Adamo}, Joseph and {Akeson}, Rachel and {Alred}, John M. and {Alibay}, Farah and {Ashby}, Matthew L.~N. and {Bach}, Yoonsoo P. and {Bleem}, Lindsey E. and {Bolton}, Douglas and {Braun}, David F. and {Bruton}, Sean and {Bryan}, Sean A. and {Chang}, Tzu-Ching and {Chen}, Shuang-Shuang and {Cheng}, Yun-Ting and {Cheshire}, IV, James R. and {Chiang}, Yi-Kuan and {Choppin de Janvry}, Jean and {Condon}, Samuel and {Cook}, Walter R. and {Cooray}, Asantha and {Crill}, Brendan P. and {Cukierman}, Ari J. and {Dor{\'e}}, Olivier and {Dowell}, C. Darren and {Dubois-Felsmann}, Gregory P. and {Eifler}, Tim and {Everett}, Spencer and {Fabinsky}, Beth E. and {Faisst}, Andreas L. and {Fanson}, James L. and {Farrington}, Allen H. and {Fatahi}, Tamim and {Fazar}, Candice M. and {Feder}, Richard M. and {Frater}, Eric H. and {Grasshorn Gebhardt}, Henry S. and {Giri}, Utkarsh and {Goldina}, Tatiana and {Gorjian}, Varoujan and {Habib}, Salman and {Hart}, William G. and {Heinrich}, Chen and {Hora}, Joseph L. and {Huai}, Zhaoyu and {Hui}, Howard and {Jo}, Young-Soo and {Jeong}, Woong-Seob and {Kang}, Jae Hwan and {Kang}, Miju and {Kecman}, Branislav and {Kim}, Chul-Hwan and {Kim}, Jaeyeong and {Kim}, Minjin and {Kim}, Young-Jun and {Kim}, Yongjung and {Kirkpatrick}, J. Davy and {Kobayashi}, Yosuke and {Korngut}, Phil M. and {Krause}, Elisabeth and {Lee}, Bomee and {Lee}, Ho-Gyu and {Lee}, Jae-Joon and {Lee}, Jeong-Eun and {Lisse}, Carey M. and {Mariani}, Giacomo and {Masters}, Daniel C. and {Mauskopf}, Philip D. and {Melnick}, Gary J. and {Minasyan}, Mary H. and {Mirocha}, Jordan and {Miyasaka}, Hiromasa and {Moore}, Anne and {Moore}, Bradley D. and {Murgia}, Giulia and {Naylor}, Bret J. and {Nelson}, Christina and {Nguyen}, Chi H. and {Nguyen}, Hien T. and {Noh}, Jinyoung K. and {Padin}, Stephen and {Paladini}, Roberta and {Park}, Sung-Joon and {Penanen}, Konstantin I. and {Putnam}, Dustin S. and {Pyo}, Jeonghyun and {Ramachandra}, Nesar and {Ramanathan}, Keshav and {Rustamkulov}, Zafar and {Reiley}, Daniel J. and {Rice}, Eric B. and {Rocca}, Jennifer M. and {Seok}, Ji Yeon and {Smith}, Roger and {Stober}, Jeremy and {Susca}, Sara and {Teplitz}, Harry I. and {Thelen}, Michael P. and {Tolls}, Volker and {Torrini}, Gabriela and {Trangsrud}, Amy R. and {Unwin}, Stephen and {Velicheti}, Phani and {Wang}, Pao-Yu and {Wen}, Robin Y. and {Werner}, Michael W. and {Williams}, Abby E. and {Williamson}, Ross and {Wincentsen}, James and {Windhorst}, Rogier A. and {Yang}, Soung-Chul and {Yang}, Yujin and {Zemcov}, Michael},
        title = "{The SPHEREx Satellite Mission}",
      journal = {\apj},
         year = 2026,
        month = mar,
       volume = {999},
       number = {1},
          eid = {139},
        pages = {139},
          doi = {10.3847/1538-4357/ae2be2},
archivePrefix = {arXiv},
       eprint = {2511.02985},
 primaryClass = {astro-ph.IM},
       adsurl = {https://ui.adsabs.harvard.edu/abs/2026ApJ...999..139B}
}

@ARTICLE{nate_2025,
       author = {{Carlson}, Nathan J. and {Bond}, J. Richard and {Chung}, Dongwoo T. and {Horlaville}, Patrick and {Morrison}, Thomas},
        title = "{The $\mathtt{WebSky}$ $\mathrm{[CII]}$ Forecasts and the search for primordial intermittent non-Gaussianity}",
      journal = {arXiv e-prints},
         year = 2025,
        month = oct,
          eid = {arXiv:2510.18312},
        pages = {arXiv:2510.18312},
          doi = {10.48550/arXiv.2510.18312},
archivePrefix = {arXiv},
       eprint = {2510.18312},
 primaryClass = {astro-ph.CO},
       adsurl = {https://ui.adsabs.harvard.edu/abs/2025arXiv251018312C}
}

@ARTICLE{horlaville2024_cii_peakPatch,
       author = {{Horlaville}, Patrick and {Chung}, Dongwoo T. and {Bond}, J. Richard and {Liang}, Lichen},
        title = "{The informativeness of [C II] line-intensity mapping as a probe of the H I content and metallicity of galaxies at the end of reionization}",
      journal = {\mnras},
         year = 2024,
        month = jul,
       volume = {531},
       number = {3},
        pages = {2958-2975},
          doi = {10.1093/mnras/stae1333},
archivePrefix = {arXiv},
       eprint = {2309.15733},
 primaryClass = {astro-ph.CO},
       adsurl = {https://ui.adsabs.harvard.edu/abs/2024MNRAS.531.2958H}
}

@ARTICLE{Kannan2022_Thesan_LIMPredictions_multiTracer,
       author = {{Kannan}, Rahul and {Smith}, Aaron and {Garaldi}, Enrico and {Shen}, Xuejian and {Vogelsberger}, Mark and {Pakmor}, R{\"u}diger and {Springel}, Volker and {Hernquist}, Lars},
        title = "{The THESAN project: predictions for multitracer line intensity mapping in the epoch of reionization}",
      journal = {\mnras},
         year = 2022,
        month = aug,
       volume = {514},
       number = {3},
        pages = {3857-3878},
          doi = {10.1093/mnras/stac1557},
archivePrefix = {arXiv},
       eprint = {2111.02411},
 primaryClass = {astro-ph.CO},
       adsurl = {https://ui.adsabs.harvard.edu/abs/2022MNRAS.514.3857K}
}

@ARTICLE{Stein_bond_webskyProject,
       author = {{Stein}, George and {Alvarez}, Marcelo A. and {Bond}, J. Richard and {van Engelen}, Alexander and {Battaglia}, Nicholas},
        title = "{The Websky extragalactic CMB simulations}",
      journal = {\jcap},
         year = 2020,
        month = oct,
       volume = {2020},
       number = {10},
          eid = {012},
        pages = {012},
          doi = {10.1088/1475-7516/2020/10/012},
archivePrefix = {arXiv},
       eprint = {2001.08787},
 primaryClass = {astro-ph.CO},
       adsurl = {https://ui.adsabs.harvard.edu/abs/2020JCAP...10..012S}
}

@ARTICLE{dongwoo_2024_COMAP_season2,
       author = {{Chung}, D.~T. and {Breysse}, P.~C. and {Cleary}, K.~A. and {Dunne}, D.~A. and {Lunde}, J.~G.~S. and {Padmanabhan}, H. and {Stutzer}, N.-O. and {Tolgay}, D. and {Bond}, J.~R. and {Church}, S.~E. and {Eriksen}, H.~K. and {Gaier}, T. and {Gundersen}, J.~O. and {Harper}, S.~E. and {Harris}, A.~I. and {Hobbs}, R. and {Ihle}, H.~T. and {Kim}, J. and {Lamb}, J.~W. and {Lawrence}, C.~R. and {Murray}, N. and {Pearson}, T.~J. and {Philip}, L. and {Readhead}, A.~C.~S. and {Rennie}, T.~J. and {Wehus}, I.~K. and {Woody}, D.~P. and {COMAP Collaboration}},
        title = "{COMAP Pathfinder {\textendash} Season 2 results: III. Implications for cosmic molecular gas content at z{\ensuremath{\sim}}3}",
      journal = {\aap},
         year = 2024,
        month = nov,
       volume = {691},
          eid = {A337},
        pages = {A337},
          doi = {10.1051/0004-6361/202451122},
archivePrefix = {arXiv},
       eprint = {2406.07512},
 primaryClass = {astro-ph.CO},
       adsurl = {https://ui.adsabs.harvard.edu/abs/2024A&A...691A.337C}
}

@ARTICLE{Cleary_2022_comap_earlyScience1,
       author = {{Cleary}, Kieran A. and {Borowska}, Jowita and {Breysse}, Patrick C. and {Catha}, Morgan and {Chung}, Dongwoo T. and {Church}, Sarah E. and {Dickinson}, Clive and {Eriksen}, Hans Kristian and {Foss}, Marie Kristine and {Gundersen}, Joshua Ott and {Harper}, Stuart E. and {Harris}, Andrew I. and {Hobbs}, Richard and {Ihle}, H{\r{a}}vard T. and {Kim}, Junhan and {Kocz}, Jonathon and {Lamb}, James W. and {Lunde}, Jonas G.~S. and {Padmanabhan}, Hamsa and {Pearson}, Timothy J. and {Philip}, Liju and {Powell}, Travis W. and {Rasmussen}, Maren and {Readhead}, Anthony C.~S. and {Rennie}, Thomas J. and {Silva}, Marta B. and {Stutzer}, Nils-Ole and {Uzgil}, Bade D. and {Watts}, Duncan J. and {Wehus}, Ingunn Kathrine and {Woody}, David P. and {Basoalto}, Lilian and {Bond}, J. Richard and {Dunne}, Delaney A. and {Gaier}, Todd and {Hensley}, Brandon and {Keating}, Laura C. and {Lawrence}, Charles R. and {Murray}, Norman and {Paladini}, Roberta and {Reeves}, Rodrigo and {Viero}, Marco P. and {Wechsler}, Risa H. and {Comap Collaboration}},
        title = "{COMAP Early Science. I. Overview}",
      journal = {\apj},
         year = 2022,
        month = jul,
       volume = {933},
       number = {2},
          eid = {182},
        pages = {182},
          doi = {10.3847/1538-4357/ac63cc},
archivePrefix = {arXiv},
       eprint = {2111.05927},
 primaryClass = {astro-ph.CO},
       adsurl = {https://ui.adsabs.harvard.edu/abs/2022ApJ...933..182C}
}

@ARTICLE{dongwoo_2022_comap_earlyscience5,
       author = {{Chung}, Dongwoo T. and {Breysse}, Patrick C. and {Cleary}, Kieran A. and {Ihle}, H{\r{a}}vard T. and {Padmanabhan}, Hamsa and {Silva}, Marta B. and {Richard Bond}, J. and {Borowska}, Jowita and {Catha}, Morgan and {Church}, Sarah E. and {Dunne}, Delaney A. and {Kristian Eriksen}, Hans and {Kristine Foss}, Marie and {Gaier}, Todd and {Ott Gundersen}, Joshua and {Harper}, Stuart E. and {Harris}, Andrew I. and {Hensley}, Brandon and {Hobbs}, Richard and {Keating}, Laura C. and {Kim}, Junhan and {Lamb}, James W. and {Lawrence}, Charles R. and {Gahr Sturtzel Lunde}, Jonas and {Murray}, Norman and {Pearson}, Timothy J. and {Philip}, Liju and {Rasmussen}, Maren and {Readhead}, Anthony C.~S. and {Rennie}, Thomas J. and {Stutzer}, Nils-Ole and {Uzgil}, Bade D. and {Viero}, Marco P. and {Watts}, Duncan J. and {Wechsler}, Risa H. and {Kathrine Wehus}, Ingunn and {Woody}, David P. and {Comap Collaboration}},
        title = "{COMAP Early Science. V. Constraints and Forecasts at z   3}",
      journal = {\apj},
         year = 2022,
        month = jul,
       volume = {933},
       number = {2},
          eid = {186},
        pages = {186},
          doi = {10.3847/1538-4357/ac63c7},
archivePrefix = {arXiv},
       eprint = {2111.05931},
 primaryClass = {astro-ph.CO},
       adsurl = {https://ui.adsabs.harvard.edu/abs/2022ApJ...933..186C}
}

@ARTICLE{Breysse_2016_highredshift_COmocks,
       author = {{Breysse}, Patrick C. and {Kovetz}, Ely D. and {Kamionkowski}, Marc},
        title = "{The high-redshift star formation history from carbon-monoxide intensity maps}",
      journal = {\mnras},
         year = 2016,
        month = mar,
       volume = {457},
       number = {1},
        pages = {L127-L131},
          doi = {10.1093/mnrasl/slw005},
archivePrefix = {arXiv},
       eprint = {1507.06304},
 primaryClass = {astro-ph.CO},
       adsurl = {https://ui.adsabs.harvard.edu/abs/2016MNRAS.457L.127B}
}

@ARTICLE{Kovetz_2019,
       author = {{Kovetz}, Ely and {Breysse}, Patrick C. and {Lidz}, Adam and {Bock}, Jamie and {Bradford}, Charles M. and {Chang}, Tzu-Ching and {Foreman}, Simon and {Padmanabhan}, Hamsa and {Pullen}, Anthony and {Riechers}, Dominik and {Silva}, Marta B. and {Switzer}, Eric},
        title = "{Astrophysics and Cosmology with Line-Intensity Mapping}",
      journal = {\baas},
         year = 2019,
        month = may,
       volume = {51},
       number = {3},
          eid = {101},
        pages = {101},
          doi = {10.48550/arXiv.1903.04496},
archivePrefix = {arXiv},
       eprint = {1903.04496},
 primaryClass = {astro-ph.CO},
       adsurl = {https://ui.adsabs.harvard.edu/abs/2019BAAS...51c.101K}
}

@ARTICLE{Bernal_2019,
       author = {{Bernal}, Jos{\'e} Luis and {Breysse}, Patrick C. and {Gil-Mar{\'\i}n}, H{\'e}ctor and {Kovetz}, Ely D.},
        title = "{User's guide to extracting cosmological information from line-intensity maps}",
      journal = {\prd},
         year = 2019,
        month = dec,
       volume = {100},
       number = {12},
          eid = {123522},
        pages = {123522},
          doi = {10.1103/PhysRevD.100.123522},
archivePrefix = {arXiv},
       eprint = {1907.10067},
 primaryClass = {astro-ph.CO},
       adsurl = {https://ui.adsabs.harvard.edu/abs/2019PhRvD.100l3522B}
}

@ARTICLE{Kovetz_2017,
       author = {{Kovetz}, Ely D. and {Viero}, Marco P. and {Lidz}, Adam and {Newburgh}, Laura and {Rahman}, Mubdi and {Switzer}, Eric and {Kamionkowski}, Marc and {Aguirre}, James and {Alvarez}, Marcelo and {Bock}, James and {Bond}, J. Richard and {Bower}, Goeffry and {Bradford}, C. Matt and {Breysse}, Patrick C. and {Bull}, Philip and {Chang}, Tzu-Ching and {Cheng}, Yun-Ting and {Chung}, Dongwoo and {Cleary}, Kieran and {Corray}, Asantha and {Crites}, Abigail and {Croft}, Rupert and {Dor{\'e}}, Olivier and {Eastwood}, Michael and {Ferrara}, Andrea and {Fonseca}, Jos{\'e} and {Jacobs}, Daniel and {Keating}, Garrett K. and {Lagache}, Guilaine and {Lakhlani}, Gunjan and {Liu}, Adrian and {Moodley}, Kavilan and {Murray}, Norm and {P{\'e}nin}, Aur{\'e}lie and {Popping}, Gerg{\"o} and {Pullen}, Anthony and {Reichers}, Dominik and {Saito}, Shun and {Saliwanchik}, Ben and {Santos}, Mario and {Somerville}, Rachel and {Stacey}, Gordon and {Stein}, George and {Villaescusa-Navarro}, Francesco and {Visbal}, Eli and {Weltman}, Amanda and {Wolz}, Laura and {Zemcov}, Micheal},
        title = "{Line-Intensity Mapping: 2017 Status Report}",
      journal = {arXiv e-prints},
         year = 2017,
        month = sep,
          eid = {arXiv:1709.09066},
        pages = {arXiv:1709.09066},
          doi = {10.48550/arXiv.1709.09066},
archivePrefix = {arXiv},
       eprint = {1709.09066},
 primaryClass = {astro-ph.CO},
       adsurl = {https://ui.adsabs.harvard.edu/abs/2017arXiv170909066K}
}

@ARTICLE{Chang_Lidz_2026_LIM,
       author = {{Chang}, Tzu-Ching and {Lidz}, Adam},
        title = "{Line-Intensity Mapping}",
      journal = {arXiv e-prints},
         year = 2026,
        month = feb,
          eid = {arXiv:2602.03011},
        pages = {arXiv:2602.03011},
          doi = {10.48550/arXiv.2602.03011},
archivePrefix = {arXiv},
       eprint = {2602.03011},
 primaryClass = {astro-ph.CO},
       adsurl = {https://ui.adsabs.harvard.edu/abs/2026arXiv260203011C}
}

@ARTICLE{Visbal2010_LIM,
       author = {{Visbal}, Eli and {Loeb}, Abraham},
        title = "{Measuring the 3D clustering of undetected galaxies through cross correlation of their cumulative flux fluctuations from multiple spectral lines}",
      journal = {\jcap},
         year = 2010,
        month = nov,
       volume = {2010},
       number = {11},
          eid = {016},
        pages = {016},
          doi = {10.1088/1475-7516/2010/11/016},
archivePrefix = {arXiv},
       eprint = {1008.3178},
 primaryClass = {astro-ph.CO},
       adsurl = {https://ui.adsabs.harvard.edu/abs/2010JCAP...11..016V}
}

@ARTICLE{Carilli_2013,
       author = {{Carilli}, C.~L. and {Walter}, F.},
        title = "{Cool Gas in High-Redshift Galaxies}",
      journal = {\araa},
         year = 2013,
        month = aug,
       volume = {51},
       number = {1},
        pages = {105-161},
          doi = {10.1146/annurev-astro-082812-140953},
archivePrefix = {arXiv},
       eprint = {1301.0371},
 primaryClass = {astro-ph.CO},
       adsurl = {https://ui.adsabs.harvard.edu/abs/2013ARA&A..51..105C}
}

@BOOK{Draine_2011,
       author = {{Draine}, Bruce T.},
        title = "{Physics of the Interstellar and Intergalactic Medium}",
         year = 2011,
    publisher = {Princeton University Press},
       adsurl = {https://ui.adsabs.harvard.edu/abs/2011piim.book.....D}
}

@ARTICLE{Habing1968_radiationFieldMeasurement,
       author = {{Habing}, H.~J.},
        title = "{The interstellar radiation density between 912 A and 2400 A}",
      journal = {\bain},
         year = 1968,
        month = jan,
       volume = {19},
        pages = {421},
       adsurl = {https://ui.adsabs.harvard.edu/abs/1968BAN....19..421H}
}

@MISC{Dullemond_2012_RADMC3D,
       author = {{Dullemond}, C.~P. and {Juhasz}, A. and {Pohl}, A. and {Sereshti}, F. and {Shetty}, R. and {Peters}, T. and {Commercon}, B. and {Flock}, M.},
        title = "{RADMC-3D: A multi-purpose radiative transfer tool}",
 howpublished = {Astrophysics Source Code Library, record ascl:1202.015},
         year = 2012,
        month = feb,
          eid = {ascl:1202.015},
        pages = {ascl:1202.015},
archivePrefix = {ascl},
       eprint = {1202.015},
       adsurl = {https://ui.adsabs.harvard.edu/abs/2012ascl.soft02015D}
}

@ARTICLE{Richings_2014_CHIMESa,
       author = {{Richings}, A.~J. and {Schaye}, J. and {Oppenheimer}, B.~D.},
        title = "{Non-equilibrium chemistry and cooling in the diffuse interstellar medium - I. Optically thin regime}",
      journal = {\mnras},
         year = 2014,
        month = jun,
       volume = {440},
       number = {4},
        pages = {3349-3369},
          doi = {10.1093/mnras/stu525},
archivePrefix = {arXiv},
       eprint = {1401.4719},
 primaryClass = {astro-ph.GA},
       adsurl = {https://ui.adsabs.harvard.edu/abs/2014MNRAS.440.3349R}
}

@ARTICLE{Richings_2014_CHIMESb,
       author = {{Richings}, A.~J. and {Schaye}, J. and {Oppenheimer}, B.~D.},
        title = "{Non-equilibrium chemistry and cooling in the diffuse interstellar medium - II. Shielded gas}",
      journal = {\mnras},
         year = 2014,
        month = aug,
       volume = {442},
       number = {3},
        pages = {2780-2796},
          doi = {10.1093/mnras/stu1046},
archivePrefix = {arXiv},
       eprint = {1403.6155},
 primaryClass = {astro-ph.GA},
       adsurl = {https://ui.adsabs.harvard.edu/abs/2014MNRAS.442.2780R}
}

@ARTICLE{Thompson2024_FIRE_CO,
       author = {{Thompson}, Oliver A. and {Richings}, Alexander J. and {Gibson}, Brad K. and {Faucher-Gigu{\`e}re}, Claude-Andr{\'e} and {Feldmann}, Robert and {Hayward}, Christopher C.},
        title = "{Predictions for CO emission and the CO-to-H$_{2}$ conversion factor in galaxy simulations with non-equilibrium chemistry}",
      journal = {\mnras},
         year = 2024,
        month = aug,
       volume = {532},
       number = {2},
        pages = {1948-1965},
          doi = {10.1093/mnras/stae1486},
archivePrefix = {arXiv},
       eprint = {2403.14325},
 primaryClass = {astro-ph.GA},
       adsurl = {https://ui.adsabs.harvard.edu/abs/2024MNRAS.532.1948T}
}

@ARTICLE{Saintonge_2017,
       author = {{Saintonge}, Am{\'e}lie and {Catinella}, Barbara and {Tacconi}, Linda J. and {Kauffmann}, Guinevere and {Genzel}, Reinhard and {Cortese}, Luca and {Dav{\'e}}, Romeel and {Fletcher}, Thomas J. and {Graci{\'a}-Carpio}, Javier and {Kramer}, Carsten and {Heckman}, Timothy M. and {Janowiecki}, Steven and {Lutz}, Katharina and {Rosario}, David and {Schiminovich}, David and {Schuster}, Karl and {Wang}, Jing and {Wuyts}, Stijn and {Borthakur}, Sanchayeeta and {Lamperti}, Isabella and {Roberts-Borsani}, Guido W.},
        title = "{xCOLD GASS: The Complete IRAM 30 m Legacy Survey of Molecular Gas for Galaxy Evolution Studies}",
      journal = {\apjs},
         year = 2017,
        month = dec,
       volume = {233},
       number = {2},
          eid = {22},
        pages = {22},
          doi = {10.3847/1538-4365/aa97e0},
archivePrefix = {arXiv},
       eprint = {1710.02157},
 primaryClass = {astro-ph.GA},
       adsurl = {https://ui.adsabs.harvard.edu/abs/2017ApJS..233...22S}
}

@ARTICLE{Cicone_2017_allsmog,
       author = {{Cicone}, C. and {Bothwell}, M. and {Wagg}, J. and {M{\o}ller}, P. and {De Breuck}, C. and {Zhang}, Z. and {Mart{\'\i}n}, S. and {Maiolino}, R. and {Severgnini}, P. and {Aravena}, M. and {Belfiore}, F. and {Espada}, D. and {Fl{\"u}tsch}, A. and {Impellizzeri}, V. and {Peng}, Y. and {Raj}, M.~A. and {Ram{\'\i}rez-Olivencia}, N. and {Riechers}, D. and {Schawinski}, K.},
        title = "{The final data release of ALLSMOG: a survey of CO in typical local low-M$_{{\ensuremath{*}}}$ star-forming galaxies}",
      journal = {\aap},
         year = 2017,
        month = aug,
       volume = {604},
          eid = {A53},
        pages = {A53},
          doi = {10.1051/0004-6361/201730605},
archivePrefix = {arXiv},
       eprint = {1705.05851},
 primaryClass = {astro-ph.GA},
       adsurl = {https://ui.adsabs.harvard.edu/abs/2017A&A...604A..53C}
}

@ARTICLE{Tacconi_2018,
       author = {{Tacconi}, L.~J. and {Genzel}, R. and {Saintonge}, A. and {Combes}, F. and {Garc{\'\i}a-Burillo}, S. and {Neri}, R. and {Bolatto}, A. and {Contini}, T. and {F{\"o}rster Schreiber}, N.~M. and {Lilly}, S. and {Lutz}, D. and {Wuyts}, S. and {Accurso}, G. and {Boissier}, J. and {Boone}, F. and {Bouch{\'e}}, N. and {Bournaud}, F. and {Burkert}, A. and {Carollo}, M. and {Cooper}, M. and {Cox}, P. and {Feruglio}, C. and {Freundlich}, J. and {Herrera-Camus}, R. and {Juneau}, S. and {Lippa}, M. and {Naab}, T. and {Renzini}, A. and {Salome}, P. and {Sternberg}, A. and {Tadaki}, K. and {{\"U}bler}, H. and {Walter}, F. and {Weiner}, B. and {Weiss}, A.},
        title = "{PHIBSS: Unified Scaling Relations of Gas Depletion Time and Molecular Gas Fractions}",
      journal = {\apj},
         year = 2018,
        month = feb,
       volume = {853},
       number = {2},
          eid = {179},
        pages = {179},
          doi = {10.3847/1538-4357/aaa4b4},
archivePrefix = {arXiv},
       eprint = {1702.01140},
 primaryClass = {astro-ph.GA},
       adsurl = {https://ui.adsabs.harvard.edu/abs/2018ApJ...853..179T}
}

@ARTICLE{Freundlich_2019,
       author = {{Freundlich}, J. and {Combes}, F. and {Tacconi}, L.~J. and {Genzel}, R. and {Garcia-Burillo}, S. and {Neri}, R. and {Contini}, T. and {Bolatto}, A. and {Lilly}, S. and {Salom{\'e}}, P. and {Bicalho}, I.~C. and {Boissier}, J. and {Boone}, F. and {Bouch{\'e}}, N. and {Bournaud}, F. and {Burkert}, A. and {Carollo}, M. and {Cooper}, M.~C. and {Cox}, P. and {Feruglio}, C. and {F{\"o}rster Schreiber}, N.~M. and {Juneau}, S. and {Lippa}, M. and {Lutz}, D. and {Naab}, T. and {Renzini}, A. and {Saintonge}, A. and {Sternberg}, A. and {Walter}, F. and {Weiner}, B. and {Wei{\ss}}, A. and {Wuyts}, S.},
        title = "{PHIBSS2: survey design and z = 0.5 - 0.8 results. Molecular gas reservoirs during the winding-down of star formation}",
      journal = {\aap},
         year = 2019,
        month = feb,
       volume = {622},
          eid = {A105},
        pages = {A105},
          doi = {10.1051/0004-6361/201732223},
archivePrefix = {arXiv},
       eprint = {1812.08180},
 primaryClass = {astro-ph.GA},
       adsurl = {https://ui.adsabs.harvard.edu/abs/2019A&A...622A.105F}
}

@ARTICLE{boogaard_2019_alma_aspecs,
       author = {{Boogaard}, Leindert A. and {Decarli}, Roberto and {Gonz{\'a}lez-L{\'o}pez}, Jorge and {van der Werf}, Paul and {Walter}, Fabian and {Bouwens}, Rychard and {Aravena}, Manuel and {Carilli}, Chris and {Bauer}, Franz Erik and {Brinchmann}, Jarle and {Contini}, Thierry and {Cox}, Pierre and {da Cunha}, Elisabete and {Daddi}, Emanuele and {D{\'\i}az-Santos}, Tanio and {Hodge}, Jacqueline and {Inami}, Hanae and {Ivison}, Rob and {Maseda}, Michael and {Matthee}, Jorryt and {Oesch}, Pascal and {Popping}, Gerg{\"o} and {Riechers}, Dominik and {Schaye}, Joop and {Schouws}, Sander and {Smail}, Ian and {Weiss}, Axel and {Wisotzki}, Lutz and {Bacon}, Roland and {Cortes}, Paulo C. and {Rix}, Hans-Walter and {Somerville}, Rachel S. and {Swinbank}, Mark and {Wagg}, Jeff},
        title = "{The ALMA Spectroscopic Survey in the HUDF: Nature and Physical Properties of Gas-mass Selected Galaxies Using MUSE Spectroscopy}",
      journal = {\apj},
         year = 2019,
        month = sep,
       volume = {882},
       number = {2},
          eid = {140},
        pages = {140},
          doi = {10.3847/1538-4357/ab3102},
archivePrefix = {arXiv},
       eprint = {1903.09167},
 primaryClass = {astro-ph.GA},
       adsurl = {https://ui.adsabs.harvard.edu/abs/2019ApJ...882..140B}
}

@ARTICLE{Bolatto_2013,
       author = {{Bolatto}, Alberto D. and {Wolfire}, Mark and {Leroy}, Adam K.},
        title = "{The CO-to-H$_{2}$ Conversion Factor}",
      journal = {\araa},
         year = 2013,
        month = aug,
       volume = {51},
       number = {1},
        pages = {207-268},
          doi = {10.1146/annurev-astro-082812-140944},
archivePrefix = {arXiv},
       eprint = {1301.3498},
 primaryClass = {astro-ph.GA},
       adsurl = {https://ui.adsabs.harvard.edu/abs/2013ARA&A..51..207B}
}

@ARTICLE{Madden_2020,
       author = {{Madden}, S.~C. and {Cormier}, D. and {Hony}, S. and {Lebouteiller}, V. and {Abel}, N. and {Galametz}, M. and {De Looze}, I. and {Chevance}, M. and {Polles}, F.~L. and {Lee}, M. -Y. and {Galliano}, F. and {Lambert-Huyghe}, A. and {Hu}, D. and {Ramambason}, L.},
        title = "{Tracing the total molecular gas in galaxies: [CII] and the CO-dark gas}",
      journal = {\aap},
         year = 2020,
        month = nov,
       volume = {643},
          eid = {A141},
        pages = {A141},
          doi = {10.1051/0004-6361/202038860},
archivePrefix = {arXiv},
       eprint = {2009.00649},
 primaryClass = {astro-ph.GA},
       adsurl = {https://ui.adsabs.harvard.edu/abs/2020A&A...643A.141M}
}

@ARTICLE{Feldmann2012_Xco_paperI,
       author = {{Feldmann}, Robert and {Gnedin}, Nickolay Y. and {Kravtsov}, Andrey V.},
        title = "{The X-factor in Galaxies. I. Dependence on Environment and Scale}",
      journal = {\apj},
         year = 2012,
        month = mar,
       volume = {747},
       number = {2},
          eid = {124},
        pages = {124},
          doi = {10.1088/0004-637X/747/2/124},
archivePrefix = {arXiv},
       eprint = {1112.1732},
 primaryClass = {astro-ph.CO},
       adsurl = {https://ui.adsabs.harvard.edu/abs/2012ApJ...747..124F}
}

@ARTICLE{Feldmann2012_Xco_paperII,
       author = {{Feldmann}, Robert and {Gnedin}, Nickolay Y. and {Kravtsov}, Andrey V.},
        title = "{The X-factor in Galaxies. II. The Molecular-hydrogen-Star-formation Relation}",
      journal = {\apj},
         year = 2012,
        month = oct,
       volume = {758},
       number = {2},
          eid = {127},
        pages = {127},
          doi = {10.1088/0004-637X/758/2/127},
archivePrefix = {arXiv},
       eprint = {1204.3910},
 primaryClass = {astro-ph.CO},
       adsurl = {https://ui.adsabs.harvard.edu/abs/2012ApJ...758..127F}
}

@ARTICLE{Lagos2012_Lco_simulatedGalaxies,
       author = {{Lagos}, Claudia del P. and {Bayet}, Estelle and {Baugh}, Carlton M. and {Lacey}, Cedric G. and {Bell}, Tom A. and {Fanidakis}, Nikolaos and {Geach}, James E.},
        title = "{Predictions for the CO emission of galaxies from a coupled simulation of galaxy formation and photon-dominated regions}",
      journal = {\mnras},
         year = 2012,
        month = nov,
       volume = {426},
       number = {3},
        pages = {2142-2165},
          doi = {10.1111/j.1365-2966.2012.21905.x},
archivePrefix = {arXiv},
       eprint = {1204.0795},
 primaryClass = {astro-ph.CO},
       adsurl = {https://ui.adsabs.harvard.edu/abs/2012MNRAS.426.2142L}
}

@ARTICLE{Blitz_Rosolowsky_2006_pressureH2_relation,
       author = {{Blitz}, Leo and {Rosolowsky}, Erik},
        title = "{The Role of Pressure in GMC Formation II: The H$_{2}$-Pressure Relation}",
      journal = {\apj},
         year = 2006,
        month = oct,
       volume = {650},
       number = {2},
        pages = {933-944},
          doi = {10.1086/505417},
archivePrefix = {arXiv},
       eprint = {astro-ph/0605035},
 primaryClass = {astro-ph},
       adsurl = {https://ui.adsabs.harvard.edu/abs/2006ApJ...650..933B}
}

@ARTICLE{Inoue2020_COUniverse,
       author = {{Inoue}, Shigeki and {Yoshida}, Naoki and {Yajima}, Hidenobu},
        title = "{The CO universe: modelling CO emission and H$_{2}$ abundance in cosmological galaxy formation simulations}",
      journal = {\mnras},
         year = 2020,
        month = nov,
       volume = {498},
       number = {4},
        pages = {5960-5971},
          doi = {10.1093/mnras/staa2744},
archivePrefix = {arXiv},
       eprint = {2008.12484},
 primaryClass = {astro-ph.GA},
       adsurl = {https://ui.adsabs.harvard.edu/abs/2020MNRAS.498.5960I}
}

@ARTICLE{Garcia2024_SLICK,
       author = {{Garcia}, Karolina and {Narayanan}, Desika and {Popping}, Gerg{\"o} and {Anirudh}, R. and {Sutherland}, Sagan and {Kaasinen}, Melanie},
        title = "{SLICK: Modeling a Universe of Molecular Line Luminosities in Hydrodynamical Simulations}",
      journal = {\apj},
         year = 2024,
        month = oct,
       volume = {974},
       number = {2},
          eid = {197},
        pages = {197},
          doi = {10.3847/1538-4357/ad642c},
archivePrefix = {arXiv},
       eprint = {2311.01508},
 primaryClass = {astro-ph.GA},
       adsurl = {https://ui.adsabs.harvard.edu/abs/2024ApJ...974..197G}
}

@ARTICLE{bisbas_2015_co_dark_clouds,
       author = {{Bisbas}, Thomas G. and {Zhang}, Zhi-Yu and {Kyrmanidou}, Maria-Christina and {Luo}, Gan and {Zhao}, Yinghe and {Topkaras}, Theodoros and {Jiang}, Xue-Jian and {Quan}, Donghui and {Li}, Di},
        title = "{Metallicity dependence of the CO-to-H$_{2}$ and the [CI]-to-H$_{2}$ conversion factors in galaxies}",
      journal = {\aap},
         year = 2025,
        month = may,
       volume = {697},
          eid = {A115},
        pages = {A115},
          doi = {10.1051/0004-6361/202554067},
archivePrefix = {arXiv},
       eprint = {2503.12073},
 primaryClass = {astro-ph.GA},
       adsurl = {https://ui.adsabs.harvard.edu/abs/2025A&A...697A.115B}
}

@ARTICLE{Liang_2023,
       author = {{Liang}, Lichen and {Feldmann}, Robert and {Murray}, Norman and {Narayanan}, Desika and {Hayward}, Christopher C. and {Angl{\'e}s-Alc{\'a}zar}, Daniel and {Bassini}, Luigi and {Richings}, Alexander J. and {Faucher-Gigu{\`e}re}, Claude-Andr{\'e} and {Chung}, Dongwoo T. and {Chan}, Jennifer Y.~H. and {Tolgay}, Do{\v{g}}a and {{\c{C}}atmabacak}, Onur and {Kere{\v{s}}}, Du{\v{s}}an and {Hopkins}, Philip F.},
        title = "{[C$_{II}$] 158 {\ensuremath{\mu}}m emission as an indicator of galaxy star formation rate}",
      journal = {\mnras},
         year = 2023,
        month = dec,
          doi = {10.1093/mnras/stad3792},
archivePrefix = {arXiv},
       eprint = {2301.04149},
 primaryClass = {astro-ph.GA},
       adsurl = {https://ui.adsabs.harvard.edu/abs/2023MNRAS.tmp.3653L}
}

@ARTICLE{Narayanan_2012,
       author = {{Narayanan}, Desika and {Krumholz}, Mark R. and {Ostriker}, Eve C. and {Hernquist}, Lars},
        title = "{A general model for the CO-H$_{2}$ conversion factor in galaxies with applications to the star formation law}",
      journal = {\mnras},
         year = 2012,
        month = apr,
       volume = {421},
       number = {4},
        pages = {3127-3146},
          doi = {10.1111/j.1365-2966.2012.20536.x},
archivePrefix = {arXiv},
       eprint = {1110.3791},
 primaryClass = {astro-ph.GA},
       adsurl = {https://ui.adsabs.harvard.edu/abs/2012MNRAS.421.3127N}
}

@ARTICLE{Accurso_2017,
       author = {{Accurso}, G. and {Saintonge}, A. and {Catinella}, B. and {Cortese}, L. and {Dav{\'e}}, R. and {Dunsheath}, S.~H. and {Genzel}, R. and {Gracia-Carpio}, J. and {Heckman}, T.~M. and {Jimmy} and {Kramer}, C. and {Li}, Cheng and {Lutz}, K. and {Schiminovich}, D. and {Schuster}, K. and {Sternberg}, A. and {Sturm}, E. and {Tacconi}, L.~J. and {Tran}, K.~V. and {Wang}, J.},
        title = "{Deriving a multivariate {\ensuremath{\alpha}}$_{CO}$ conversion function using the [C II]/CO (1-0) ratio and its application to molecular gas scaling relations}",
      journal = {\mnras},
         year = 2017,
        month = oct,
       volume = {470},
       number = {4},
        pages = {4750-4766},
          doi = {10.1093/mnras/stx1556},
archivePrefix = {arXiv},
       eprint = {1702.03888},
 primaryClass = {astro-ph.GA},
       adsurl = {https://ui.adsabs.harvard.edu/abs/2017MNRAS.470.4750A}
}

@ARTICLE{Keating_2020,
       author = {{Keating}, Laura C. and {Richings}, Alexander J. and {Murray}, Norman and {Faucher-Gigu{\`e}re}, Claude-Andr{\'e} and {Hopkins}, Philip F. and {Wetzel}, Andrew and {Kere{\v{s}}}, Du{\v{s}}an and {Benincasa}, Samantha and {Feldmann}, Robert and {Loebman}, Sarah and {Orr}, Matthew E.},
        title = "{Reproducing the CO-to-H$_{2}$ conversion factor in cosmological simulations of Milky-Way-mass galaxies}",
      journal = {\mnras},
         year = 2020,
        month = nov,
       volume = {499},
       number = {1},
        pages = {837-850},
          doi = {10.1093/mnras/staa2839},
archivePrefix = {arXiv},
       eprint = {2001.08197},
 primaryClass = {astro-ph.GA},
       adsurl = {https://ui.adsabs.harvard.edu/abs/2020MNRAS.499..837K}
}

@BOOK{Rybicki_Lightman_1986,
       author = {{Rybicki}, George B. and {Lightman}, Alan P.},
        title = "{Radiative Processes in Astrophysics}",
         year = 1986,
    publisher = {John Wiley \& Sons},
      address = {New York},
       adsurl = {https://ui.adsabs.harvard.edu/abs/1986rpa..book.....R}
}

@ARTICLE{SPHEREx_2014,
       author = {{Dor{\'e}}, Olivier and {Bock}, Jamie and {Ashby}, Matthew and {Capak}, Peter and {Cooray}, Asantha and {de Putter}, Roland and {Eifler}, Tim and {Flagey}, Nicolas and {Gong}, Yan and {Habib}, Salman and {Heitmann}, Katrin and {Hirata}, Chris and {Jeong}, Woong-Seob and {Katti}, Raj and {Korngut}, Phil and {Krause}, Elisabeth and {Lee}, Dae-Hee and {Masters}, Daniel and {Mauskopf}, Phil and {Melnick}, Gary and {Mennesson}, Bertrand and {Nguyen}, Hien and {{\"O}berg}, Karin and {Pullen}, Anthony and {Raccanelli}, Alvise and {Smith}, Roger and {Song}, Yong-Seon and {Tolls}, Volker and {Unwin}, Steve and {Venumadhav}, Tejaswi and {Viero}, Marco and {Werner}, Mike and {Zemcov}, Mike},
        title = "{Cosmology with the SPHEREX All-Sky Spectral Survey}",
      journal = {arXiv e-prints},
         year = 2014,
        month = dec,
          eid = {arXiv:1412.4872},
        pages = {arXiv:1412.4872},
          doi = {10.48550/arXiv.1412.4872},
archivePrefix = {arXiv},
       eprint = {1412.4872},
 primaryClass = {astro-ph.CO},
       adsurl = {https://ui.adsabs.harvard.edu/abs/2014arXiv1412.4872D}
}

@ARTICLE{Daddi_2015_COexcitationLadder,
       author = {{Daddi}, E. and {Dannerbauer}, H. and {Liu}, D. and {Aravena}, M. and {Bournaud}, F. and {Walter}, F. and {Riechers}, D. and {Magdis}, G. and {Sargent}, M. and {B{\'e}thermin}, M. and {Carilli}, C. and {Cibinel}, A. and {Dickinson}, M. and {Elbaz}, D. and {Gao}, Y. and {Gobat}, R. and {Hodge}, J. and {Krips}, M.},
        title = "{CO excitation of normal star-forming galaxies out to z = 1.5 as regulated by the properties of their interstellar medium}",
      journal = {\aap},
         year = 2015,
        month = may,
       volume = {577},
          eid = {A46},
        pages = {A46},
          doi = {10.1051/0004-6361/201425043},
archivePrefix = {arXiv},
       eprint = {1409.8158},
 primaryClass = {astro-ph.GA},
       adsurl = {https://ui.adsabs.harvard.edu/abs/2015A&A...577A..46D}
}

@ARTICLE{Goldreich_and_kwan_1974_turbulence,
       author = {{Goldreich}, Peter and {Kwan}, John},
        title = "{Molecular Clouds}",
      journal = {\apj},
         year = 1974,
        month = may,
       volume = {189},
        pages = {441-454},
          doi = {10.1086/152821},
       adsurl = {https://ui.adsabs.harvard.edu/abs/1974ApJ...189..441G}
}

@ARTICLE{Rachford_2002_Nco_Nh2,
       author = {{Rachford}, Brian L. and {Snow}, Theodore P. and {Tumlinson}, Jason and {Shull}, J. Michael and {Blair}, William P. and {Ferlet}, Roger and {Friedman}, Scott D. and {Gry}, Cecile and {Jenkins}, Edward B. and {Morton}, Donald C. and {Savage}, Blair D. and {Sonnentrucker}, Paule and {Vidal-Madjar}, Alfred and {Welty}, Daniel E. and {York}, Donald G.},
        title = "{A Far Ultraviolet Spectroscopic Explorer Survey of Interstellar Molecular Hydrogen in Translucent Clouds}",
      journal = {\apj},
         year = 2002,
        month = sep,
       volume = {577},
       number = {1},
        pages = {221-244},
          doi = {10.1086/342146},
archivePrefix = {arXiv},
       eprint = {astro-ph/0205415},
 primaryClass = {astro-ph},
       adsurl = {https://ui.adsabs.harvard.edu/abs/2002ApJ...577..221R}
}

@ARTICLE{Crenny_Federman_2004_Nco_Nh2,
       author = {{Crenny}, T. and {Federman}, S.~R.},
        title = "{Reanalysis of Copernicus Measurements of Interstellar Carbon Monoxide}",
      journal = {\apj},
         year = 2004,
        month = apr,
       volume = {605},
       number = {1},
        pages = {278-284},
          doi = {10.1086/382231},
archivePrefix = {arXiv},
       eprint = {astro-ph/0312650},
 primaryClass = {astro-ph},
       adsurl = {https://ui.adsabs.harvard.edu/abs/2004ApJ...605..278C}
}

@ARTICLE{Sheffer_2008_Nco_Nh2,
       author = {{Sheffer}, Y. and {Rogers}, M. and {Federman}, S.~R. and {Abel}, N.~P. and {Gredel}, R. and {Lambert}, D.~L. and {Shaw}, G.},
        title = "{Ultraviolet Survey of CO and H$_{2}$ in Diffuse Molecular Clouds: The Reflection of Two Photochemistry Regimes in Abundance Relationships}",
      journal = {\apj},
         year = 2008,
        month = nov,
       volume = {687},
       number = {2},
        pages = {1075-1106},
          doi = {10.1086/591484},
archivePrefix = {arXiv},
       eprint = {0807.0940},
 primaryClass = {astro-ph},
       adsurl = {https://ui.adsabs.harvard.edu/abs/2008ApJ...687.1075S}
}

@ARTICLE{Burgh_2010_Nco_Nh2,
       author = {{Burgh}, Eric B. and {France}, Kevin and {Jenkins}, Edward B.},
        title = "{Atomic and Molecular Carbon as a Tracer of Translucent Clouds}",
      journal = {\apj},
         year = 2010,
        month = jan,
       volume = {708},
       number = {1},
        pages = {334-341},
          doi = {10.1088/0004-637X/708/1/334},
archivePrefix = {arXiv},
       eprint = {0911.1812},
 primaryClass = {astro-ph.GA},
       adsurl = {https://ui.adsabs.harvard.edu/abs/2010ApJ...708..334B}
}

@ARTICLE{Grevesse2010_solarMetallicity,
       author = {{Grevesse}, N. and {Asplund}, M. and {Sauval}, A.~J. and {Scott}, P.},
        title = "{The chemical composition of the Sun}",
      journal = {\apss},
         year = 2010,
        month = jul,
       volume = {328},
       number = {1-2},
        pages = {179-183},
          doi = {10.1007/s10509-010-0288-z},
       adsurl = {https://ui.adsabs.harvard.edu/abs/2010Ap&SS.328..179G}
}
\bibliographystyle{aasjournal}

\end{document}